\documentclass[useAMS,usenatbib]{mn2e}
\usepackage{amsmath}
\usepackage{amssymb}

\usepackage{graphics}
\usepackage{graphicx}
\usepackage{epstopdf}
\usepackage{footnote}
\usepackage{tablefootnote}
\usepackage{txfonts}
\usepackage{lipsum}
\usepackage{float}
\usepackage[draft]{hyperref}
\usepackage{macros}
\usepackage{fixltx2e}

\makesavenoteenv{tabular}

\DeclareSymbolFont{cmletters}{OML}{cmm}{m}{it}
\DeclareMathSymbol{v}{\mathalpha}{cmletters}{"76}

\usepackage[usenames,dvipsnames,svgnames,table]{xcolor}
\usepackage{hyperref}
\definecolor{darkblue}{rgb}{0.0,0.0,0.3}
\hypersetup{colorlinks,breaklinks,
            linkcolor=darkblue,urlcolor=darkblue,
            anchorcolor=darkblue,citecolor=darkblue}

\newcommand{\rg}{r_\mathrm{g}}
\newcommand{\betaw}{\beta_{\rm w}}

\begin{document}
  
\title[]{Wind-Fed GRMHD Simulations of Sagittarius A*: Tilt and Alignment of Jets and Accretion Discs,  Electron Thermodynamics, and Multi-Scale Modeling of the Rotation Measure
 }
\author[S. M. Ressler, C. J. White, E. Quataert]{S. M. Ressler$^{1,2},$ C. J. White$^{3,4}$, E. Quataert$^{4}$\\
$^{1}$Canadian Institute for Theoretical Astrophysics, University of Toronto, Toronto, On, Canada M5S 3H8 \\
$^{2}$Kavli Institute for Theoretical Physics, University of California Santa Barbara, Santa Barbara, CA 93107 \\
$^{3}$Center for Computational Astrophysics, Flatiron Institute, Simons Foundation, New York, NY 10010 \\
$^{4}$Department of Astrophysical Sciences, Princeton University, Princeton, NJ 08544} 


\maketitle

\begin{abstract}
 Wind-fed models offer a unique way to form predictive models of the accretion flow surrounding Sagittarius A*.
We present 3D, wind-fed MHD and GRMHD simulations spanning the entire dynamic range of accretion from parsec scales to the event horizon.
We expand on previous work by including nonzero black hole spin and dynamically evolved electron thermodynamics.
Initial conditions for these simulations are generated from simulations of the observed Wolf-Rayet stellar winds in the Galactic Centre.
The resulting flow tends to be highly magnetized ($\beta \approx 2$) with an $\sim$ $r^{-1}$ density profile independent of the strength of magnetic fields in the winds.
Our simulations reach the MAD state for some, but not all cases.  
In tilted flows, SANE jets tend to align with the angular momentum of the gas at large scales, even if that direction is perpendicular to the black hole spin axis. 
Conversely, MAD jets tend to align with the black hole spin axis.
The gas angular momentum shows similar behavior: SANE flows tend to only partially align while MAD flows tend to fully align.
With a limited number of dynamical free parameters, our models can produce accretion rates, 230 GHz flux, and unresolved linear polarization fractions roughly consistent with observations for several choices of electron heating fraction. 
Absent another source of large-scale magnetic field, winds with a higher degree of magnetization (e.g., where the magnetic pressure is 1/100 of the ram pressure in the winds) may be required to get a sufficiently large RM with consistent sign.  
 
\end{abstract}


\section{Introduction}

General relativistic magnetohydrodynamic (GRMHD) simulations have been established as the standard for modeling black hole accretion \citep{Komiss1999,Gammie2003,IllinoisGRMHD,White2016,BHAC,HAMR}. 
For the low-luminosity supermassive black hole in the centre of our Galaxy, Sagittarius A* (Sgr A*), the accretion flow is believed to be both optically thin and geometrically thick, lending itself well to numerical simulation without inclusion of radiative losses.
Because of this and Sgr A*'s significance as an observational source, there is an ever growing library of simulations that have been directly applied to the Galactic Centre (e.g., \citealt{EHT_SGRA_5}).
This includes torus-based models aligned with the spin of the black hole (e.g., \citealt{Fishbone1976,Porth2019}) including both magnetically arrested (MAD, \citealt{Narayan2003,Igumenshchev2003}) and standard and normal evolution (SANE, \citealt{Narayan2012}) flows, as well as tilted torus models \citep{Liska2018,Chatterjee2020}, and even models that takes into account wind-feeding from nearby Wolf-Rayet (WR) stars \citep{Ressler2020b}.  
Discerning among these models can be done by comparison to observations and/or by improving the theoretical understanding that go into the calculations, both of which we seek to do in this work.

The first resolved image of the 230 GHz emission of Sgr A* taken by the Event Horizon Telescope (EHT, \citealt{EHT_SGRA_1,EHT_SGRA_2,EHT_SGRA_3,EHT_SGRA_4,EHT_SGRA_5}) and the observation of near-infrared (NIR) flares moving on horizon scales \citep{GRAVITYFlare} has recently provided exciting new constraints on accretion models.
These horizon-scale missions, however, are only one piece of the observational picture.
Some of the strongest constraints on the accretion flow's emission comes from multi-wavelength, unresolved measurements, especially when the time variability of the emission is considered.
This includes, e.g.,  the low-frequency radio slope and the sub-mm ``bump'' \citep{Falcke1998,An2005,Doeleman2008,Bower2015}, NIR/X-ray flaring statistics \citep{Neilsen2015,Do2019b,Witzel2021}, linear/circular polarization fractions, and the rotation measure (RM) of the sub-mm emission \citep{Marrone2007,Bower2018,ALMARM2021}.  
The rotation measure of Sgr A* provides a particularly unique constraint on the flow's density and magnetic field.
The dominant scale setting the rotation measure is unknown, but it could be as large as a few 10--100 Schwarzschild radii from the black hole \citep{Ressler2019,Dexter2020,ALMARM2021} or larger.  In contrast, the sub-mm emission is likely dominated by scales $\lesssim$ 10 Schwarzschild radii.  
The RM has been used in the past to estimate the accretion rate onto Sgr A*, but more recent work has questioned the direct relationship between the two quantities \citep{Ricarte2020}. 
This is because there is potentially a complicated interplay between ``internal'' Faraday rotation (rotation from within the emission source) and ``external'' Faraday rotation (rotation from an external medium) that can only be properly treated with a full radiative transfer calculation covering a large dynamic range in radii.
The mean-value of Sgr A*'s RM is $\approx -5 \times 10^5$ rad/m$^2$ and its magnitude has been observed to vary by more than a factor of 10 over $\sim$ hour timescales \citep{Marrone2007,Bower2018}.  
Significantly, though, this variation has never been observed to include a sign change.

Another important consideration for accretion models of the Galactic Centre is the behavior of the relativistic jet.  
It has long been debated whether or not the Galactic Centre is host to such a jet, as there is no obvious signature of a well-collimated outflow reaching large distances from the black hole like there is in other galactic nuclei (e.g., M87, \citealt{Reid1989,Junor1999,Biretta1999,Marshall2002,Abdo2009,Perlman2011,Walker2018}).
However, some have argued that certain cavity-like features and disordered but bipolar structures may provide indirect evidence for a weak jet \citep{Royster2019,YZ2020}.  
Models with jets have also often been invoked to explain features of the emission from Sgr A*, for example, the approximately flat slope of $F_\nu$ in the low-frequency radio emission \citep{Blandford1979,Mosci2013}.
One interesting possibility is that there is a jet present at smaller radii where the radio emission is unresolved but that it cannot escape to large radii due to either weak power or instability (e.g., the kink instability, \citealt{Bromberg2016,Sasha2016,Ressler2021}).  
Weak power could be caused by a central black hole that is spinning slowly, as jet power scales quadratically with spin \citep{BZ1977}, or a limited supply of magnetic flux.
In any case, the lack of direct evidence for a jet is a constraint unto itself that can be an important discriminator between accretion models, particularly at large radii.

A significant additional complication connecting accretion models to observations of Sgr A* in that the horizon-scale plasma is nearly collisionless.  
Because of this, the electron temperature predicted by assuming electron-ion equilibrium is unreliable and there is significant uncertainty in calculating the predicted emission from a given GRMHD dynamical model.  
The two existing methods for overcoming this obstacle are 1) assigning electron temperatures to the simulations in post-processing based on local plasma conditions \citep{Peek2005,Mosci2009,CK2015,Mosci2016} or 2) directly evolving the electron temperature (or entropy) alongside the GRMHD simulations \citep{Ressler2015,Sadowski2017,Mizuno2021}. 
1) is much more flexible and fast, thus making it manageable to apply in a large library of simulations and/or a parameter survey (e.g., \citealt{EHT5,EHT_SGRA_5,Anantua2020}).
2) can more directly connect to fundamental plasma physics by using heating fraction calculated from collisionless particle-in-cell (PIC) simulations \citep{Howes2011,Rowan2017,Werner2018,Zhdankin2019} and is arguably more predictive.

In \citet{Ressler2018,Ressler2019,Ressler2020}, we presented our magnetized wind-fed accretion models of Sgr A* (building on the work of \citealt{Cuadra2005,Cuadra2006,Cuadra2008,Calderon2020}) that directly incorporate observational constraints on the WR stars  \citep{Paumard2006,Martins2007,YZ2015}.  
In \citet{Ressler2020b}, we extended the larger-scale wind simulations to event horizon scales using a nested three-simulation technique for a non-spinning black hole.  We showed that these ``wind-fed'' GRMHD simulations result in accretion rates consistent with previous estimates and that they can match the 230 GHz flux for a particular choice of post-processing electron temperature model.  
This is non-trivial because the physical units and absolute orientation of the gas in these simulations are not free but instead fixed by the properties of the stellar winds at large radii. 
Moreover, \citet{Murchikova2022} showed that the same simulations can better represent the observed, high-cadence 230 GHz variability than torus-initialized simulations.
In this work we seek to further our overarching goal of improving the predictive power of GRMHD models of Sgr A*.
 To do this we incorporate the self-consistent treatment of electron thermodynamics inspired by fundamental plasma physics to eliminate the need for a post-processing electron temperature model.  
We also consider the effects of nonzero black hole spin on the accretion flow and jet formation/propagation.
Finally, we compute several unresolved emission properties from the simulations, including the RM, 230 GHz flux, and polarization fractions.

This work is organized as follows. \S \ref{sec:methods} describes our computational methods, \S \ref{sec:results} details our results, \S \ref{sec:disc} discusses them in more detail, and \S \ref{sec:conc} concludes.




\section{Methods}
\label{sec:methods}
All simulations in this work make use of {\tt Athena++} \citep{White2016,Athenapp}, a multi-purpose fluid dynamics grid-based solver of the MHD and GRMHD equations in conservative form.

In order to study the wide range of spatial and temporal scales in the Galactic Centre accretion flow, we employ a three-simulation technique described in \citeauthor{Ressler2020b} (\citeyear{Ressler2020b}, see also \citealt{Yuan2012}).  This technique takes the results of a large-scale MHD simulation of WR stellar winds orbiting Sgr A* (encompassing a radial range of $\approx 300 $--$6 \times 10^6 r_{\rm g}$ or $\approx 9 \times 10^{-4}$-- $1 $ pc) and then uses them for the initial conditions of a smaller scale MHD simulation (encompassing a radial range of $\approx 2$--$2 \times 10^5 r_{\rm g}$ or $\approx 7 \times 10^{-7}$-- $3 \times 10^{-2} $ pc).  The results of this intermediate-scale simulation are then used as initial conditions for a horizon-scale GRMHD simulation (encompassing a radial range of $\approx 1$--$2 \times 10^3 r_{\rm g}$). 

The two MHD wind simulations are initialized at 1.1 kyr in the past and run until at least the present day, using wind-speeds, mass-loss rates, and orbits for the WR stellar winds that are derived from observations \citep{Belo2006,Paumard2006,Lu2009,Martin2014,Gillessen2017}.  
Note that for some of these stars, updated orbits computed from the ever-growing data set are available (namely, for E20/IRS 16C, E23/IRS 16SW, E32/16SE1, E40/16SE2, and E56/34W, \citealt{vonFellenberg2022}), but we choose to use the older orbital solutions for ease of comparison to previous simulations.  
The difference in orientation between these new orbits and the ones used in our simulations are small, $\lesssim 10^\circ$, so we don't expect there to be significant qualitative effects of this choice.  
The orbital orientations, mass-loss rates, and wind speeds used for the three WR stellar winds expected to be the most important for accretion (see, e.g., \citealt{Cuadra2008,Ressler2020}) are listed in Table \ref{tab:parameters}.
The magnetic fields in the winds are toroidal (the spin axis is chosen randomly for each star) with strengths parameterized by $\beta_{\rm w}$, the ratio between the ram pressure and magnetic pressure in the winds. 
A detailed description and analysis of these simulations can be found in \citet{Ressler2020}.  Here we focus on two choices of $\beta_{\rm w}$, $10^2$ and $10^6$. 

The two intermediate-scale MHD simulations are run for 0.24 yr or $\sim$ $3.4 \times 10^5 M$, long enough to reach a quasi-steady state in the inner regions.  They are initialized using data taken from $t= 0.15$ kyr and $t=0.05$ kyr  in the large-scale wind wind-fed MHD simulation (where $t=0$ is defined as the present day) for $\betaw=10^2$ and $\betaw=10^6$ simulations, respectively.  
We also run an additional $\betaw=10^2$ simulation using $t=0.03$ kyr as the initialization time to study how our results depend on this parameter.

The GRMHD simulations are run for $\gtrsim 70000 M$.  We fix the direction of black hole spin to point away from earth, consistent with both the GRAVITY \citep{GRAVITYFlare}, EHT \citep{EHT_SGRA_5}, and ALMA \citep{Wielgus2022} interpretations of their observations.  We choose black hole spin values of $a = 0$ and $a = 0.9375$ for both choices of $\beta_{\rm w}$, making four GRMHD simulations in total.

All MHD simulations are performed in Cartesian coordinates and all GRMHD simulations are performed in Cartesian Kerr-Schild (CKS,\citealt{Kerr1963}) coordinates.

The MHD simulations use piecewise linear method (plm) reconstruction while the GRMHD simulations use piecewise parabolic method (ppm) reconstruction. All simulations use the Harten-Lax-van Leer+Einfeldt (HLLE, \citealt{Einfeldt1988}) Riemann solver.   Levels of static mesh refinement (SMR) are added every factor of $\sim$ 2 in radius to mimic the behavior of a logarithmic radial grid. In this work we primarily focus on the GRMHD simulations, which are performed on a $(1600 r_{\rm g})^3$ box with 128$^3$ cells and 11 additional levels of SMR.  The final level places a 128$^3$ grid within $- 3.125 r_{\rm g} \le x,y,z \le 3.125 r_{\rm g} $, so that there are $\sim$ 40 (27) cells between $r= 0$ and the horizon in each coordinate direction for $a=0$ ($a=0.9375$). Within $r = r_{\rm H}/2$, where $r_{\rm H}$ is the event horizon radius, we set the density and pressure to the floors and the velocity to free-fall. The magnetic field is allowed to freely evolve.    

In both the intermediate-scale MHD and GRMHD simulations we add to the equation of fluid dynamics an electron entropy equation in the form:
\begin{equation}
  \label{eq:electron_entropy}
    \frac{\partial}{\partial t} \left( \rho s_e u^t \right) + \frac{\partial}{\partial x^i} \left( \rho s_e u^i\right)  = f_e Q,
\end{equation}
in GRMHD or 
\begin{equation}
  \frac{\partial}{\partial t} \left( \rho s_e\right) + \frac{\partial}{\partial x^i} \left( \rho s_e v^i\right)  = f_e Q,
\end{equation}
in MHD, where $s_e$ is the electron entropy per particle, $\rho$ is the mass density, $u^\mu$ is the fluid four-velocity, $v^i$ is the non-relativistic three-velocity, $Q \equiv \rho T_{\rm g} u^\mu \partial_\mu s_{\rm g}$ is the irreversible heating rate per unit volume, $T_{\rm g}$ is the total gas temperature, $s_{\rm g}$ is the total gas entropy per particle, and $f_e$ is the fraction of this heat that goes to electrons.  $Q$ is computed at each time step by comparing the advected total entropy with the actual entropy of the simulation \citep{Ressler2015}, while $f_e$ depends on local plasma conditions.  We adopt three different choices of $f_e$, one representing turbulent-based heating (\citealt{Howes2010}, H10) and two representing magnetic reconnection-based heating (\citealt{Rowan2017,Werner2018}, R17 and W18).  Since the electrons in our model do not back-react on the total fluid, we can evolve multiple different electron entropies in the same simulation.

We use an electron adiabatic index, $\gamma_e$, that is a function of electron temperature, smoothly transition between the non-relativistic (typically large radii) and the ultra-relativistic (typically small radii) limits.   In the non-relativistic limit, $\gamma_e \rightarrow 5/3$ and in the relativistic limit $\gamma_e \rightarrow 4/3$.  We adopt the approximation of \citet{Sadowski2017} for the relation between $s_e$ and the electron temperature given the temperature-dependent (electron) adiabatic index.  The total gas adiabatic index is set to $\gamma=5/3$ since the gas energy density is dominated by the non-relativistic ions.

\subsection{Ray Tracing}
For computing the emission from the GRMHD simulations, we use {\tt ipole}
\citep{ipole}, that integrates the full set of polarized radiative transfer equations along null geodesics for synchrotron emission and absorption due to thermal electrons.  CKS data from {\tt Athena++} is first interpolated onto a spherical grid in modified Kerr-Schild coordinates \citep{McKinney2004} in order to be read in by {\tt ipole}. This process makes use of open source software that is a part of the EHT imaging pipeline.\footnote{\href{https://github.com/AFD-Illinois/EHT-babel}{https://github.com/AFD-Illinois/EHT-babel}}
For the purposes of ray tracing, the camera is located at $r=1600$ $r_{\rm g}$ (the edge of the GRMHD simulations) and $\theta=0^\circ$ (the physical line of sight appropriate for Galactic Center coordinates).  
 Regions with $\sigma=b^2/\rho>1$ are excluded from the calculation, where $b^2 = b^\mu b_\mu$ is the square magnitude of the four-magnetic field (i.e., the square magnitude of the fluid frame magnetic field three-vector \citealt{Gammie2003}).
 This is done in order to prevent regions where the hydrodynamic/thermodynamic quantities of the simulations become inaccurate (due to the magnetic energy dominating the conserved energy) from contributing to the emission.  
 For a detailed discussions on how ray-tracing cuts on $\sigma$ can affect the results see, e.g., \citet{Chael2018} and \citet{Jia2023}.

\section{Results}
\label{sec:results}
\subsection{Intermediate Scale MHD Simulations: Angular Momentum and Orientation}

In this subsection we briefly highlight the key properties of the intermediate-scale MHD simulations used to generate initial conditions for the GRMHD simulations.

\begin{figure*}
\includegraphics[width=0.45\textwidth]{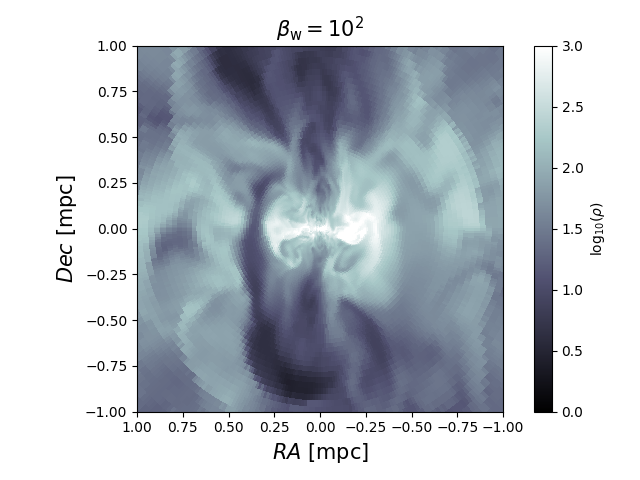}
\includegraphics[width=0.45\textwidth]{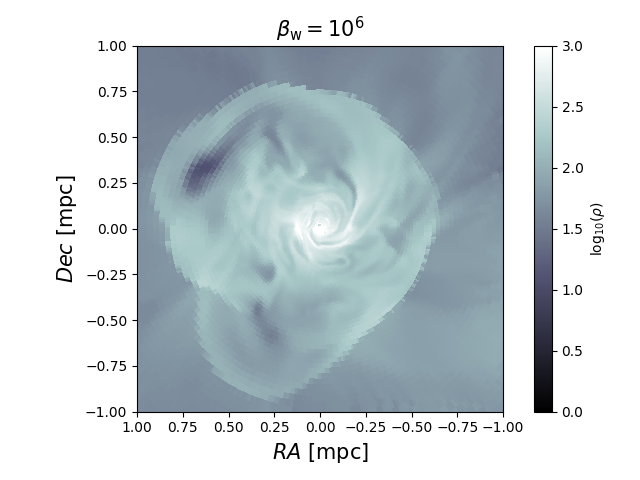}
\includegraphics[width=0.45\textwidth]{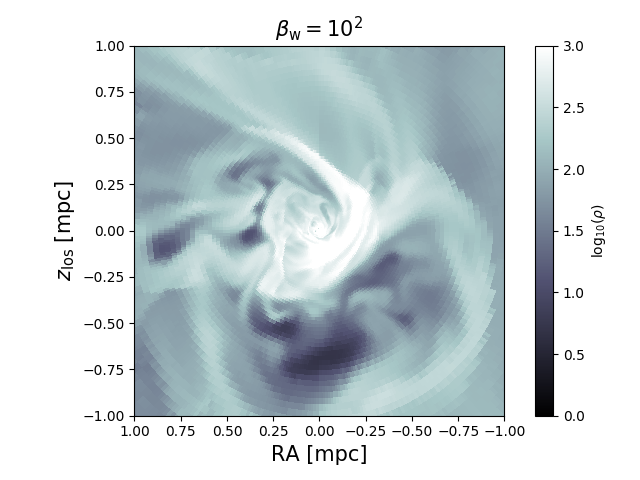}
\includegraphics[width=0.45\textwidth]{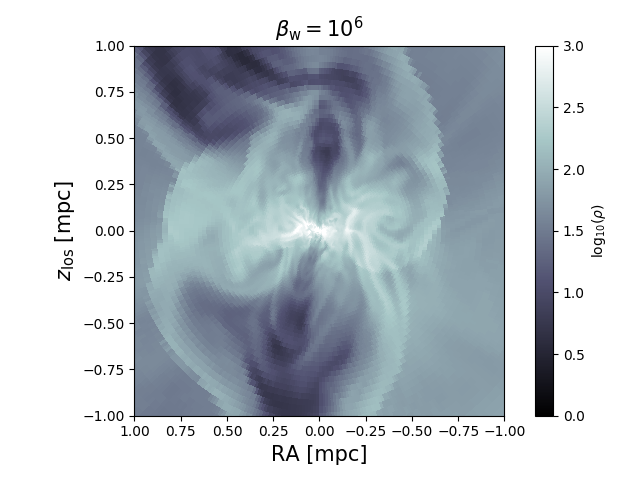}
\caption{2D slices  in the plane of the sky (top) and in the plane of the line of sight (bottom) of mass density in our intermediate-scale MHD simulations for $\beta_{\rm w}=10^2$ (left) and $\beta_{\rm w}=10^6$ (right). Here $z_{\rm los}$ is a coordinate along the lign of sight with $z_{\rm los}=0$ at the location of the black hole.  The $\beta_{\rm w}=10^2$ simulation is roughly edge-on, while the $\beta_{\rm w}=10^6$ simulation is roughly face-on.  These orientations are determined by a combination of the stellar wind properties, the strength of the magnetic field in the winds, and stochasticity.  Both simulations have a cavity aligned with the average angular momentum of the flow that is a factor of $\sim$ 4--10 less dense than the midplane. } 
\label{fig:intermediate}
\end{figure*}

Figure \ref{fig:intermediate} shows two-dimensional contours of mass density in Galactic Center R.A. and Dec. coordinates as well as in the plane of the line of sight at the final time of each of the two MHD simulations.  
Due to the different relative strengths of the magnetic field at large radii, the accretion flows have very different orientations after being evolved.  
The gas in the $\beta_{\rm w}=10^2$ simulation has an average angular momentum vector pointing nearly perpendicular to the line of sight (i.e., edge-on), while the angular momentum vector of the gas in the $\beta_{\rm w}=10^6$ simulation is nearly aligned with the line of sight (i.e., face-on). 

\begin{figure}
\includegraphics[width=0.45\textwidth]{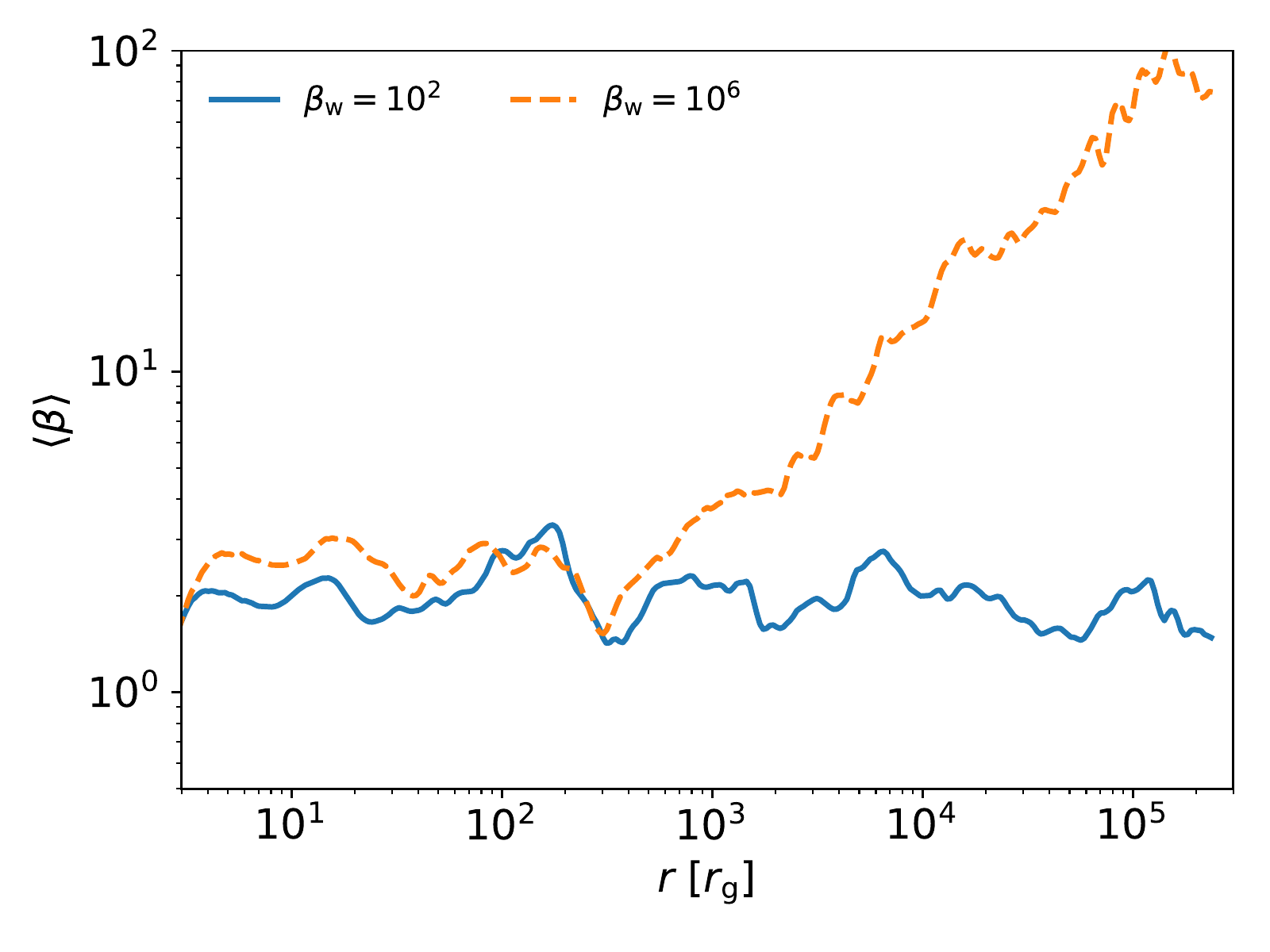}
\caption{Average plasma $\beta$, $\langle \beta\rangle = \langle P\rangle /\langle P_{\rm B} \rangle$, as a function of radius in our two intermediate-scale MHD simulations.  $\beta$ in the $\beta_{\rm w}=10^6$ simulations steadily decreases with decreasing radius until it saturates at $\beta \approx 2$.  The $\beta_{\rm w}=10^2$ simulation, on the other hand, has $\beta \approx 2$ at all radii.  This means that while the field in both simulations is dynamically important at small radii, only the $\beta_{\rm w}=10^2$ simulation has a strong enough magnetic field to affect the flow at large radii.}
\label{fig:beta_comp_intermediate}
\end{figure}

Unfortunately, there is not a simple connection between the value of $\betaw$ and the resulting orientation of the accretion flow.  
Naively, one might think that the $\betaw=10^6$ simulation would have a similar orientation to the purely hydrodynamics result of an accretion flow roughly in the same plane as the clockwise stellar disc in the Galactic Center \citep{Ressler2018}.  
This is true for larger radii where the magnetic field is weak, but at small radii ($r\lesssim 5 \times 10^3 r_{\rm g}$) the field becomes dynamically important, as shown in Figure \ref{fig:beta_comp_intermediate}, which plots the relative strength (i.e., plasma $\beta$) of the magnetic field as a function of radius for the two MHD simulations. 
Once the field becomes dynamically important, magnetic forces can significantly torque the accretion flow and alter its angular momentum.  
The direction of the field for $\betaw=10^6$ at these small radii is essentially random because for all larger radii it did not have the strength to maintain any initial geometry that may have been imparted by the stellar winds.  
In fact, the orientation of the $\betaw=10^2$ field (which is dynamically important at all radii) is more correlated with the stellar wind source terms, so that simulations of more strongly magnetized winds more often display disc inclinations similar to that of the clockwise stellar disc.
Note that this is because the feeding is dominated by a handful of WR stellar winds (those with small pericenter distances and relatively slow winds) that happen to be located in the stellar disc \citep{Ressler2018}.
The correlation between the stellar wind source terms and the resulting accretion flow orientation at small radii in $\betaw=10^2$ simulations is still not perfect, however, as there are often times when the latter can be significantly tilted with respect to the former (see, e.g., Figure 9 of \citealt{Ressler2020}) by as much as $\sim$ $90^\circ$ or greater.


As we have noted, both simulations have dynamically important magnetic fields in the inner regions of the domain, $r \lesssim 700 r_{\rm g} $.  
This is because even when the stellar winds are weakly magnetized, flux freezing causes the magnetic pressure to grow faster with decreasing radius than the gas pressure.  Given the large dynamic radial range of the Galactic Center accretion flow, this means that even for an initial $\beta=\beta_{\rm w} = 10^6$, $\beta$ can reach $\sim $ 2 at radii well outside the event horizon (see \citealt{Ressler2020}). As a consequence of this, the GRMHD simulations that we will shortly describe are initialized with plasma in which the magnetic field is already fairly strong.

Another important feature of the accretion flow in these simulations is the presence of moderately evacuated polar regions.  This can be seen in the left panel of Figure \ref{fig:intermediate} for the $\beta_{\rm w} = 10^2$ simulation but is also present in the $\beta_{\rm w} = 10^6$ simulation (both have similar structures if you rotate them so that the angular momentum direction vectors align).  A combination of centrifugal force and magnetic pressure result in these regions having a factor of $\sim$ 3--10 times less density than the respective midplanes.  This contrast is much more mild than in aligned torus-initialized simulations because of the greater presence of low angular momentum gas provided by the stellar winds.

To put the orientation and sizes of these simulations in context, we have listed the angular momentum direction vectors and distances from Sgr A* for several key Galactic Centre structures in Table \ref{tab:parameters}.  
These include the Circumnuclear Disc (not included in our simulations), the clockwise stellar disc\footnote{This is referred to as the ``inner'' clockwise disc in \citet{vonFellenberg2022}.  At larger distances from the black hole $r \gtrsim 0.3$ pc, they also report an ``outer'' clockwise disc with different orienation and a counterclockwise disc.  Since the WR stars of interest to accretion lie mostly at distances interior to these features we neglect them in the table.}, the orbits of the three WR stellar expected to be most important for accretion (namely, E20/IRS 16C, E23/IRS 16SW, and E39/IRS 16NW, \citealt{Cuadra2008,Ressler2020}), the accretion flow in our intermediate scale simulations, and our chosen black hole spin axis.
For simplicity, observational and/or model uncertainties are not shown and so the quoted values should be taken as approximate.
We discuss the (often) time-varying orientations of the GRMHD accretion flows on horizon scales in detail in  \S \ref{sec:accretion}.

\begin{table*}
  \begin{center}
    \label{tab:parameters}
    \def\arraystretch{1.75}
    \begin{tabular}{|c|c|c|c|c|c|} 
           \multicolumn{5}{c}{Angular Momentum Directions}\\
           \hline
            Structure &Distance from Sgr A* & $n_x$ & $n_y$ & $n_z$ & References \\
      \hline
      Circumnuclear Disc$^\dagger$ & $\gtrsim 2$ pc &  0.83 & -0.39 & -0.41& 1 \\ 
      \hline
      Clockwise Stellar Disc$^{\ddagger}$ &0.032--0.13 pc  & 0.16& -0.89& 0.42 & 2,3,4,5 \\ 
      \hline
      WR Star E20/IRS 16C$^{*}$ & 0.055 --0.074 pc & 0.38 & -0.68 & 0.62 & 2,6,7 \\
      \hline
      WR Star E23/IRS 16SW$^{*}$ &  0.060--0.12 pc& -0.36 & -0.85 & 0.39 & 2,6,7 \\
      \hline
      WR Star E39/IRS 16NE$^{**}$& 0.12--0.16 pc& 0.022 & 0.019 & -1.00 & 2,6,7 \\
      \hline
      $\betaw=10^2$ Intermediate-Scale Accretion Flow & $10^{-5}$--10$^{-3}$ pc$^{\dagger \dagger}$ & -0.021 & -1.00 & 0.056 & -\\
      \hline
      $\betaw=10^6$ Intermediate-Scale Accretion Flow & $10^{-5}$--10$^{-3}$ pc$^{\dagger \dagger}$&-0.098 & -0.0048 & 0.94 & -\\
      \hline
      $\betaw=10^2$ Intermediate-Scale Accretion Flow (diff. $t_{\rm restart}$) & $10^{-5}$--10$^{-3}$ pc$^{\dagger \dagger}$ & 0.40 & -0.18 & 0.85 & -\\
       \hline
             Black Hole Spin Axis& - & 0 & 0 & 1 & -\\
       \hline
       \end{tabular}
        \begin{tabular}{|c|c|c|c|}
       \multicolumn{4}{c}{WR Stellar Wind Properties}\\
\hline
Star & $\dot M_{\rm wind}$ & $v_{\rm wind}$  & References \\
      \hline
      WR Star E20/IRS 16C &2.24$\times 10^{-5}$ $M_\odot$/yr & 650 km/s & 7,8 \\
      \hline
      WR Star E23/IRS 16SW$^{\P}$ & 1.12$\times 10^{-5}$ $M_\odot$/yr & 600 km/s & 7,8 \\
      \hline
      WR Star E39/IRS 16NE$^{\P}$ & 2.24$\times 10^{-5}$ $M_\odot$/yr & 650 km/s & 7,8 \\
      \hline
    \end{tabular}
  \end{center}
   \caption{ \\
   $^\dagger$ For reference only; the circumnuclear disc is not included in our simulations.  See, e.g., \citet{Blank2016,Solanki2023} for simulations that do include the circumnuclear disc.
      \\
   $^{\ddagger}$ AKA the ``inner'' clockwise disc in \citet{vonFellenberg2022}. \\
   $^{*}$Likely disc star.  Line-of-sight coordinate calculated in \citet{Belo2006}. \\
   $^{**}$Non-disc star.  Line-of-sight coordinate calculated by minimizing eccentricity of orbit as in \citet{Cuadra2008}. \\
   $^{\dagger\dagger}$ Denotes the region used to measure angular momentum direction, not necessarily the radial extent of any disc-like structure. \\
      $^{\P}$ Not sampled in \citet{Martins2007}.  Properties assumed to be the same as similar nearby WR stars as in \citet{Cuadra2008}. \\
   $^1$ \citealt{Genzel2010} and references therein \\
   $^2$ \citealt{Paumard2006} \\
   $^3$ \citealt{Lu2009} \\
   $^4$ \citealt{Yelda2014} \\
   $^5$ \citealt{vonFellenberg2022} \\
   $^6$ \citealt{Belo2006} \\
   $^7$ \citealt{Cuadra2008} \\
   $^8$ \citealt{Martins2007} (see also \citealt{YZ2015})\\}
\end{table*}

\subsection{GRMHD Simulations}

We now describe the four GRMHD simulations that were initialized from the two intermediate-scale MHD simulations, focusing in turn on the accretion flow, electron thermodynamics, the relativistic jet, and the emission properties. In \S \ref{sec:alt_real}, we describe two additional pairs of intermediate-scale MHD/GRMHD simulations (for $a=0$ and $a=0.9375$) initialized at a somewhat different time in the large-scale wind-fed MHD simulation.  This allows us to assess the robustness of the GRMHD results.

\subsubsection{Accretion Flow}
\label{sec:accretion}

\begin{figure}
\includegraphics[width=0.45\textwidth]{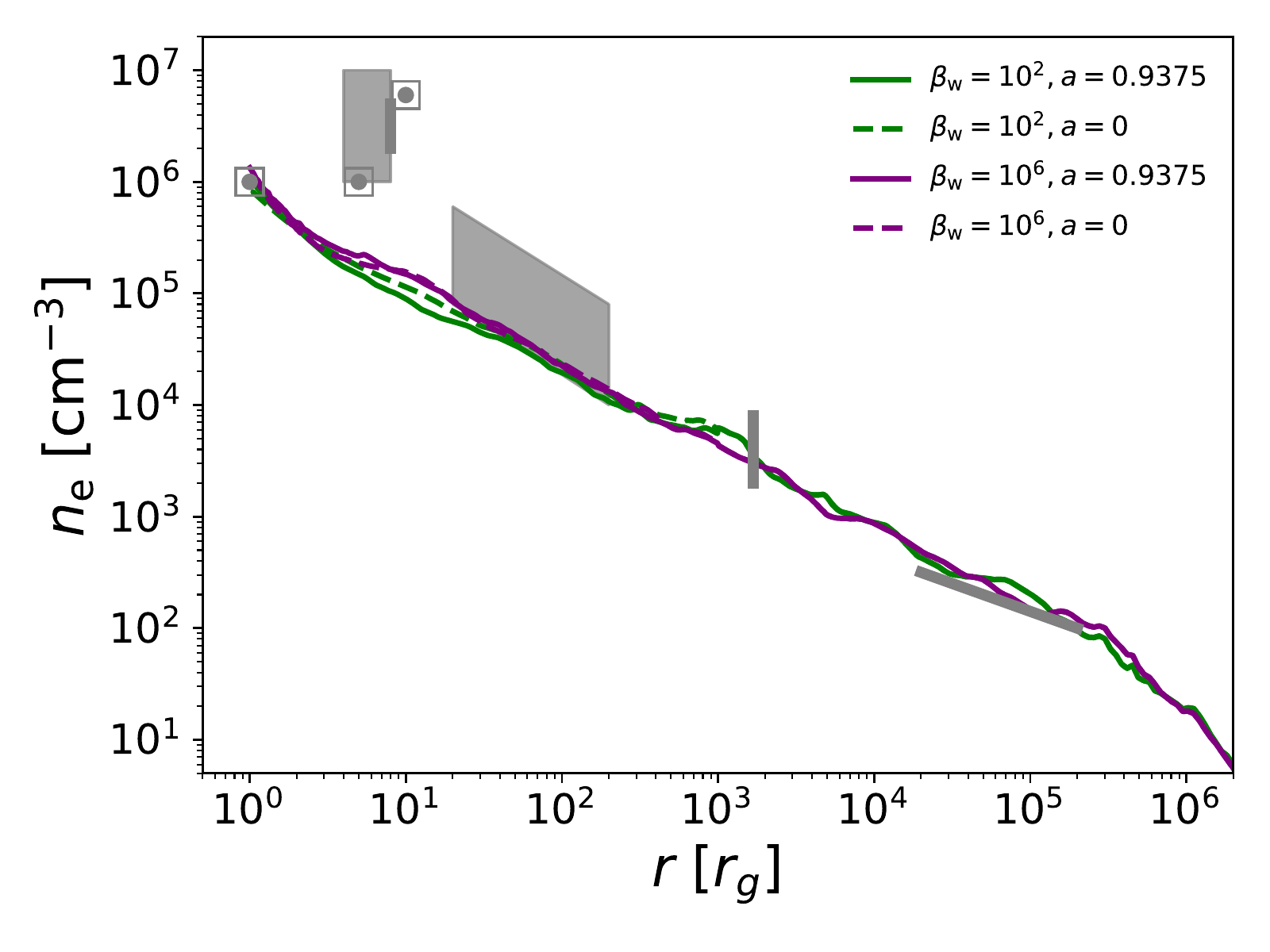}
\caption{Angle-averaged electron number density versus radius in our four simulations compared to several model-dependent observational estimates at different scales.
The latter are mostly reproduced with permission from \citet{Gillessen2019}, and are based on measurements of the sub-millimeter size and flux (shaded gray rectangle, \citealt{Bower2015,vonFellenberg2018}), the sub-millimeter polarization and flux (gray dots surrounded by gray squares, \citealt{Quataert2000} at $r=r_{\rm g}$ and \citealt{Agol2000} at $r=10 r_{\rm g}$), the rotation measure (shaded gray parallelogram, \citealt{Marrone2007}), the drag force on G2 (vertical gray line near $r=2\times10^3 r_{\rm g}$ \citealt{Gillessen2019}), and the X-ray spectral energy distribution ($r^{-1/2}$ gray line at large radii, \citealt{Wang2013}). 
To these we add the estimate based on the EHT-observed flux (gray dot surrounded by gray square at $r=5 \rg$, \citealt{EHT_SGRA_5}) and sub-millimeter spectral fitting (vertical line at $r=8 \rg$, \citealt{Bower2019}).
We emphasize that these are model-dependent estimates of the number density, many of which are based on one-zone or one-dimensional models, are strongly dependent on certain assumptions, and/or are degenerate with other parameters, e.g., the magnetic field strength.
That said, the number densities from our simulations are remarkably consistent with most of these observational-derived estimates.  
The only exceptions are those around $r\sim 10 r_{\rm g}$ that are also outliers from the other estimates.
 } 
\label{fig:gillessen_comp}
\end{figure}

To study the accretion flow, we first highlight the radial, angle-averaged profiles of MHD quantities, which have particular importance in accretion and jet evolution \citep{Gruzinov2013,Bromberg2011}.
Figure \ref{fig:gillessen_comp} in particular shows how our simulated number density profiles compare to previous observationally-derived estimates at several different radial distances from Sgr A*.
These estimates (see Figure 6 of \citealt{Gillessen2019}) are mostly independent of each other and extend from the horizon scale out to the Bondi radius. 
Although each of these estimates is model-dependent and sensitive to specific assumptions and parameters, they are generally consistent with an $r^{-1}$ density profile below the Bondi radius. 
For comparison we piece together our MHD and GRMHD simulations.  When two simulations overlap with each other we select the smaller scale simulation to generate these number density profiles.
Given the uncertainty of the observationally-derived estimates and the uncertainty of the mean molecular weight of the gas, the simulations and the estimates are remarkably consistent with one another except for the region around $r\sim 10 r_{\rm g}$.
The estimates there (based off of the sub-millimeter size, flux, and polarization), however, also seem to be outliers from the estimates at smaller and larger radii and are in reality more uncertain than indicated on this figure given the systematic uncertainties. 

The other angle-averaged radial profiles are consistent with the power-laws found in the large-scale MHD wind simulations \citep{Ressler2020,Ressler2020b}, with $\langle T_{\rm g}\rangle \  \tilde \propto \  r^{-1}$, $ \sqrt{\langle b^2 \rangle}\ \tilde \propto \  r^{-1}$, and $v_\varphi \approx 0.5 v_{\rm kep}$.  
The sub-Keplerian nature of the flow is due to the relatively strong pressure support from a combination of thermal, magnetic, and radial ram pressure (see, e.g., Figure 13 in \citealt{Ressler2018}).

\begin{figure*}
\includegraphics[width=0.95\textwidth]{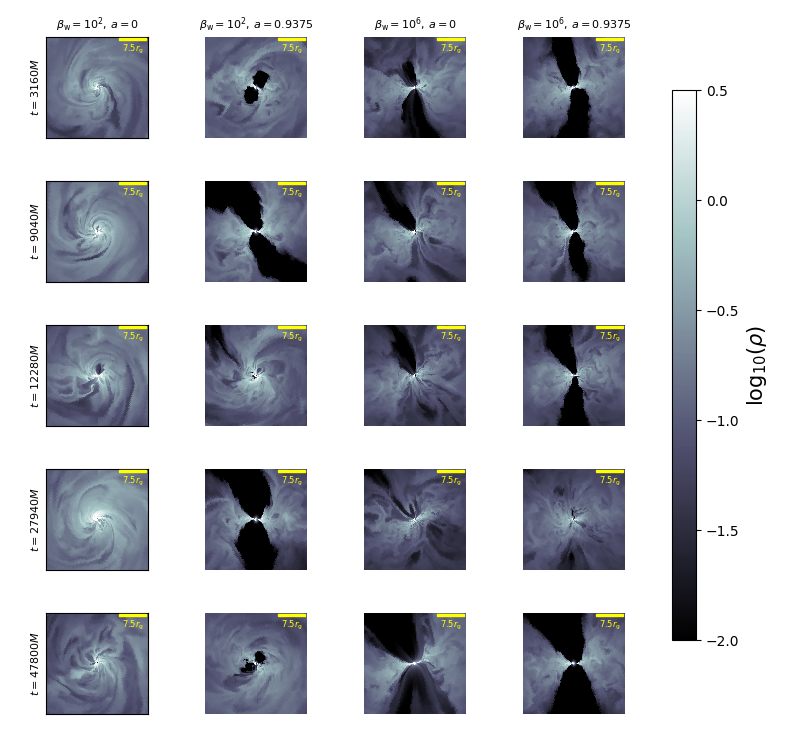}
\caption{Slices of mass density in the $x$-$z$ plane at 5 different times in our four simulations.  For simulations with $a=0.9375$, this means that black hole spin points upwards.  For $\beta_{\rm w}=10^2$ (where the $a=0$ flow is rotating $\sim$ in the plane of these images), the black hole spin at times completely reorients the gas to have $\sim$ aligned angular momentum.  For $\beta_{\rm w}=10^6$ the gas orientation remains relatively unchanged with spin because it already had angular momentum $\sim$ aligned with black hole spin axis.
Animations of this figure are available at \href{https://youtube.com/playlist?list=PL3pLmTeUPcqSd4jVBnRubYQpa-Dma25ir}{https://youtube.com/playlist?list=PL3pLmTeUPcqSd4jVBnRubYQpa-Dma25ir} } 
\label{fig:rho_contour}
\end{figure*}

Turning now to more detailed analysis, Figure \ref{fig:rho_contour} shows the mass density at 5 different times in each of the four simulations. The slices are in the $x$-$z$ plane on the scale of $r\lesssim 30 r_{\rm g}$, where $z$ corresponds to both the black hole spin axis and the line of sight from earth ($x$, $y$, and $z$, correspond to Galactic Centre coordinates).  Note that this plane is the same as that shown in the bottom row of Figure \ref{fig:intermediate}.  Unsurprisingly, the gas in each of the $a=0$ simulations has a similar orientation to the gas in the corresponding intermediate-scale MHD simulation: rotation in the $\beta_{\rm w}=10^2$ is primarily in the $x$-$z$ plane and rotation in the $\beta_{\rm w}=10^6$ is primarily in the $x$-$y$ plane.    
Black hole spin can significantly alter this picture, however.  For  $\beta_{\rm w}=10^6$, where the angular momentum of the gas is already mostly aligned with the spin axis, not much changes in terms of morphology.  In fact, the contours in the two rightmost columns in Figure \ref{fig:rho_contour} comparing $a=0$ and $a=0.9375$ for $\beta_{\rm w}=10^6$ are remarkably similar.  The main difference is that the $a=0.9375$ simulations tend to have more evacuated polar regions due to the \citet{BZ1977} jet (see \S \ref{sec:jet} for more discussion of the jet in our simulations).
For $\beta_{\rm w}=10^2$, on the other hand, the gas can get completely reoriented at certain times by the black hole spin.  For instance, at $t=9040\  M$ and $t=27940\ M$, the angular momentum of the gas has essentially aligned with the spin axis.  At other times (e.g., $t=12280\ M$) the $a=0$ and $a=0.9375$ simulations look fairly similar.  More often the $a=0.9375$ simulation is somewhere in between these two extremes (e.g., at $t=3160\ M$ and $t=47800\ M$), where the angular momentum of the gas in the inner regions ($r \lesssim$ a few $r_{\rm g}$) have partially aligned but the angular momentum of the gas at larger distances is still orthogonal to the spin axis.   
Throughout the duration of the simulation the black hole rotation and jet are continually fighting against the infalling gas with varying success, resulting in these distinct phases of evolution. 

\begin{figure}
\includegraphics[width=0.45\textwidth]{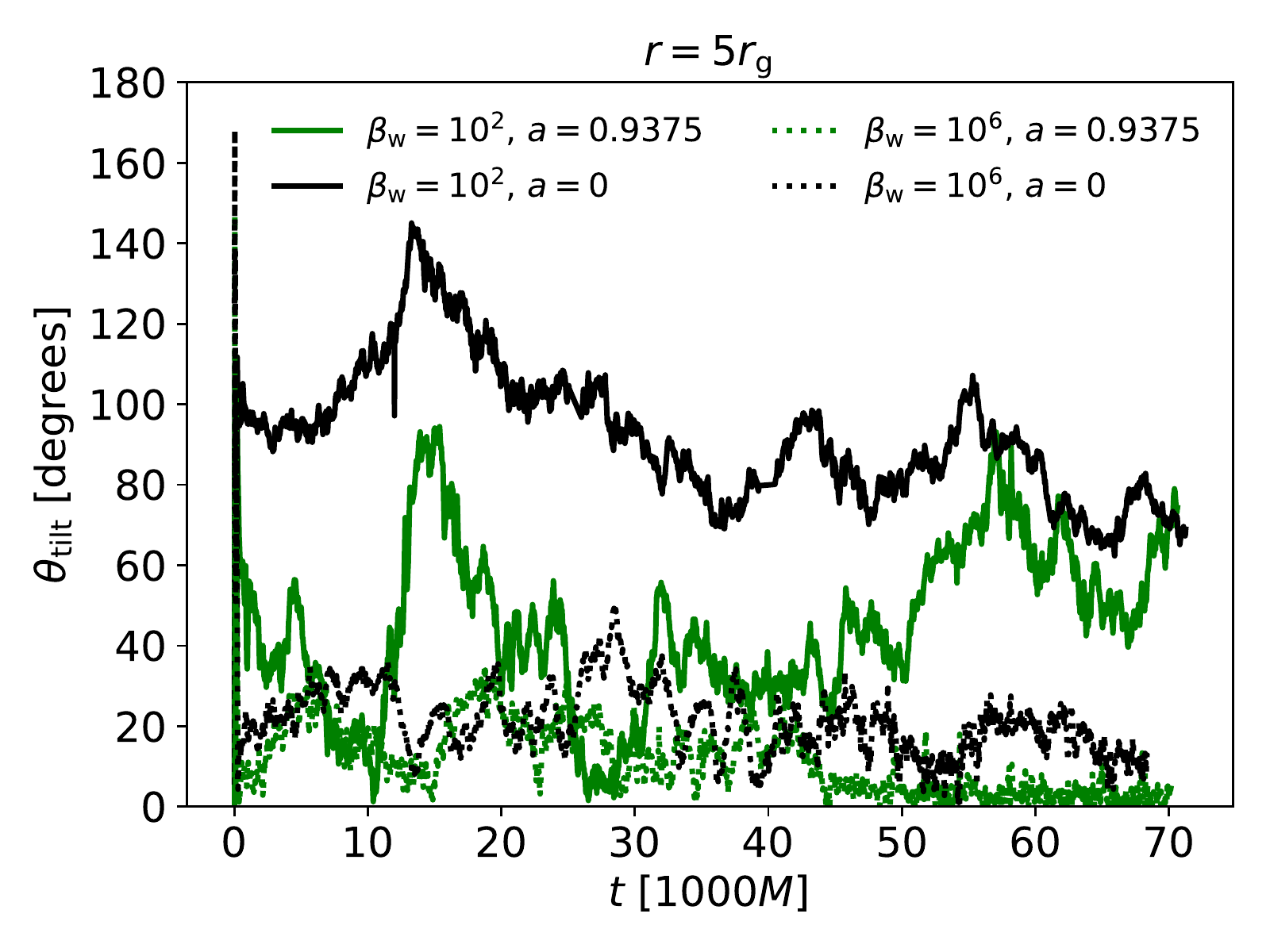}
\includegraphics[width=0.45\textwidth]{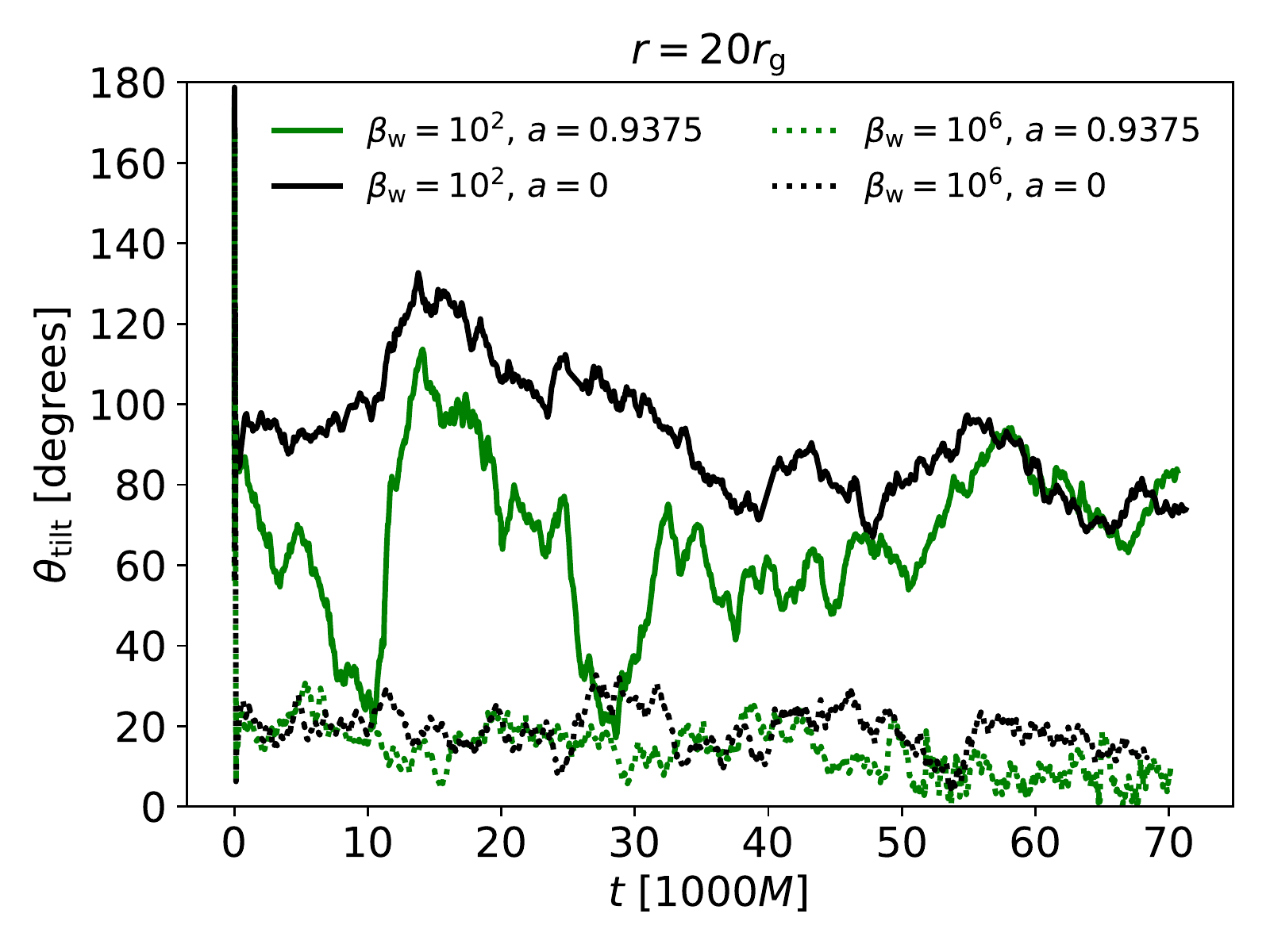}
\includegraphics[width=0.45\textwidth]{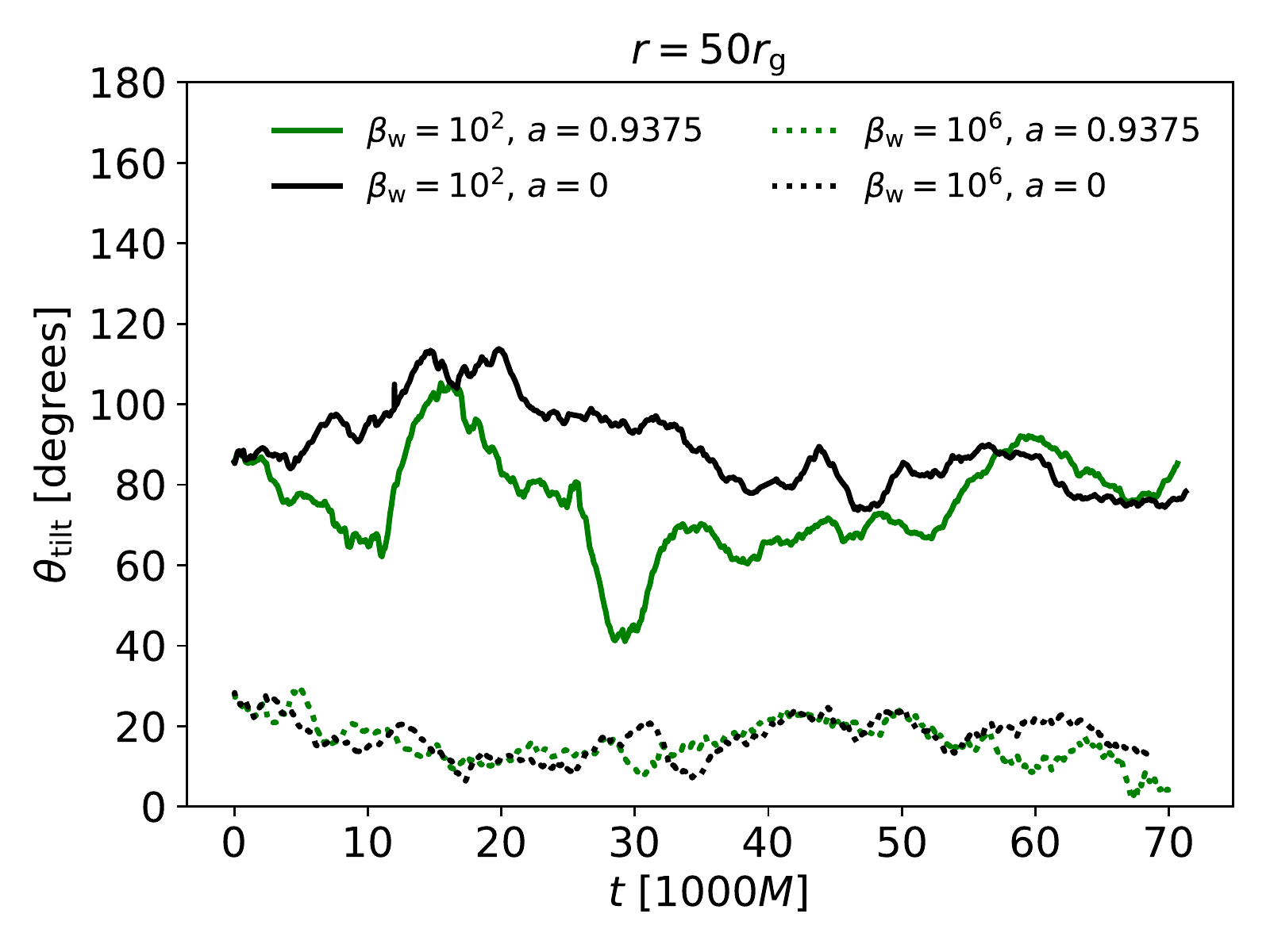}
\caption{Angle between the angle-averaged angular momentum of the gas and the black hole spin axis, $\theta_{\rm tilt}$, measured at $r=5 r_{\rm g}$ (top), $r=20 r_{\rm g}$ (middle), and $r=50 r_{\rm g}$ (bottom) for our four simulations.  For $a=0$, $\theta_{\rm tilt}$ represents the angle with the $z$-axis.
The angular momentum of the gas in both $\beta_{\rm w}=10^6$ simulations is always nearly aligned with the spin axis because its angular momentum at large radii was already pointing approximately along the $z$ axis.  The gas in the $\beta_{\rm w}=10^2$ simulations, however, starts out nearly perpendicular to the spin axis as seen by the $a=0$ curves (solid black lines). In contrast, for $a=0.9375$ (solid green lines), the black hole spin can align the gas angular momentum with the black hole spin, but only temporarily.  This process is most effective at the smallest radii (e..g, $r=5 r_{\rm g}$) and reasonably effective at slightly larger radii (e.g., $25 r_{\rm g}$), but not very effective at larger radii (e.g., $50 r_{\rm g}$).
 } 
\label{fig:tilt_angles}
\end{figure}

We can further quantify this behavior by measuring the tilt angle with respect to the black hole spin axis (or just the $z$-axis in the $a=0$ cases)  at different radii as a function of time. 
If we define $L_x = \rho (y u^z- z u^y)$, $L_y = \rho (z u^x- x u^z)$ and $L_z = \rho (x u^y- y u^x)$,  then we can calculate
\begin{equation}
  \theta_{\rm tilt} = \arctan\left(\frac{\langle L_z\rangle}{\sqrt{\langle L_x\rangle^2 + \langle L_y\rangle^2 + \langle L_z\rangle^2}}\right).
  \end{equation}
  We plot this quantity in Figure \ref{fig:tilt_angles} for $r = 5 r_{\rm g}$, $r = 20 r_{\rm g}$, and $r = 50 r_{\rm g}$.  
  As expected, the tilt angles at all radii for both $\beta_{\rm w}=10^6$ simulations are small (typically $\sim 20^\circ$) and similar to each other at all times.
  On the other hand, due to the orientation of the gas being fed from large radii (Figure \ref{fig:intermediate}), tilt angles in the $a=0$, $\beta_{\rm w}=10^2$ simulations are, on average, around $\sim$ $100^\circ$ at all radii, with the most variation seen at $r=5r_{\rm g}$.  
When an $a=0.9375$ black hole is introduced, then, the tilt angle is strongly effected.  
At $r=5 r_{\rm g}$, at certain times the tilt angle approaches $\approx$ 0, though this alignment is typically short-lived ($\lesssim$ 5000 $M$).  
$\theta_{\rm tilt}$ tends to cycle from being strongly misaligned ($\theta_{\rm tilt} \gtrsim60^\circ$) to strongly aligned ($\theta_{\rm tilt} \lesssim 20^\circ$), with the majority of time spent in an intermediate state ($20^\circ \lesssim\theta_{\rm tilt}\lesssim 60^\circ$).
The $r=25 r_{\rm g}$ gas shows similar behavior but the alignment is never as strong as it is for smaller radii, with $\theta_{\rm tilt}$ angles systematically larger by $\sim$ 20$^\circ$ than the $r=5r_{\rm g}$ gas.
The larger radii gas at $r=50r_{\rm g}$ never reaches significant alignment, with $\approx 45^\circ$ being the smallest value of $\theta_{\rm tilt}$ seen during the entire simulation.
At all times and all three radii in the $\beta_{\rm w}=10^2$ simulations, $\theta_{\rm tilt}$ is smaller in the $a=0.9375$ case than the $a=0$ case.

\begin{figure}
\includegraphics[width=0.45\textwidth]{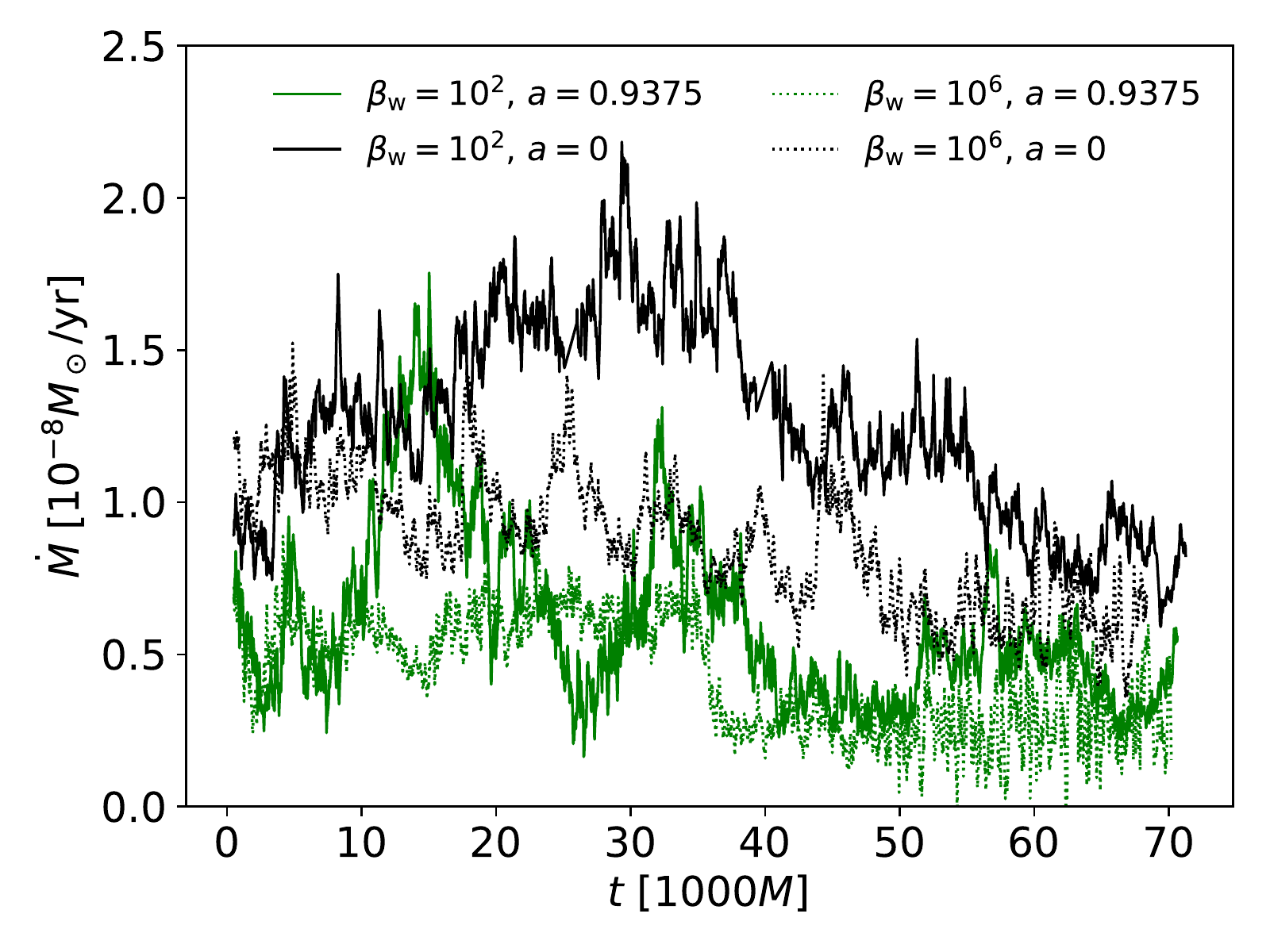}
\includegraphics[width=0.45\textwidth]{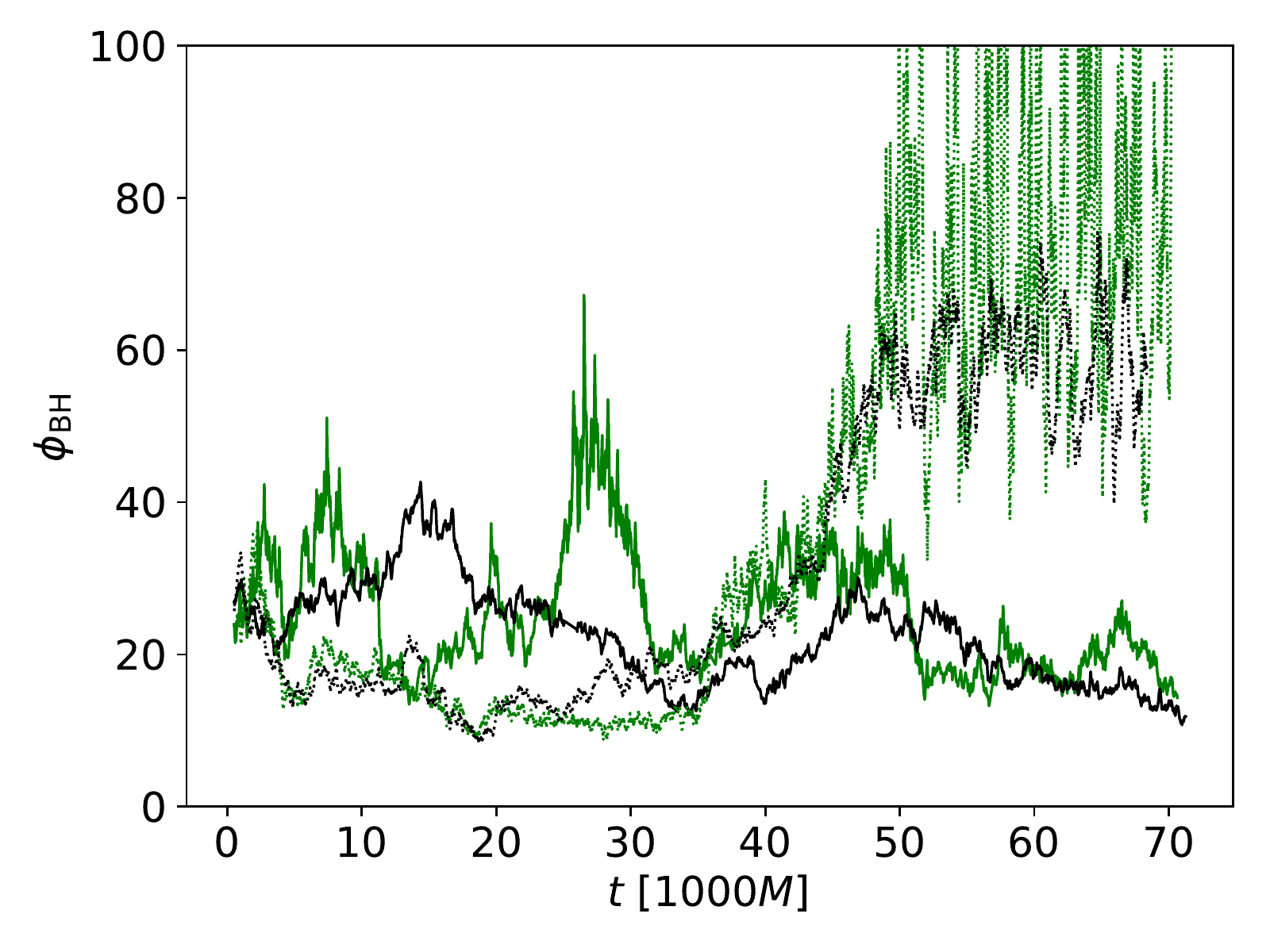}
\caption{ Accretion rate and magnetic flux threading the event horizon as a function of time for the $\beta_{\rm w}=10^2$ (solid) and $\beta_{\rm w}=10^6$ (dashed) simulations for $a=0.9375$ (green) and $a=0$ (black). In these units the MAD state corresponds to $\phi_{\rm BH}\sim$ 50--70.  Accretion rates fall in the range of 0.5--2 $\times 10^{-8}$ $M_\odot$/yr and can vary by as much as a factor of $3$--$4$ over the course of the simulations.
 } 
\label{fig:time_plots}
\end{figure}

To further analyze temporal variability, we measure two key quantities: the accretion rate onto the black hole, $\dot M$, and the magnetic flux threading the event horizon, $\phi_{\rm BH}$.  These are displayed in Figure \ref{fig:time_plots} as a function of time. Here we define $\phi_{\rm BH}$ in units such that the typical MAD state is reached at $\sim$ 50.   
The accretion rate for all simulations tends to fall within the range $2.5 \times 10^{-9}$--$2\times 10^{-8}$ $M_\odot$/yr, consistent with previous estimates of the accretion rate onto Sgr A* \citep{Marrone2007,Shcherbakov2012,Ressler2017}.
$\phi_{\rm BH}$ in both of the $\betaw=10^6$ simulations shows clear indications of a magnetically arrested flow at late times ($t\gtrsim 45{,}000\ M$), saturating at $\approx 60$--80. 
The $\betaw=10^2$, $a=0.9375$ simulation reaches a peak value of $\phi_{\rm BH}$ $\sim$ 50--60 at $\approx$ $28{,}000\ M$ but is otherwise below the MAD value for the rest of the duration. 
The $\betaw=10^2$, $a=0$ simulation remains SANE throughout, with $\phi_{\rm BH} \lesssim 40$.  
There is significant variability in both $\dot M$ and $\phi_{\rm BH}$, with an accretion rate that is typically anti-correlated with the magnetic flux;  the peaks in magnetic flux are associated with valleys in accretion rate. 
By comparison with the $\beta_{\rm w}=10^2$, $a=0.9775$ simulation in Figure \ref{fig:tilt_angles}, we see that the strongest alignment (smallest $\theta_{\rm tilt}$) of the accretion flow occurs at peaks of $\phi_{\rm BH}$ (and valleys of $\dot M$). This suggests that the strong magnetic fields threading the black hole provide the torque necessary to significantly alter the angular momentum of the gas.

It is interesting to note that our fiducial $\betaw=10^6$ simulations go MAD but the fiducial $\betaw=10^2$ simulations don't even though the latter have significantly stronger magnetic fields at large radii (Figure \ref{fig:beta_comp_intermediate}).
This shows that strong fields at larger radii are not required to ultimately reach the MAD state on event-horizon scales.  
As we have shown, compression and advection of the field towards the black hole can amplify it to the point of dynamical importance, even to the point that there is a larger supply of net flux in this particular realization of the $\betaw=10^6$ case than this particular realization of  $\betaw=10^2$ case.
In \S \ref{sec:alt_real}, we study a $\betaw=10^2$ GRMHD simulation initialized at a different time in the original large-scale wind-fed MHD simulation that is more consistently MAD in which the black hole spin axis and gas angular momentum become much more closely aligned.
In light of these findings, we conclude that there is not necessarily a direct relation between the field strength at large radii and whether or not the horizon-scale flow is SANE or MAD.
Instead it also depends on the particular details of field and flow geometry.


\subsubsection{Electron Thermodynamics}
\label{sec:electrons}

\begin{figure}
\includegraphics[width=0.45\textwidth]{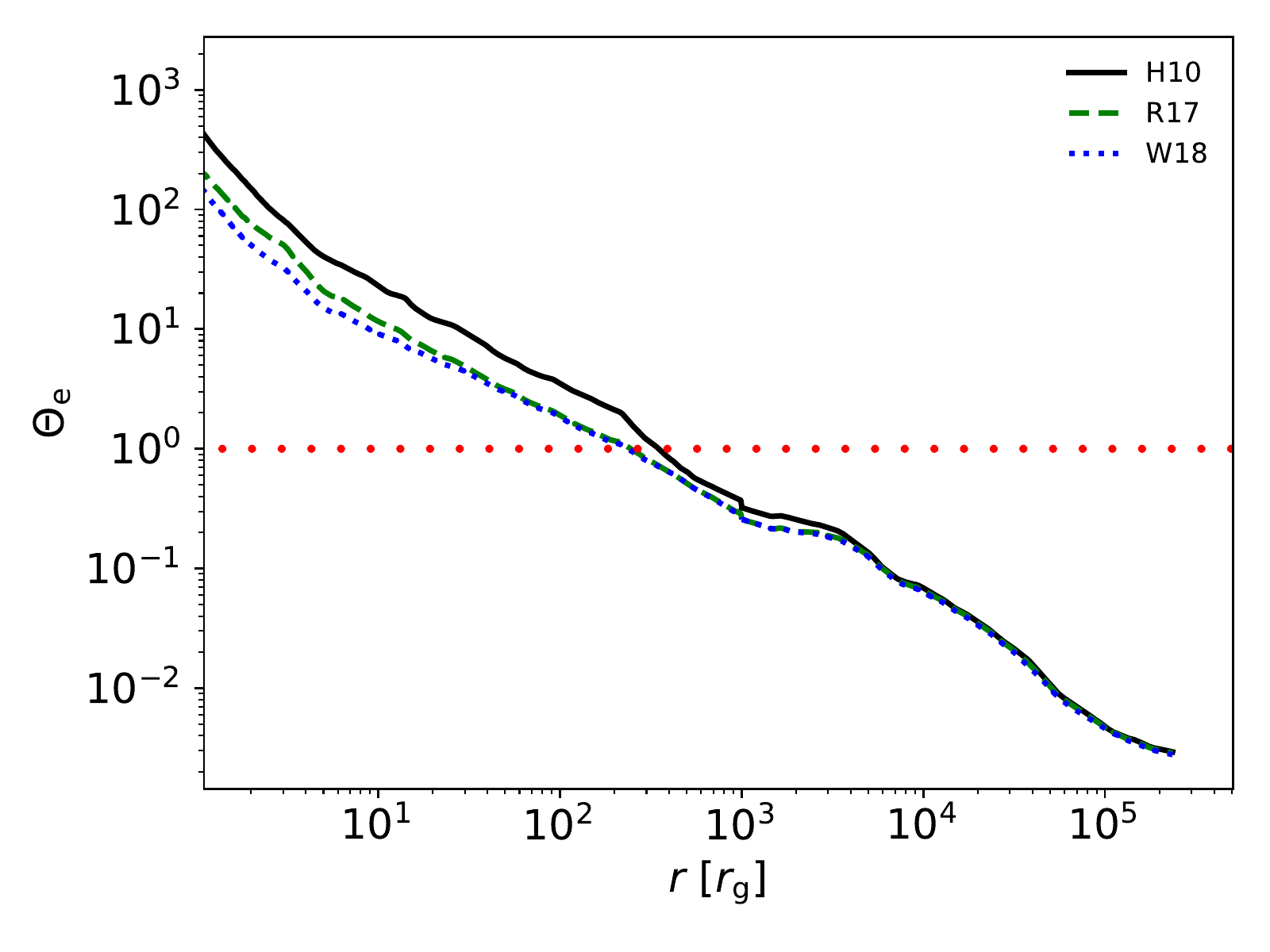}
\includegraphics[width=0.45\textwidth]{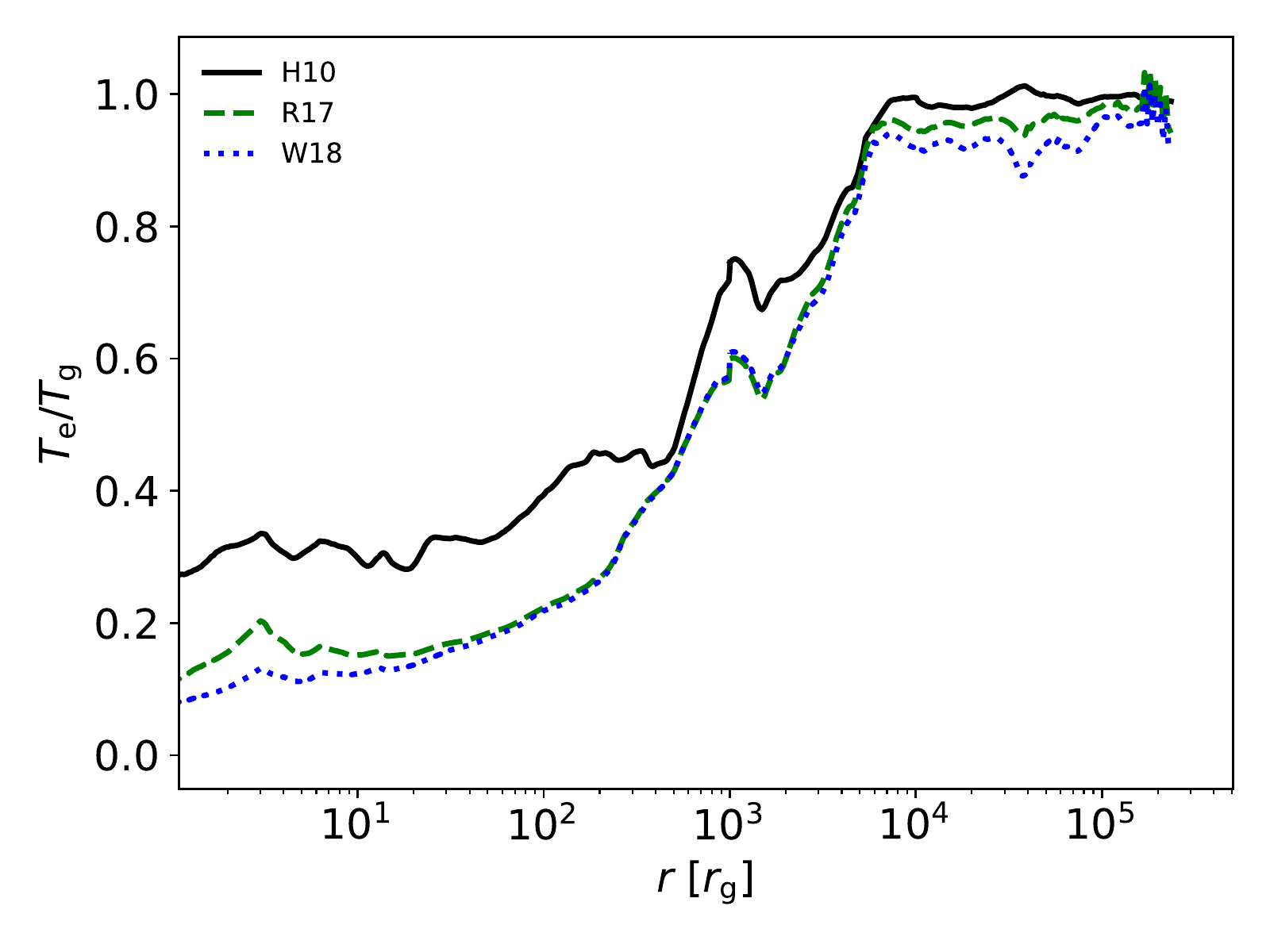}
\caption{Top: Dimensionless electron temperature, $\Theta_e$, in our $a=0.9375$, $\beta_{\rm w} = 10^2$ simulations for the three different electron temperature models. The line of red circles represents $\Theta_e = 1$, which is reached at $r \approx 260 r_{\rm g}$ in the reconnection heating models (R17 and W18) and $r \approx 380  r_{\rm g}$ in the turbulent heating models (H10).  Bottom: Electron to total gas temperature, $T_e/T_{\rm g}$ for the same three models.  At large radii the electrons (by construction) are in thermal equilibrium with the gas temperature but become relatively colder as gas falls inwards towards the black hole and ions are preferentially heated.  Turbulent heating generally results in electron temperatures higher by a factor of $\sim$ 2.  }
\label{fig:Te_Tg}
\end{figure}

In addition to black hole spin, the other key new feature of the simulations presented here compared to \citet{Ressler2020b} is the presence of an electron entropy evolution equation.  The result is self-consistent electron temperatures across $\sim$ 4 orders of magnitude in distance from the event horizon.  
In Figure \ref{fig:Te_Tg} we plot the mass-weighted angle average of dimensionless electron temperature, $\Theta_{\rm e}$, and electron-to-total gas temperature ratio, $T_{\rm e}/T_{\rm g}$, as function of radius for all three electron heating models at a particular time.  We show only the $
\beta_{\rm w}=10^2$, $a=0.9375$ simulation; the electron temperature profiles in the other simulations are similar.  
Starting out from large radii ($r\gtrsim 10^4 r_{\rm g}$) where $T_{\rm e}$ is initialized as equal to $T_{\rm g}$, the electron temperature decreases relative to the gas temperature as a function of decreasing radius, reaching $\sim$ 0.1 $T_{\rm g}$ at the event horizon in ``reconnection'' heating models R17 and W18 and $\sim$ 0.3 $T_{\rm g}$ in the ``turbulent'' heating model H10. 
In terms of $\Theta_{\rm e}$, this corresponds to a transition from non-relativistic to relativistically hot electron temperatures at $\approx$ 260 $r_{\rm g}$ and $\approx$ 380 $r_{\rm g}$ in the ``reconnection'' and ``turbulent'' heating models, respectively.  
Near the event horizon, the average electron temperatures can be quite hot, $\Theta_{\rm e} \gtrsim 100$.  
Overall, within the inner $r \lesssim 10^3 r_{\rm g}$ ``reconnection'' heating models are about a factor of 2 colder than ``turbulent'' reconnection heating models.  This is because the H10 $f_{\rm e}$ function reaches $\approx 1$ in regions of low $\beta$ while the R17 and W18 models $f_{\rm e}$ functions peak at $\approx 0.5$.  

Figure \ref{fig:Te_Tg} also implies that initializing GRMHD simulations of Sgr A* with $T_{\rm e}=T_{\rm g}$ is not well motivated and is likely to overestimate the electron temperature at large radii.

\subsubsection{Relativistic Jet}
\label{sec:jet}

\begin{figure*}
\includegraphics[width=0.95\textwidth]{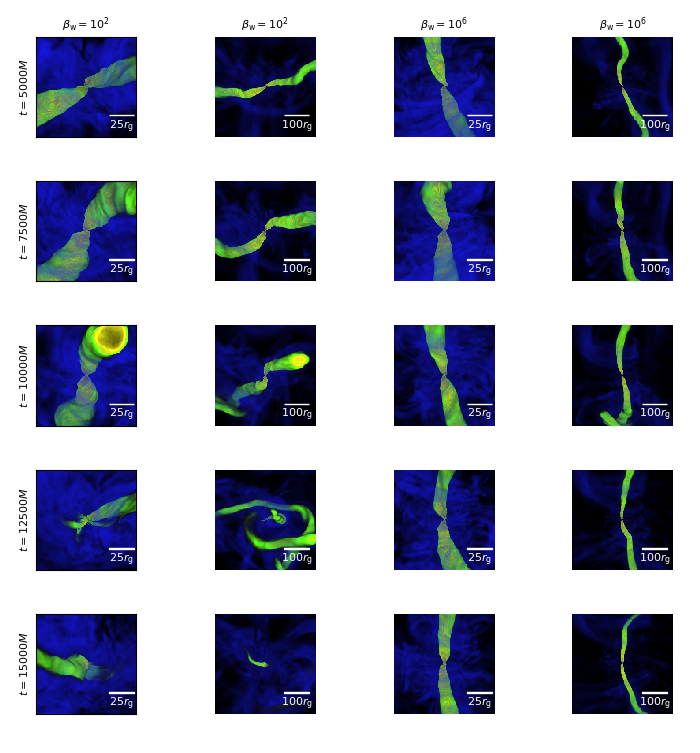}
\caption{3D volume renderings of the jets using the variable $\sigma = b^2/\rho$ in our two $a=0.9375$ simulations at 5 different times and two different spatial scales. 
To create these renderings, we perform plane-parallel ray-tracing on the data cube of $\sigma$ and concentrate opacity at $\sigma=10^{-2}$ (blue), $\sigma=1$ (green), and $\sigma=10$ (red) following \href{https://medium.com/swlh/create-your-own-volume-rendering-with-python-655ca839b097}{https://medium.com/swlh/create-your-own-volume-rendering-with-python-655ca839b097}.
In these figures the black hole spin points up.  
 In the $\beta_{\rm w}=10^6$ simulation, the angular momentum of the accretion flow is initially aligned with the spin axis of the black hole (see Figure \ref{fig:intermediate}) so that the jet readily propagates in this same direction.  Conversely, for the $\betaw=10^2$ simulations the angular momentum of the accretion flow is initially $\sim$ perpendicular to the spin axis resulting in a more complicated propagation.  At small distances from the the black hole, the jet tends to align with the spin axis, particularly during periods of high magnetic flux (compare with Figure \ref{fig:time_plots}), but at larger radii the jet aligns with the angular momentum of the larger-scale accretion flow, perpendicular to the spin axis.  
 Animations of this figure are available at \href{https://youtube.com/playlist?list=PL3pLmTeUPcqSd4jVBnRubYQpa-Dma25ir}{https://youtube.com/playlist?list=PL3pLmTeUPcqSd4jVBnRubYQpa-Dma25ir}.} 
\label{fig:jet_3D_comp}
\end{figure*}

In both of the $a=0.9375$ simulations the rotating black hole produces electromagnetically dominated jets via the \citet{BZ1977} process.  To visualize their structure and orientation we use 3D volume rendering on the two simulations, highlighting regions of high $\sigma = b^2/\rho$.  
To create these renderings, we perform plane-parallel ray-tracing on the data cube of $\sigma$ and concentrate opacity at $\sigma=10^{-2}$, $\sigma=1$, and $\sigma=10$. 
Figure \ref{fig:jet_3D_comp} shows this for the jets in our $a=0.9375$ simulations at 5 different times on $r \lesssim 50 r_{\rm g}$ and $r \lesssim 200 r_{\rm g}$ scales. 
The behavior of the $\beta_{\rm w} = 10^6$ jet is relatively simple.
While some asymmetries and tilt are present, overall it seems to efficiently propagate to large radii near the $z$-axis (the line of sight to earth) while remaining collimated and stable.  
In contrast, the jet in the $\beta_{\rm w}=10^2$ simulation has a much more complicated evolution.
On the smallest scales ($r\lesssim  10 r_{\rm g}$), the jet tends to be aligned with the black hole spin axis some of the time, but at larger scales it tends to propagate perpendicular to that axis most of the time.  
Occasionally the jet aligns out to $\sim$ 50 $r_{\rm g}$ when $\phi_{\rm BH}$ is high (compare with Figure \ref{fig:time_plots}) but this is typically short lived (see Figure \ref{fig:tilt_angles}).
The jet in this simulation is also more intermittent than the $\beta_{\rm w}=10^6$ simulation and at times can be quite weak (e.g., the last two rows in Figure \ref{fig:jet_3D_comp}).  This often occurs after a period of strong alignment (large $\phi_{\rm BH}$ during which the jet is pushing directly against the infalling accretion flow that is the primary source of magnetic flux powering the jet.  
That is, in a sense, the jet `bites the hand that feeds it'.

\begin{figure}
\includegraphics[width=0.45\textwidth]{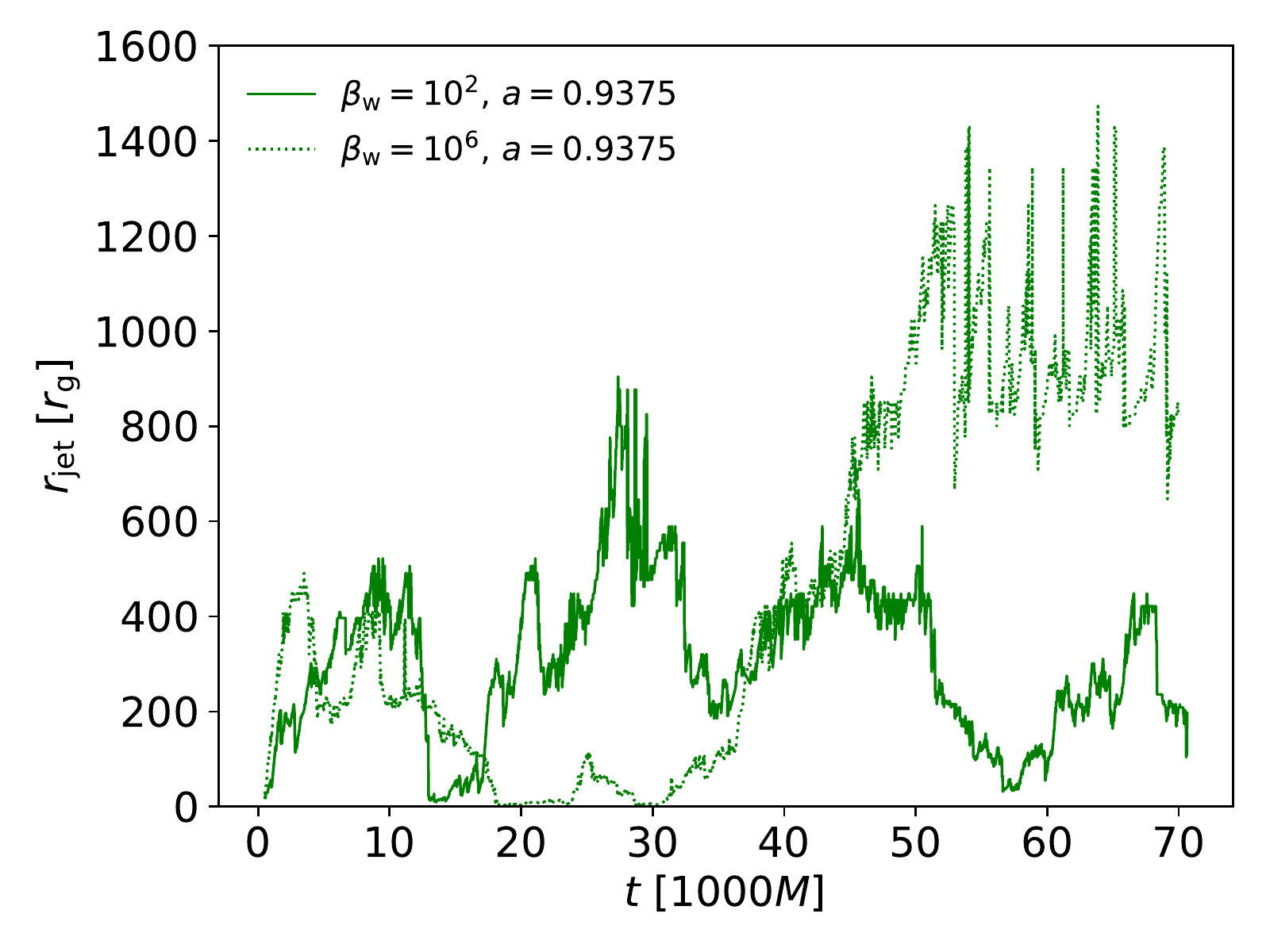}
  \includegraphics[width=0.45\textwidth]{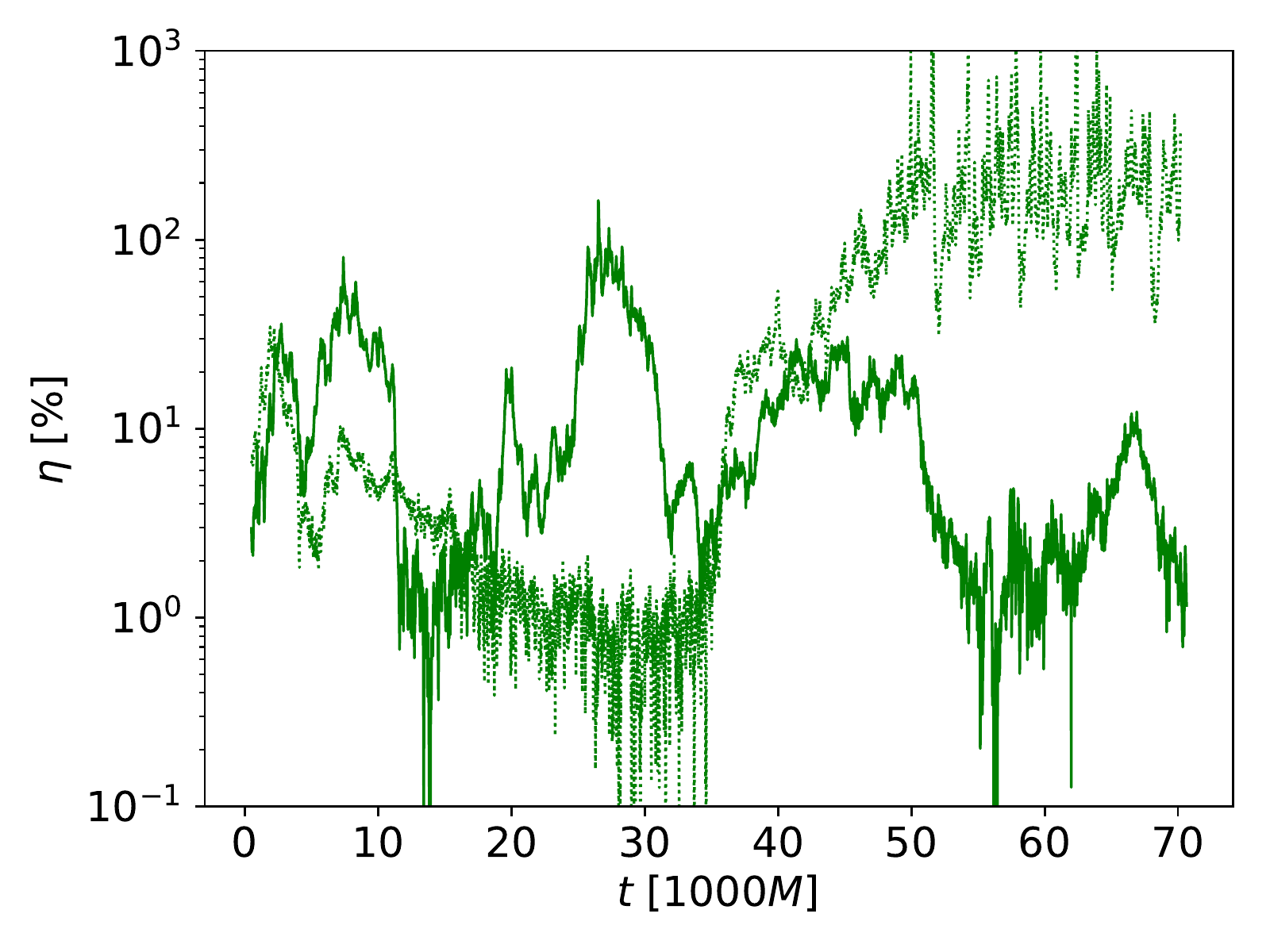}
\caption{ 
Jet quantities as a function of time in the two $a=0.9375$ simulations. 
 Bottom:  Upper jet radius, $r_{\rm jet}$ ($\theta < {\rm \pi}/2$).  
 Top: Outflow efficiency, $\eta$.While both simulations show quiescent periods with weak or no jets (e.g., $\sim$ 10--25${, }000$ $M$ for $\beta_{\rm w}=10^2$ and 12--18${, }000$ $M$ for  $\beta_{\rm w}=10^6$), eventually the $\betaw=10^6$ jet reaches the edge of the simulation box at $r\approx 1600 M$ once the MAD state is reached (compare with Figure \ref{fig:time_plots}).  The $\betaw=10^2$ jet never reaches past $\sim$ 800 $r_{\rm g}$ and tends to recede to small radii during periods of low efficiency.  
 }
\label{fig:rjet}
\end{figure}

The reason the jets in the two simulations propagate in such different directions is because of the large scale orientation of the accretion flow.  As  seen previously in Figure \ref{fig:intermediate}, the $\beta_{\rm w}=10^2$ simulation has a low density region caused by a combination of centrifugal and magnetic forces mostly aligned with the $y$-axis (that is, perpendicular to the black hole spin axis in the $z$ direction), while the $\beta_{\rm w}=10^6$ simulation has this low density region mostly aligned with the $z$-axis.
These low density cavities provide paths of least resistance in which the jets can propagate, or, in other words, the higher density accretion flows contain and collimate the electromagnetically powered outflows. 
As a result, the jets are not strongly hindered by the infalling accretion flow and can reach larger radii. 
To quantify this, we follow \citet{Liska2019} in defining the jet as all regions with $b^2/2 > 1.5 \rho $ and then call $ r_{\rm jet}^{+/-}$ the maximum radii for which $b^2/2 > 1.5 \rho $ and $\theta<{\rm \pi}/2$ (+) or $\theta>{\rm \pi}/2$ (-).
$r_{\rm jet}^{+}$, which has similar behavior as $ r_{\rm jet}^{-}$ is shown as a function of time in the top panel of Figure \ref{fig:rjet} for both $a=0.9375$ simulations. 

Both jets initially reach $\approx$ $400\ r_{\rm g}$ and then essentially disappear for a time before being reignited.  The $\betaw=10^6$ jet then proceeds to approach the outer edge of the box (located at $1600\ \rg$) at which point it stalls at $\approx$ $1000\ \rg$. Conversely, the $\betaw=10^2$ jet never reaches those outer radii, instead going through phases of pushing outwards and then falling back over $20{,}000$--$30{,}000\ M$ cycles.  The difference is associated with a difference in magnetic flux threading the black hole, $\phi_{\rm BH}$ (bottom panel of Figure \ref{fig:time_plots}), and thus jet power. 
This is clearly seen in the bottom panel of Figure \ref{fig:rjet}, which plots the relative outflow efficiency, $\eta \equiv \dot E /|\dot M|$, vs. time for the two jets. 
The $\betaw=10^2$ does reach $\sim$ 100\% efficiency at $30{, }000$ $M$ but then falls to $\lesssim10\%$ at later times. 
Conversely, the $\betaw=10^6$ jet develops a particularly high efficiency at later times, $\eta\approx$ 200--300\% for $t \gtrsim 45{,}000\ M$ once the MAD state is reached.  

The fixed outer boundary conditions of our GRMHD simulations do not allow us to study what happens to the jet once it reaches $r\gtrsim 1600\ \rg$.  
Once the jet reaches the boundary it effectively runs into a wall and abruptly stalls.  
The larger radii flow in the intermediate-scale MHD simulation is not affected, as we discuss in \S \ref{sec:no_feed}.
Future work can study how the jet might transport energy, momentum, and magnetic flux to larger radii and potentially ``feedback'' onto the large-scale flow and field structure.

\subsubsection{Emission Properties}
\label{sec:emission}

\begin{figure}
\includegraphics[width=0.44\textwidth]{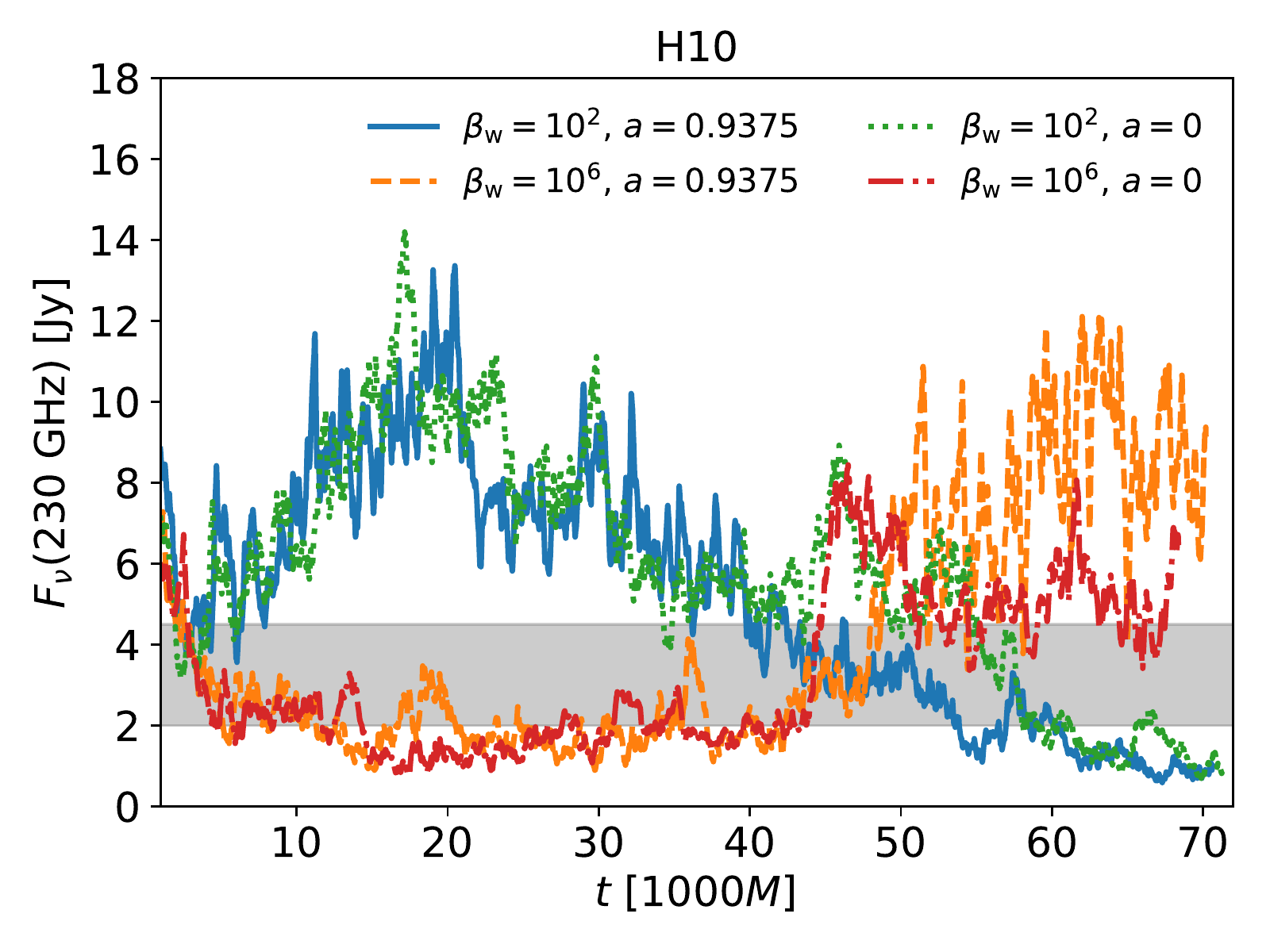}
\includegraphics[width=0.44\textwidth]{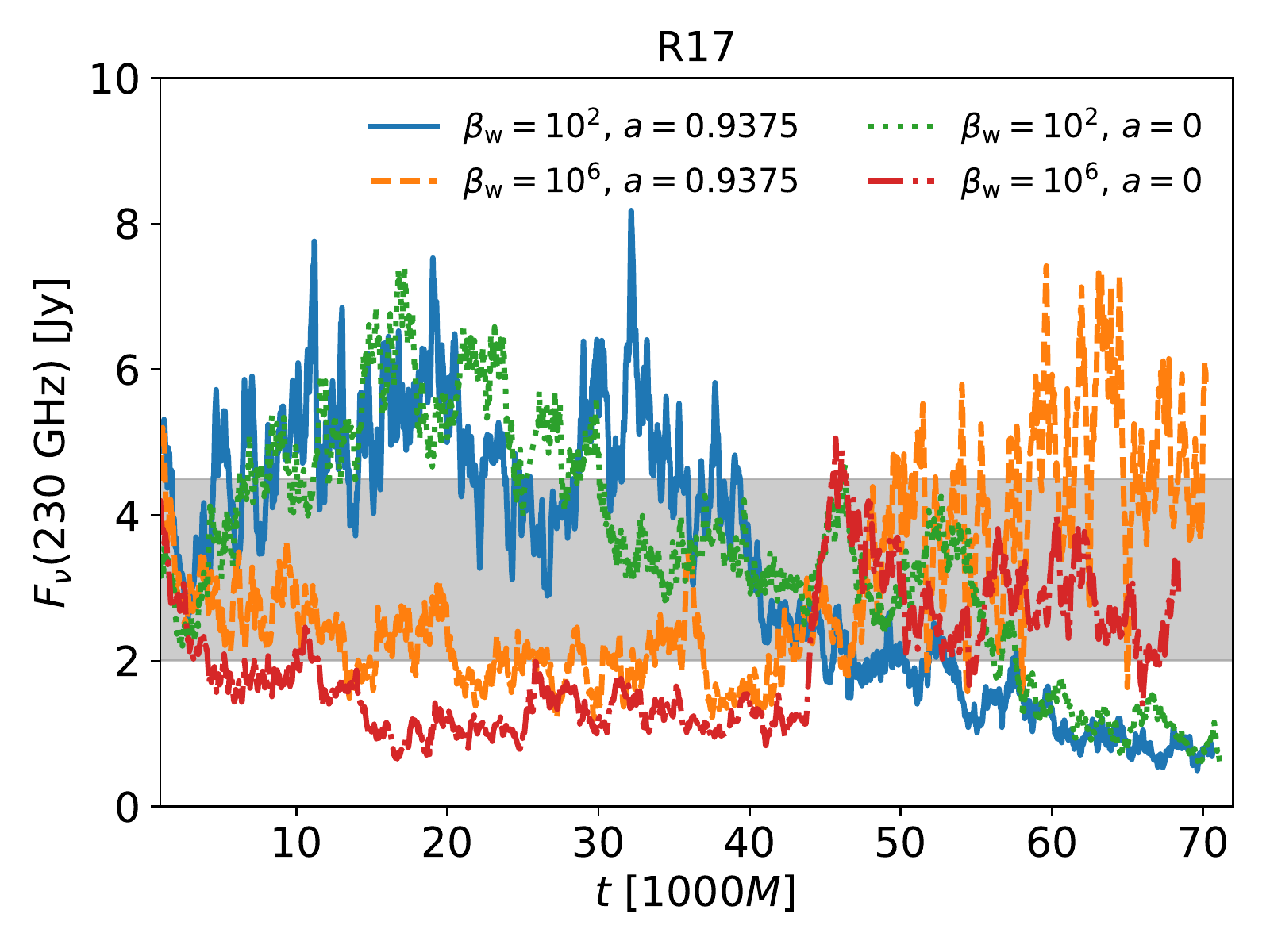}
\includegraphics[width=0.44\textwidth]{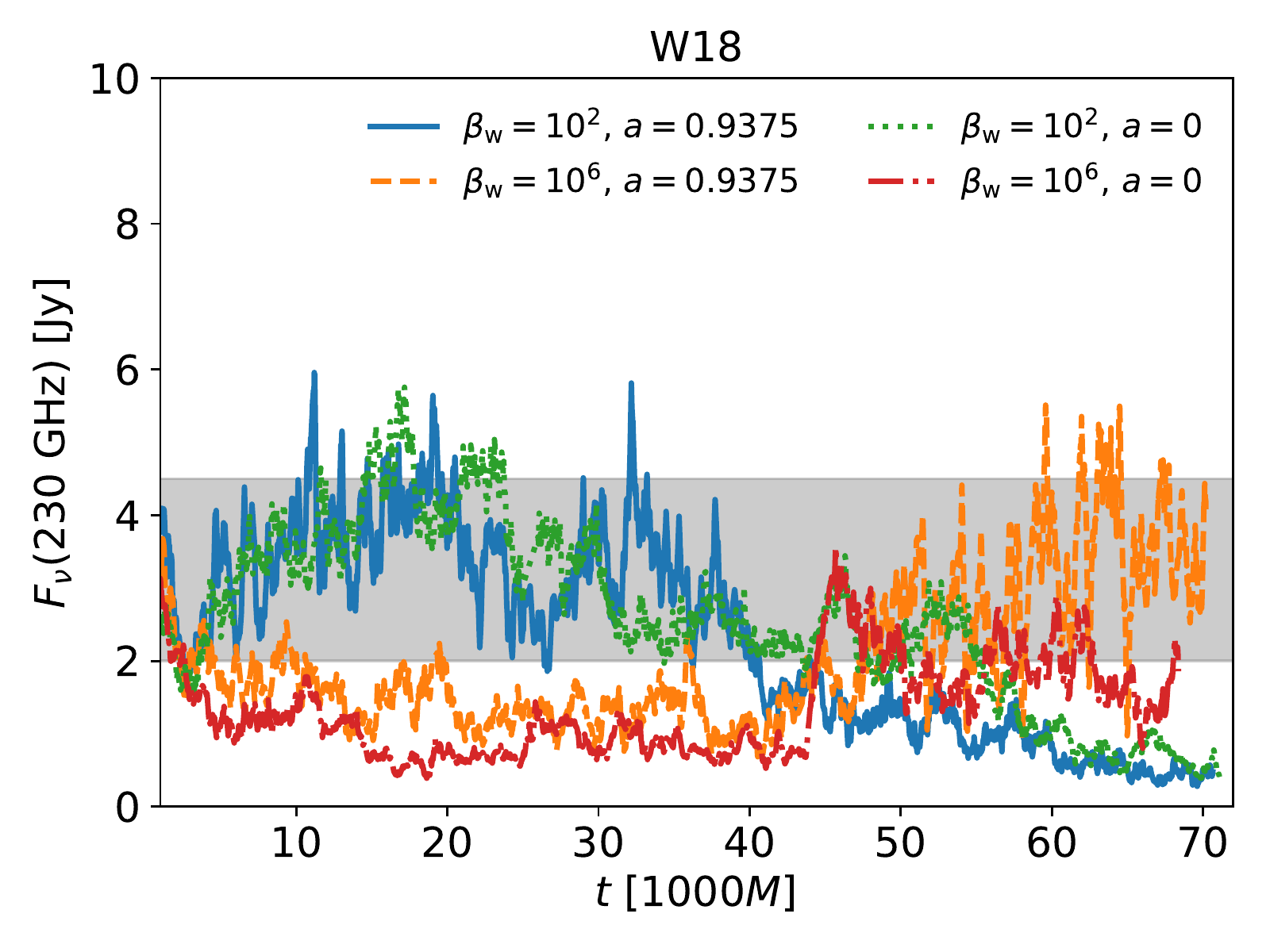}
\caption{230 GHz flux as a function of time in our four simulations for the H10 ``turbulent'' heating model (top) and the R17 (middle) and W18 (bottom) ``reconnection'' heating models.  The shaded region represents the observed range of values for Sgr A* \citep{Doeleman2008,Dexter2014,Bower2015,Iwata2020,Murchikova2022,Wielgus2022_light_curve}, roughly 2--4.5 Jy.  
The $\beta_{\rm w}=10^6$ simulations with R17 and W18 heating predict fluxes close to the observed range, particularly once the MAD state is reached at $t\gtrsim 45{, }000\ M$.
H10 heating models tend to result in flux values signicantly higher than observed at certain times in every simulation.   For all heating models the $\beta_{\rm w}=10^2$ simulations tend to have fluxes lower than observed at late times ($t\gtrsim 50{, }000\ M$).  }
\label{fig:flux_vt}
\end{figure}

Here we focus on three particular unresolved observational probes of the accretion flow: the 230 GHz flux, the unresolved rotation measure, and the unresolved linear polarization fraction.  We save a detailed analysis of 230 GHz images, spectral energy distribution, and polarization maps to a later work.  

The 230 GHz fluxes, $F_\nu(\textrm{230\ GHZ})$, predicted by our simulations are shown in Figure \ref{fig:flux_vt} for the three heating models.  
For comparison, the observed range of values for Sgr A* are shown as a shaded region \citep{Doeleman2008,Dexter2014,Bower2015,Iwata2020,Murchikova2022,Wielgus2022_light_curve}.
2.4 Jy \citep{Doeleman2008} is often taken as the ``canonical'' mean 230 GHz flux value in Sgr A* accretion modeling for the purposes of normalization of physical units in scale-free simulations.
However, the true mean appears closer to 3--4 Jy with variation in the range of 2--4.5 Jy. 
We therefore use the latter range as representative of the data, corresponding to roughly 2$\sigma$ variability \citep{Wielgus2022_light_curve}.
The differences between the $\beta_{\rm w}=10^2$ light curves and the $\beta_{\rm w}=10^6$ light curves for a given $a$ (factors of as much as $\sim$ 10) are much more significant than the differences between the $a=0$ and $a=0.9375$ light curves for a given $\beta_{\rm w}$ (at most factors of $\sim$ 2). 
The H10 heating model (with a maximum electron heating fraction $f_{\rm e} =1$ in regions of low $\beta$) overall tends to result in higher fluxes than the R17 and W18 models (with a maximum $f_{\rm e}\approx 0.5$), with the W18 model resulting in slightly lower fluxes than R17 by a factor of $\sim$ 50\%.
At early times ($t \lesssim 40{, }000\ M$), the fluxes in the $\beta_{\rm w}=10^2$ simulations fall nicely within the observed range for the W18 heating model.  
During the same time the R17 model for the same simulations falls sometimes within the observed range but is often too high (by a factor of $\lesssim 2$), while the H10 heating model produces fluxes higher than observed by a factor of $\gtrsim$ a few. 
At later times ($t\gtrsim 50{, }000\ M$) $F_\nu$ in the $\betaw=10^2$ simulations dips slightly below the observed range to $\approx$ 1 Jy for all heating models.  
The fluxes in the $\beta_{\rm w}=10^6$ simulations show the opposite trend.  At early times ($t \lesssim 45{, }000 \ M$), the values for $F_\nu$ using H10 and R17 are just under the minimum observed fluxes around 1-3 Jy while for W18 they are slightly lower at 1--2 Jy.  
Once the MAD state is reached at later times ($t\gtrsim 45{, }000\ M$) in the $\betaw=10^6$ simulations, the 230 GHz fluxes increase.  
For the H10 heating model the simulations reach 6--10 Jy (tending to be well above observations), for the R17 heating model they reach 2--6 Jy (tending to fall within the observed range), and for the W18 model they reach 1--5 Jy (also tending to fall within the observed range).   
In terms of the 230 GHz flux, the W18 model with the $\betaw=10^6$, $a=0.9375$ simulation best matches the observations once the MAD state is reached at $t\gtrsim 45\ M$, while the W18 model with the $\betaw=10^2$ simulations best match the observations at earlier times .  
Considering the whole light curve, the R17 model with the $\betaw=10^6$, $a=0.9375$ simulation tends to best match the observations, falling well within $\sim$ $50\%$ of the observed range.
Conversely, the H10 model with the $\betaw=10^2$ simulations fairs poorest, with $F_\nu$ either well above or moderately below the observed range most of the time.

Computing the RMS variability fractions as the standard deviation of the fluxes shown in Figure \ref{fig:flux_vt} divided by the mean, we find that all simulation/heating model combinations are variable at the 50--60\% level (except for the $\betaw=10^6$, $a=0.9375$ H10 model with $>70\%$ variability) while the observed value is closer to 20--40\% \citep{Dexter2014,Bower2015,Dexter2020,Wielgus2022_light_curve}.  
On the other hand, if we exclude the initial SANE portion of the light curve for $\betaw=10^6$ ($t \lesssim 45{, }000\ M$), the resulting RMS variability fractions are lower, for instance, the $\betaw=10^6$, $a=0.9375$ simulation with R17 and W18 has values between 20--30\%.
If the MAD state continues indefinitely once reached, then these values may be a more accurate representation of the light curve variability over longer periods of time.
Even for the $\betaw=10^2$ simulations that do not go MAD, it is not clear how much, if any, of the initial evolution of the system is transient and perhaps skewing the measured flux variability to higher values. 
Addressing this possibility requires simulations run for significantly longer times.  
In any case, we emphasize that for a given electron heating model and simulation, there are no free parameters to adjust.  
Therefore it was no guarantee that \emph{any} simulation/heating model combination correspond with observations at all.  
Given the uncertainty in electron heating physics and the large dynamical range simulated to reach the event horizon from the WR stellar winds at large radii, we regard the reasonable agreement in Figure \ref{fig:flux_vt} between models and simulations as quite encouraging.

\begin{figure}
\includegraphics[width=0.44\textwidth]{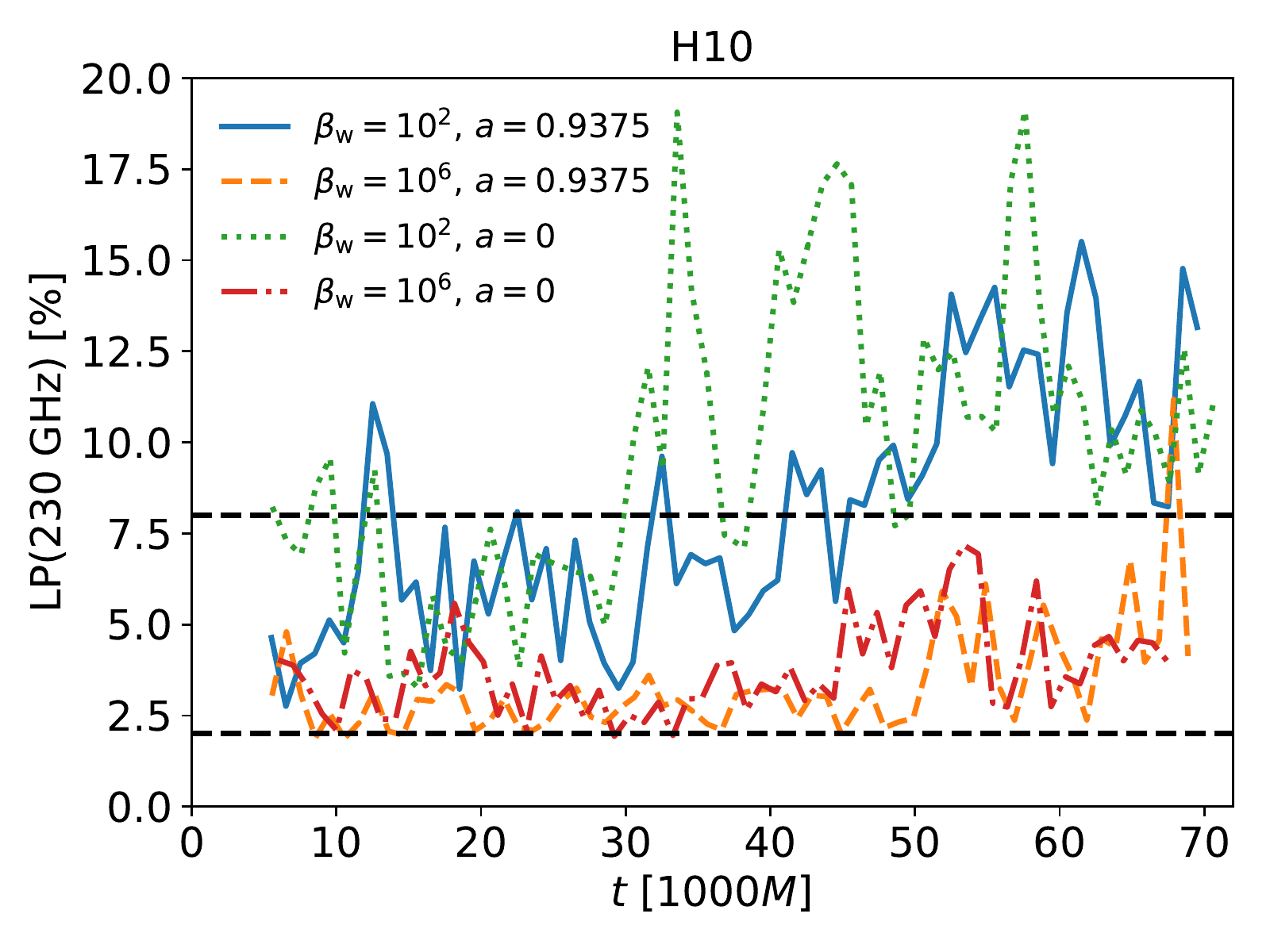}
\includegraphics[width=0.44\textwidth]{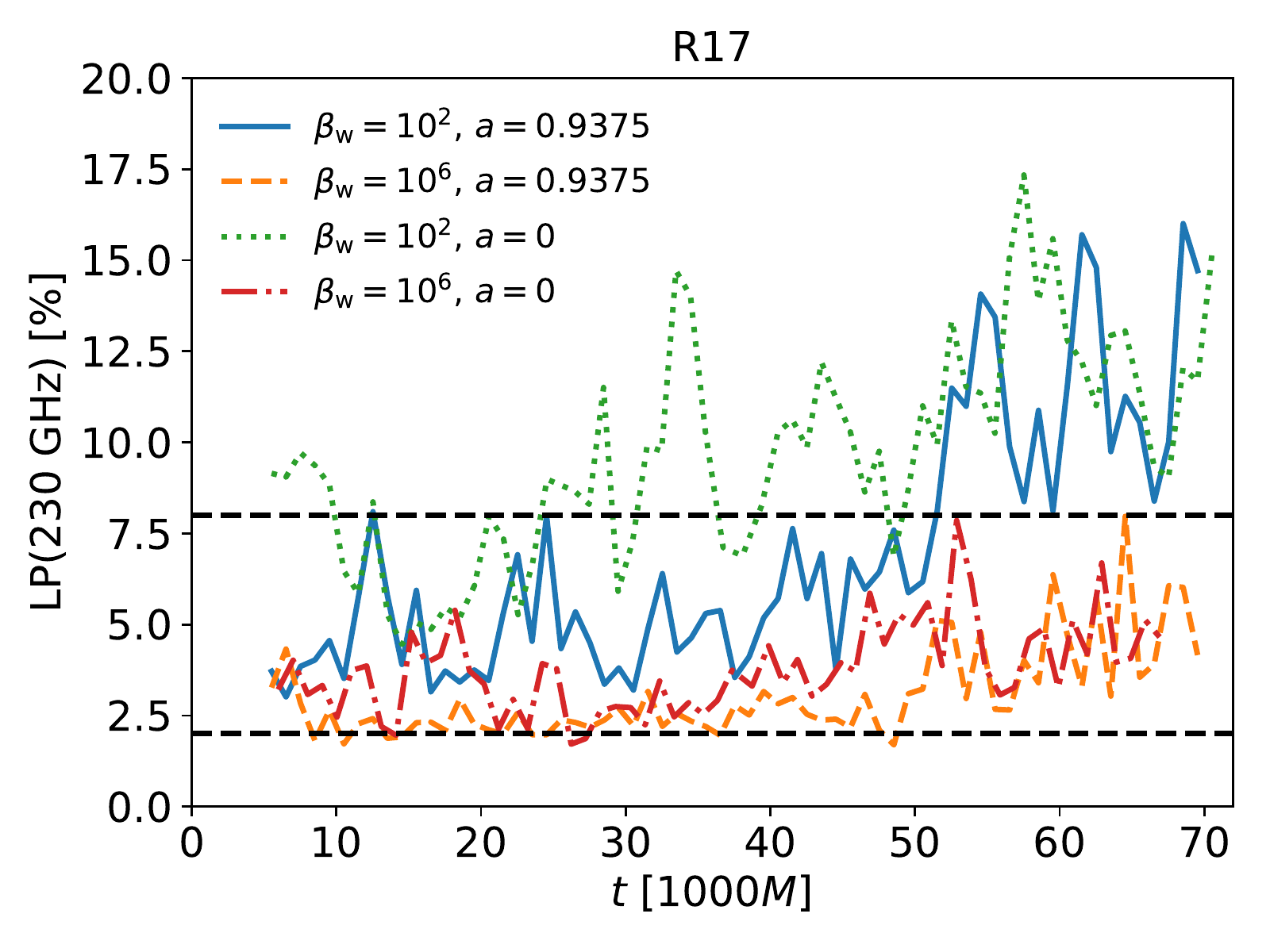}
\includegraphics[width=0.44\textwidth]{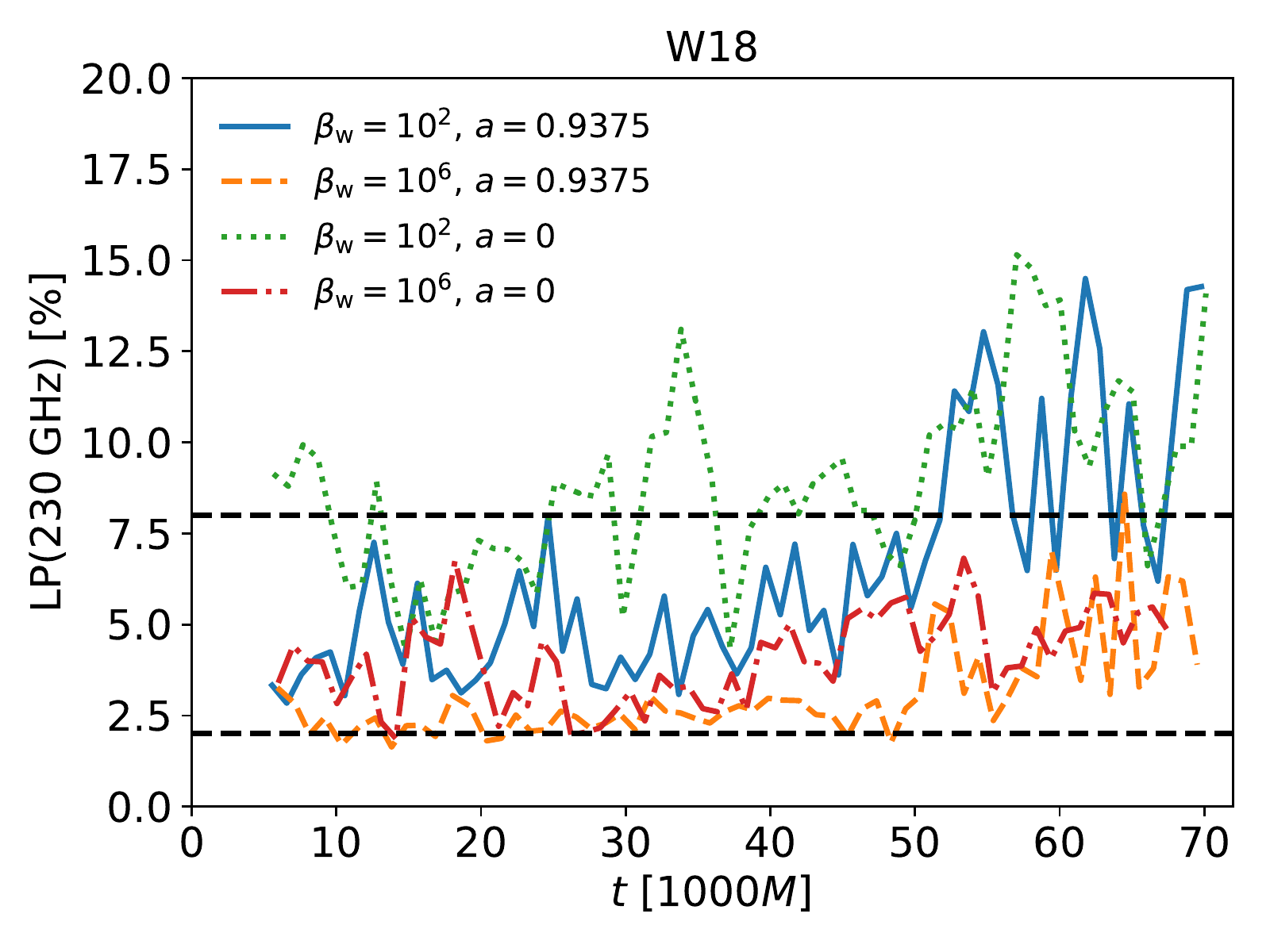}
\caption{Image-integrated linear polarization fraction as a function of time in our four simulations for the H10 (top), R17 (middle), and W18 (bottom) heating models. The dotted black lines demarcate the observed range for Sgr A* \citep{Bower2018}.  Data is averaged over 1000 $M$ ($\approx$ 6 hr) before being plotted for the purposes of readability. 
For all heating models, the LP of the $\betaw=10^6$ simulations falls neatly within the observed range almost all of the time, while for $\betaw=10^2$ the LPs often are too large compared to observations.  }
\label{fig:LP_vt}
\end{figure}

We compute unresolved linear polarization fractions using 
\begin{equation}
  \langle \textrm{LP}\rangle = \frac{\sqrt{\left(\sum Q\right)^2 + \left(\sum U\right)^2}}{\sum I},
\end{equation}
where $I$, $Q$, and $U$, are the Stokes parameters defined in the usual way and $\sum$ represents a sum over all pixels in the image.  
This quantity is shown for both heating models and all four simulations in Figure \ref{fig:LP_vt}.  
Instead of plotting the LP at each data point from the simulations, we average the data in time over 1000 M ($\approx$ 6 hr) intervals to make the figure more readable.
For comparison, the observed range of this quantity for Sgr A* is $\sim$ 2--8\% \citep{Bower2018}.  
The LPs for the $\beta_{\rm w}=10^6$ simulations fall within this range for essentially their entire duration, tending to stay around 2--5\% and are thus both consistent with observations for either electron heating model.
The LPs in the $\beta_{\rm w}=10^2$ simulations tend to be higher with higher magnitude variability. 
For $a=0.9375$ and H10 it occasionally rises above 8\% but most of the time varies between $\approx$ 3--8\%.  The same simulation with the R17 heating model has slightly lower LP values so that it falls within the observed range essentially all of the time.
The LP in the $\beta_{\rm w}=10^2$, $a=0$ simulation, however, is often well above 8\% for both heating models, reaching maximum values of $\gtrsim 14$\%.

To compute the unresolved rotation measure predicted by our simulations, we first determine the net electric vector polarization angle (EVPA) using 
\begin{equation}
\langle \textrm{EVPA}\rangle =  \frac{1}{2}\arctan \left(\frac{\sum U}{\sum Q}\right).
\end{equation} 
The rotation measure is then 
\begin{equation}
  \textrm{RM} = \frac{d \textrm{EVPA}}{d\lambda^2},
  \label{eq:RM_exact}
\end{equation}
which we evaluate at frequencies (230 GHz, 232 GHz), that is,
\begin{equation}
  \textrm{RM} = \frac{\textrm{EVPA}(232 \textrm{ GHz}) -\textrm{EVPA}(230 \textrm{ GHz}) }{ [c/(232\textrm{ GHz})]^2 - [c/(230\textrm{ GHz})]^2}.
  \label{eq:RM_num}
\end{equation}

If the RM is consistent with external Faraday rotation, EVPA $\propto$ $\lambda^2$ and Equation \eqref{eq:RM_exact} will give the same result at all wavelengths.  However, if there is significant \emph{internal} Faraday rotation (where emission and rotation are happening at similar locations/spatial scales), there can be departure from $\lambda^2$ dependence.  
The observed EVPA in Sgr A* above $\gtrsim $ 220 GHz is consistent with EVPA $\propto$ $\lambda^2$ \citep{Marrone2007,Bower2018} with an associated RM of $\sim$ $-5.6 \times 10^5 $ rad m$^{-2}$.  
In our simulations we generally find EVPA $\propto \lambda^2$ dependence for frequencies $\gtrsim$ 200 GHz but see departures at lower frequencies (consistent with the simulations in \citealt{Dexter2020}). Since there is no measurements of the EVPA at those frequencies, this is consistent with observations.  

Since the GRMHD simulations only extend to $r\approx 1600 \ r_{\rm g}$, we also add to this RM the RM computed from the intermediate-scale MHD simulation assuming point-source emission (a very good approximation at that scale).  We describe this portion of the calculation in more detail in \S \ref{sec:disc_rm} (see Equation \ref{eq:pseudo_RM}).  For the fiducial four GRMHD simulations this larger scale contribution is sub-dominant. 

\begin{figure*}
\includegraphics[width=0.42\textwidth]{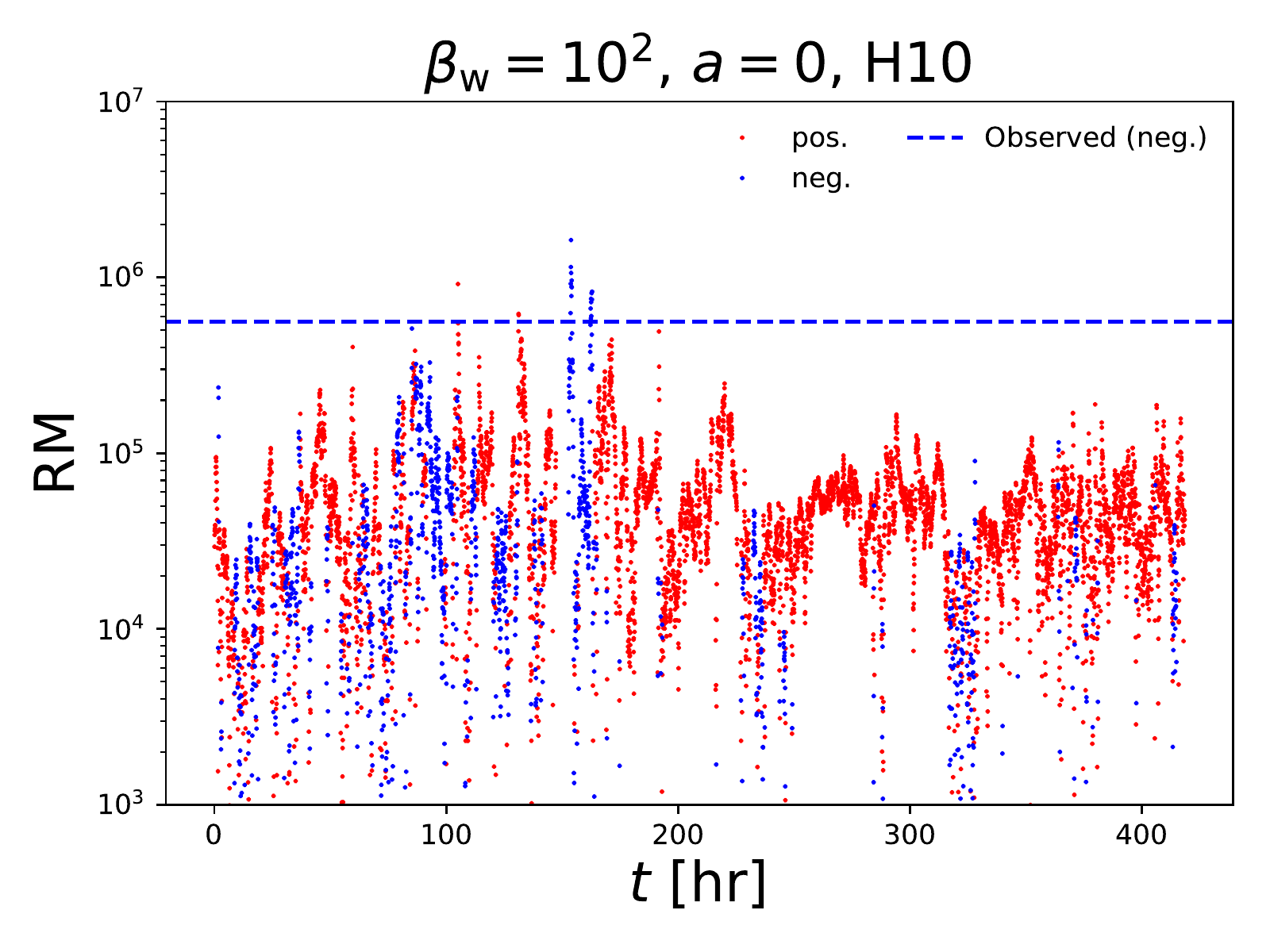}
\includegraphics[width=0.42\textwidth]{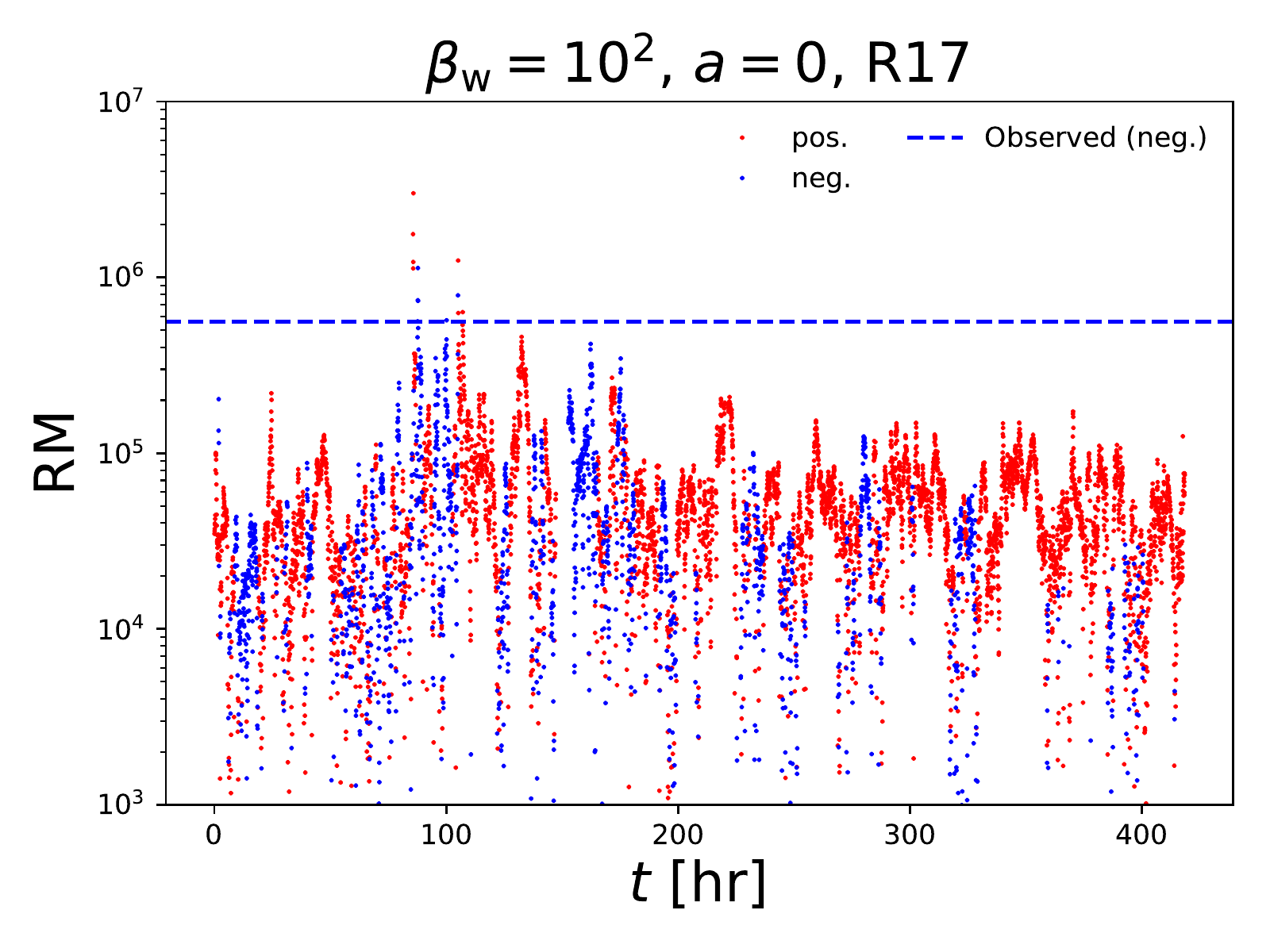}
\includegraphics[width=0.42\textwidth]{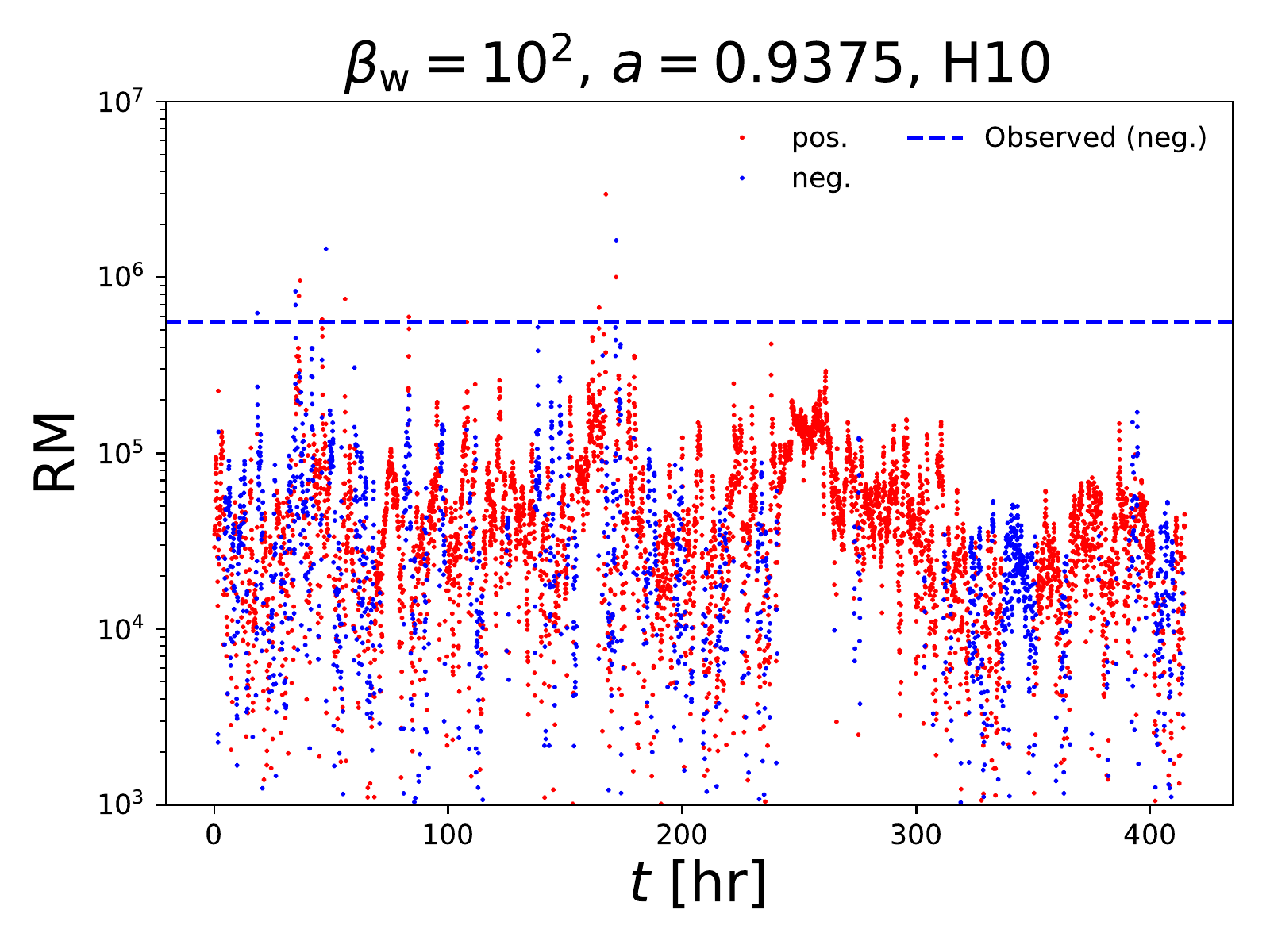}
\includegraphics[width=0.42\textwidth]{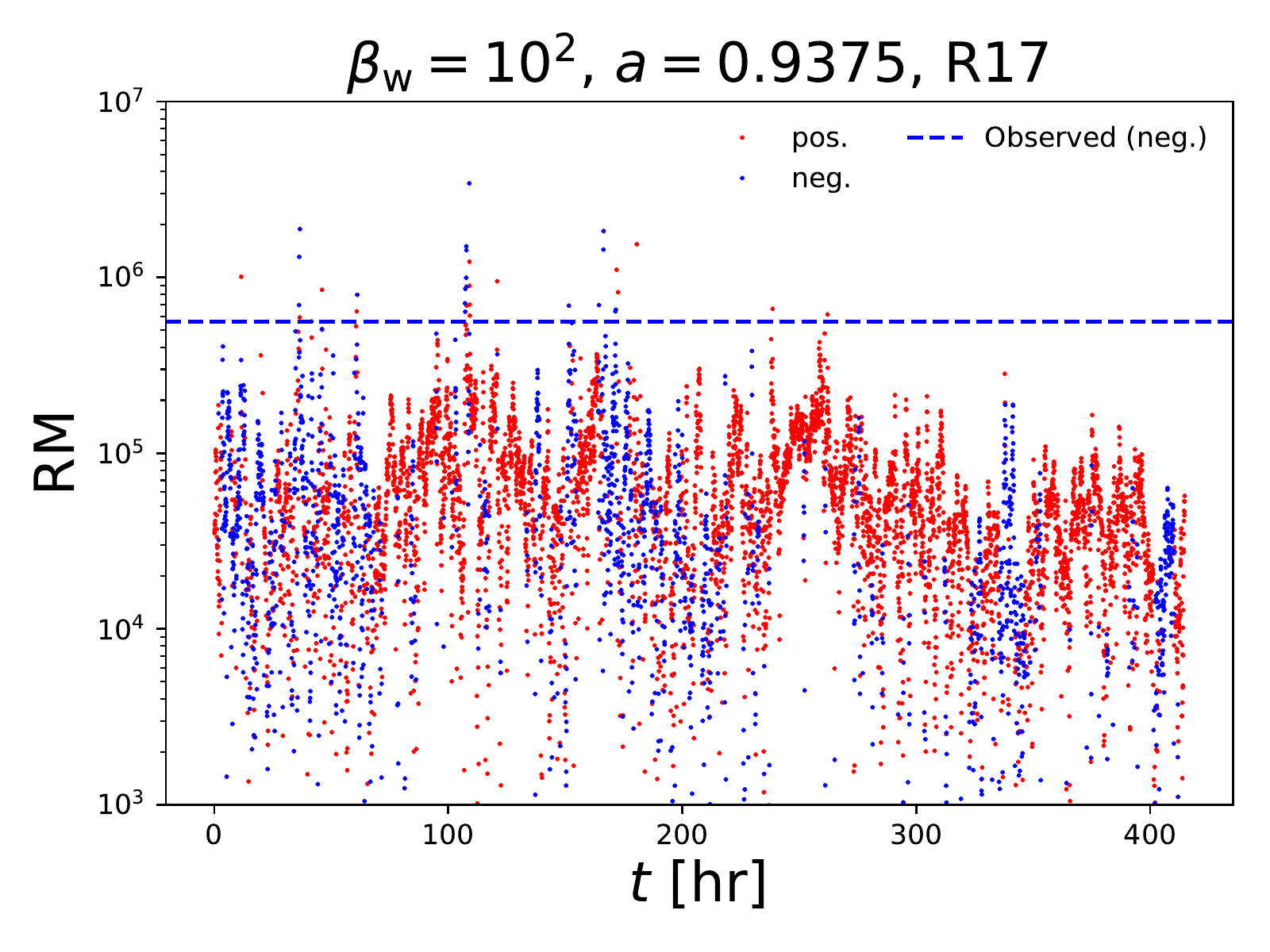}
\includegraphics[width=0.42\textwidth]{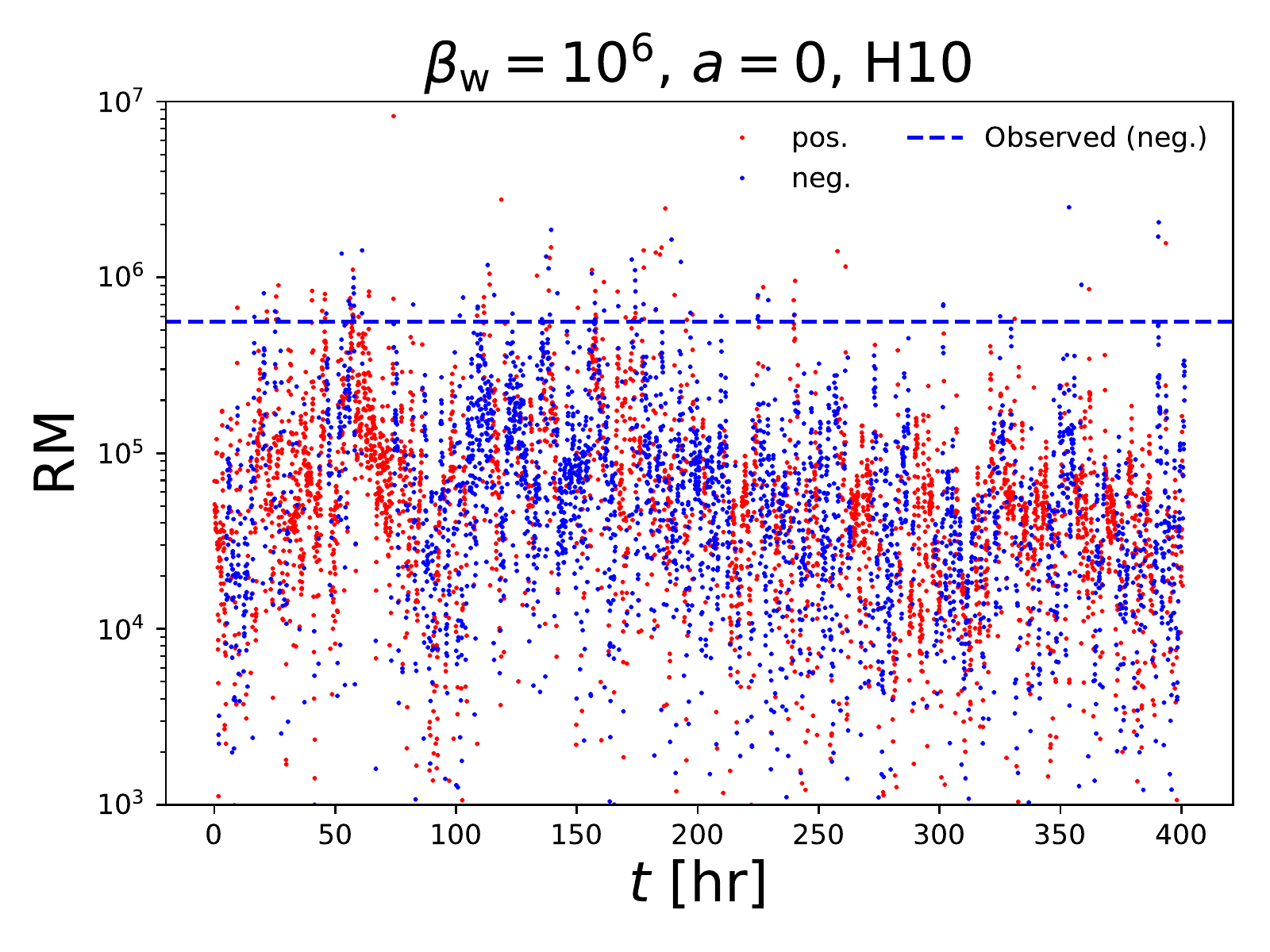}
\includegraphics[width=0.42\textwidth]{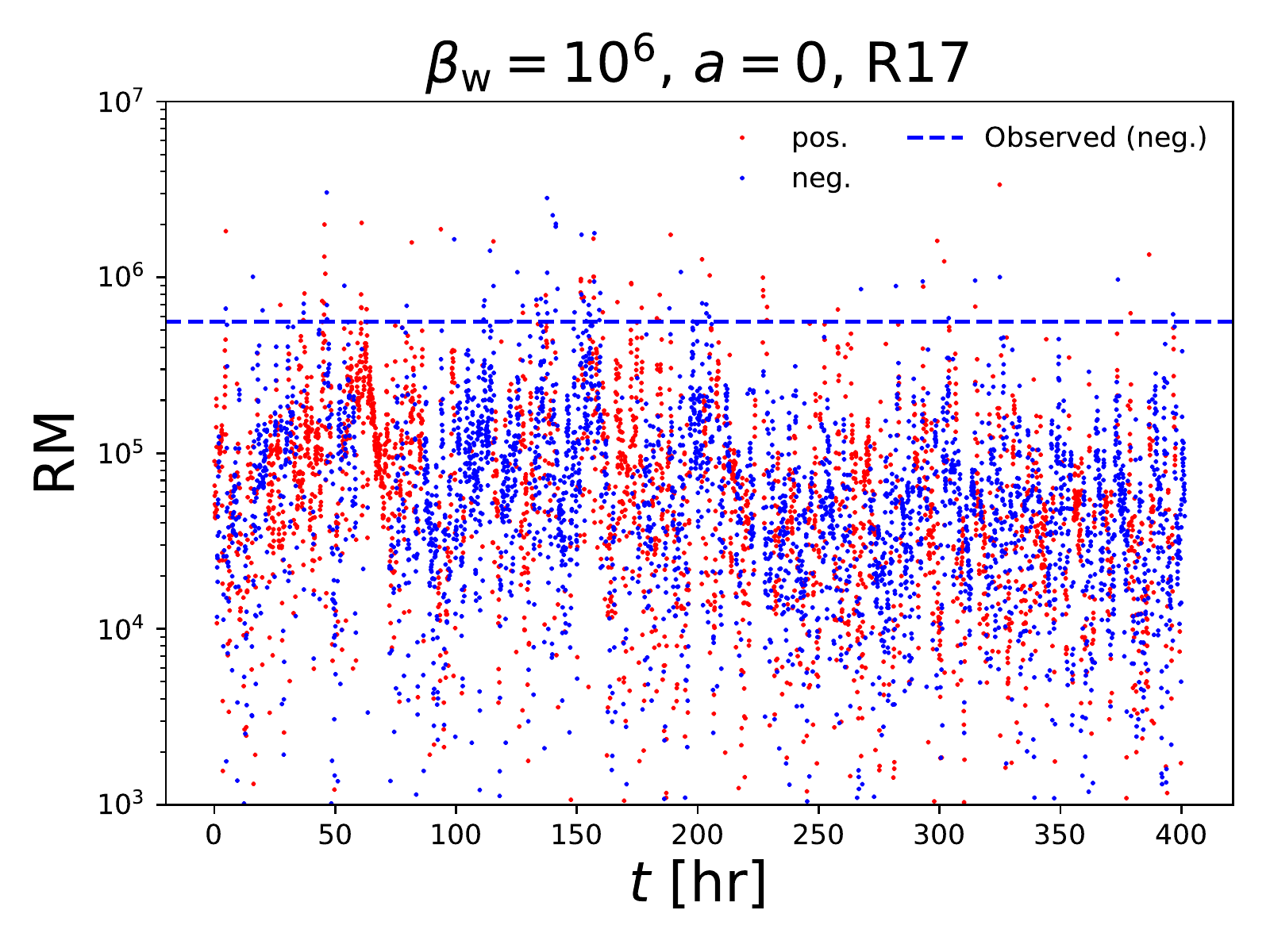}
\includegraphics[width=0.42\textwidth]{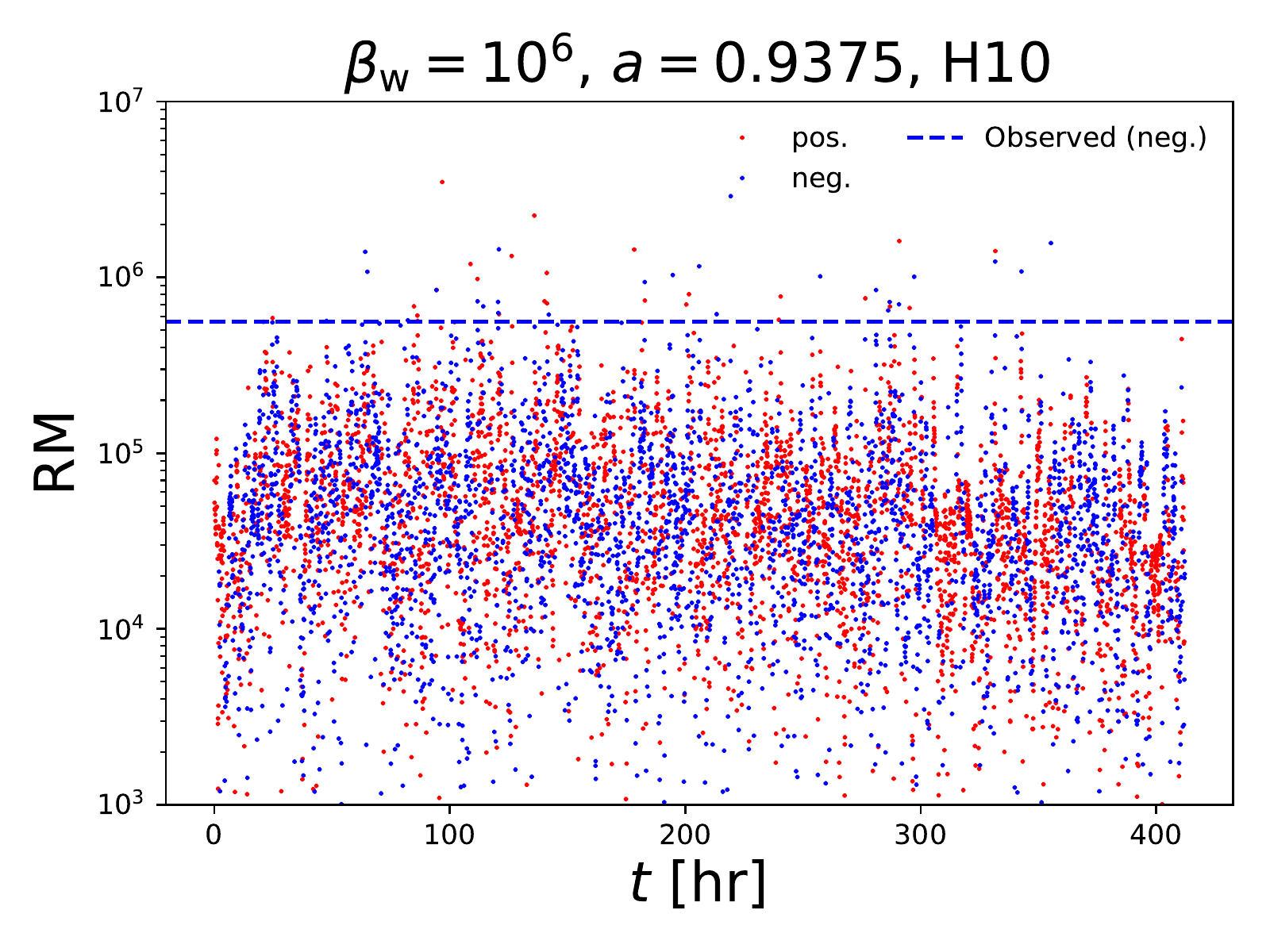}
\includegraphics[width=0.42\textwidth]{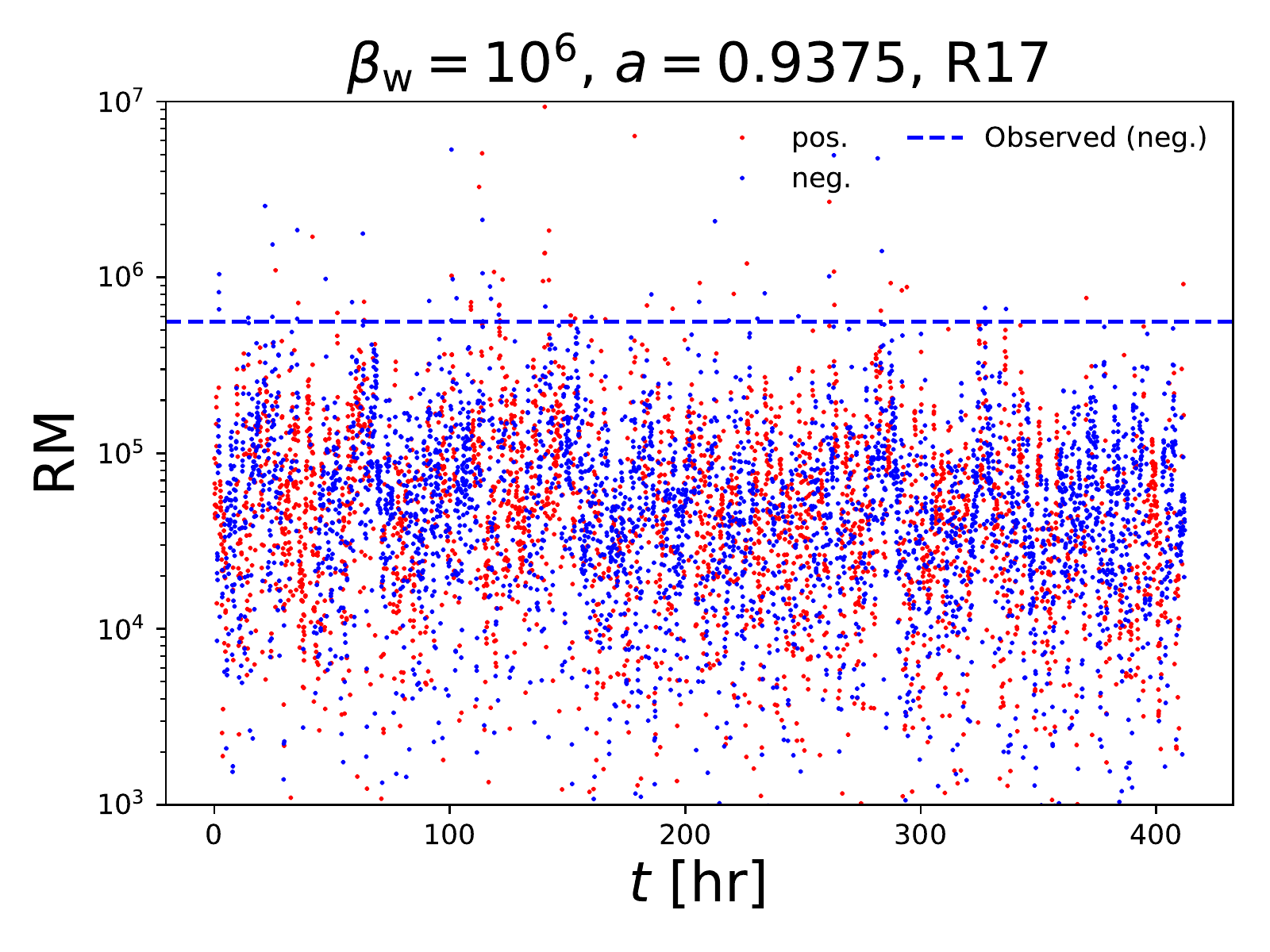}
\caption{ 
Rotation measure as a function of time for H10 and R17 heating models in the four simulations (the RMs for the W18 heating model look similar).  Blue points represent negative RM values and red points represent positive RM values.  The blue dashed line is the observed value (though it varies significantly in time). The RMs of the four simulations tend to be a factor of at least a few lower in magnitude than the mean observed value and change sign on the order of $\sim$ 10 hours. }
\label{fig:rm}
\end{figure*}

In Figure \ref{fig:rm}, we show the rotation measure computed in this way as a function of time for our four simulations and the H10 and R17 heating models (the W18 heating model produces a very similar RM).  
The observed mean value of $-5 \times 10^5$ rad/m$^2$ is also plotted for comparison.
Qualitatively, the differences in RM between the two electron heating models for a given simulation is small, which makes sense because the RM is only indirectly dependent on electron temperature in that it becomes suppressed for relativistically hot electrons.
That is, the nonrelativistic expression for external Faraday rotation depends only on the electron number density and magnetic field parallel to the line of sight and not the electron temperature.
Spin has a stronger effect on the RM in the $\beta_{\rm w}=10^2$ simulations (compared to the $\beta_{\rm w}=10^6$ simulations) because it can significantly alter the structure of the accretion flow and jet (as discussed in \S \ref{sec:accretion} and \S \ref{sec:jet}).
In terms of magnitude, the $\betaw=10^2$ simulations significantly under-produce the RM by a factor of $\gtrsim 10$.  
The $a=0.9375$ simulations tend to have a slightly higher RM by a factor of $\sim$ a few, but still average around $5 \times 10^4$ rad/m$^2$.
The RM values in the $\betaw=10^6$, $a=0.9375$ simulation are comparable to the values in the $\betaw=10^2$, $a=0.9375$ simulation, if not slightly higher.  
The $\betaw=10^6$, $a=0$ RM values are higher still and closer to observations, at rare times reaching magnitudes $\gtrsim 5 \times 10^5$ rad/m$^2$.  
However, even in this simulation the RM tends to be $\lesssim 10^5$ rad/m$^2$ in magnitude.
In terms of variability, the predicted RM from all simulations changes sign far too rapidly to account for the fact that Sgr A*'s RM has consistently been measured as negative, including multiple observing campaigns between 2002--2007 and then again in 2018 (see Figure 10 in  \citealt{Bower2018}).   
The RMs in our simulations, in contrast, change sign on hour timescales, with the longest intervals of consistent sign lasting only $\approx$ 20 hours.
On the other hand, the $\emph{magnitude}$ of the observed RM does vary by factors of at least a few over the course of three hours \citep{Bower2018}, which is qualitatively consistent with the level of variability seen in the RMs plotted from our simulations in Figure \ref{fig:rm}.  
It is difficult to make a quantitative comparison of RM variability due to the relatively small sample size of measurements of RM measurements in Sgr A*.  

\subsubsection{Alternative Realizations of the Flow}
\label{sec:alt_real}

In addition to the four fiducial simulations, we add two $\betaw=10^2$ simulations with $a=0$ and $a=0.9375$ in which the intermediate simulation is initialized from $t=30$ yr data in the original large-scale MHD simulation (as opposed to $t=150$ yr). 
This particular time was chosen for having a relatively large amount of net magnetic flux for over half a century (see \S \ref{sec:disc_rm} for how we assess this via an approximation to the RM).
The GRMHD simulations are initialized from the intermediate-scale simulation at 0.24 yr (the same time used for the fiducial set of intermediate-scale simulations described) and run for $\sim$ $30{, }000\ M$.

Instead of repeating the entire analysis presented in the previous few subsections, we highlight a few key quantities that display the most interesting differences from those of the fiducial simulations.

\begin{figure}
\includegraphics[width=0.45\textwidth]{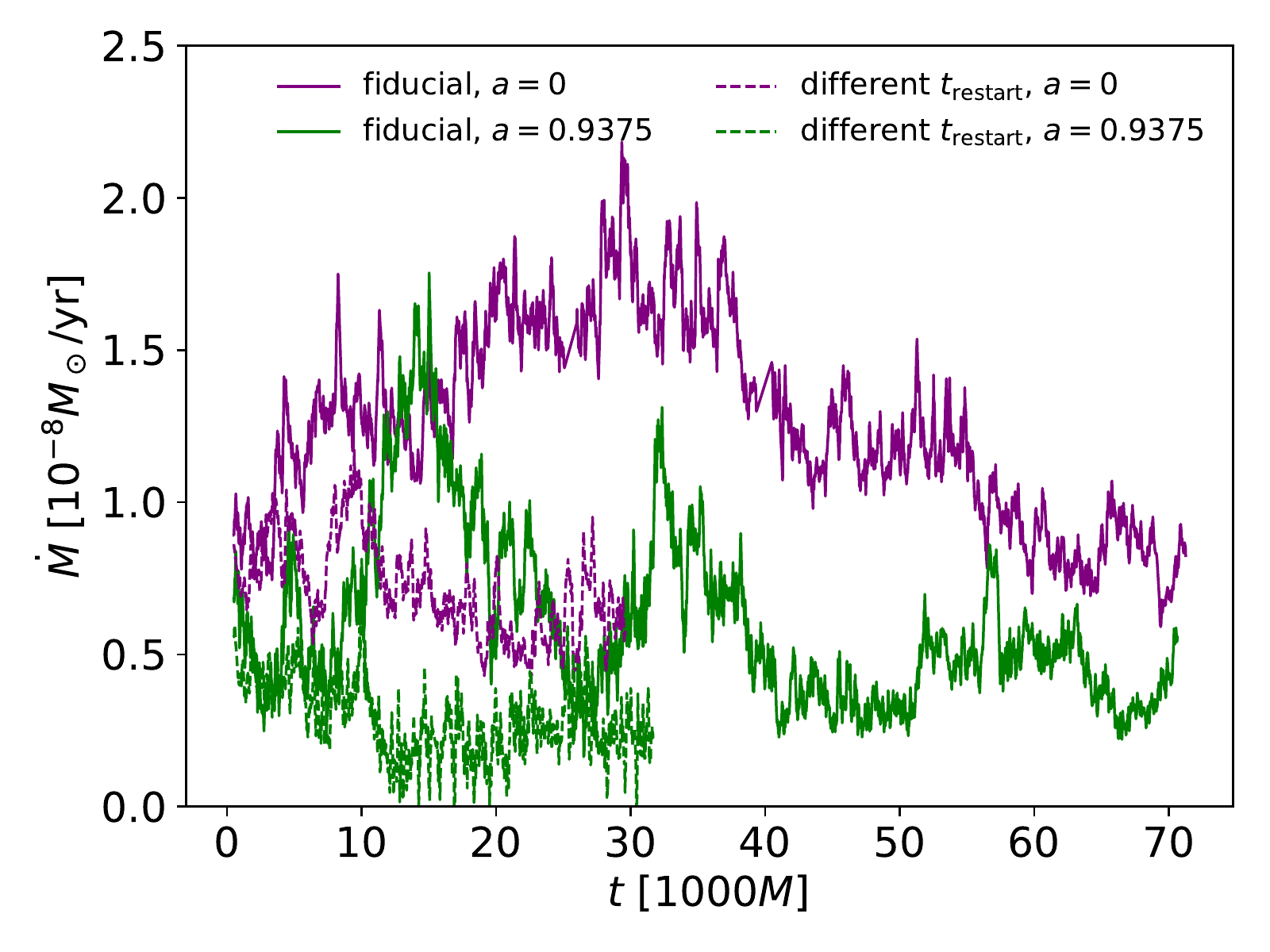}
\includegraphics[width=0.45\textwidth]{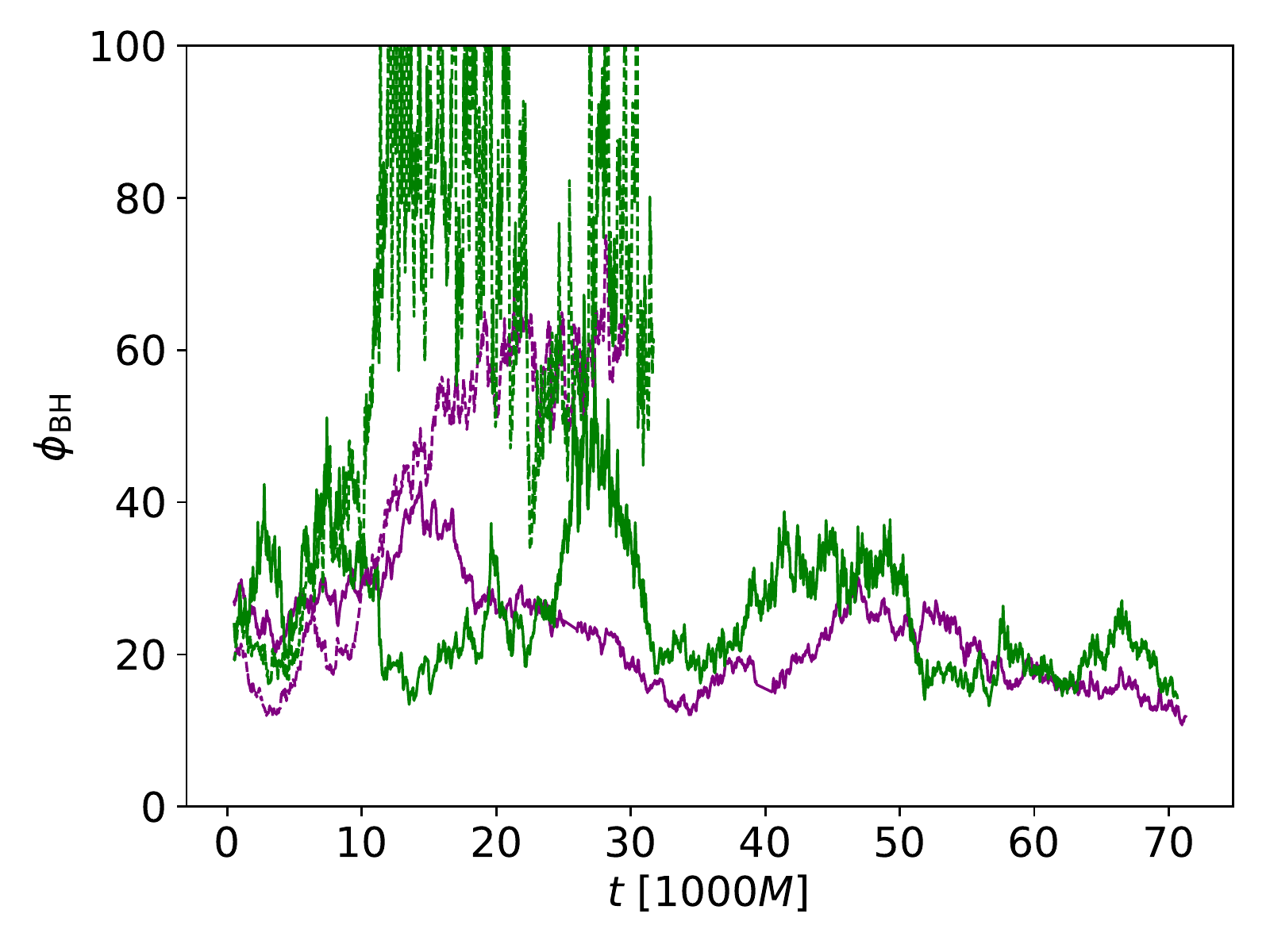}
\caption{ Accretion rate, $\dot M$, and magnetic flux threading the event horizon, $\phi_{\rm BH}$, as a function of time for the additional $\betaw=10^2$ simulations (dashed) initialized using data from a different time ($t_{\rm restart}$) compared to the fiducial $\betaw=10^2$ simulations (solid) discussed in \S \ref{sec:disc_rm} and \S \ref{sec:R20}.  $a=0.9375$ is green while $a=0$ is purple. 
In these units the MAD state corresponds to $\phi_{\rm BH}\sim$ 50--70.
The new simulations go MAD by $\sim$ $11{, }000$ $M$ for $a=0.9375$ and $\sim$ $20{, }000$ $M$ for $a=0$ and therefore also have  a lower accretion rate than the fiducial simulations by factors of $\lesssim 2$--4.   } 
\label{fig:time_plots_113}
\end{figure}

For instance, the resulting accretion rate and the horizon-penetrating flux are plotted as a function of time in Figure \ref{fig:time_plots_113} alongside the same quantities from the fiducial $\betaw=10^2$ simulations.  
The larger supply of net magnetic flux in the new simulations lead to them going MAD at $\sim$ $t=11{, }000\ M$ for $a=0.9375$ and $t=20{, }000\ M$ for $a=0$, saturating around $\phi_{\rm BH}\approx$ 60--70 (compared to $\sim$ 20 in the fiducial simulations).  Correspondingly, the values for $\dot M$ are smaller by a typical factor of $\sim 2$--4 during the time sampled due to the increased outward magnetic forces.
Note that these two sets of simulations are initialized from the same large-scale wind-fed MHD simulation, just at different times, yet only one set reaches the MAD state.

\begin{figure}
\includegraphics[width=0.45\textwidth]{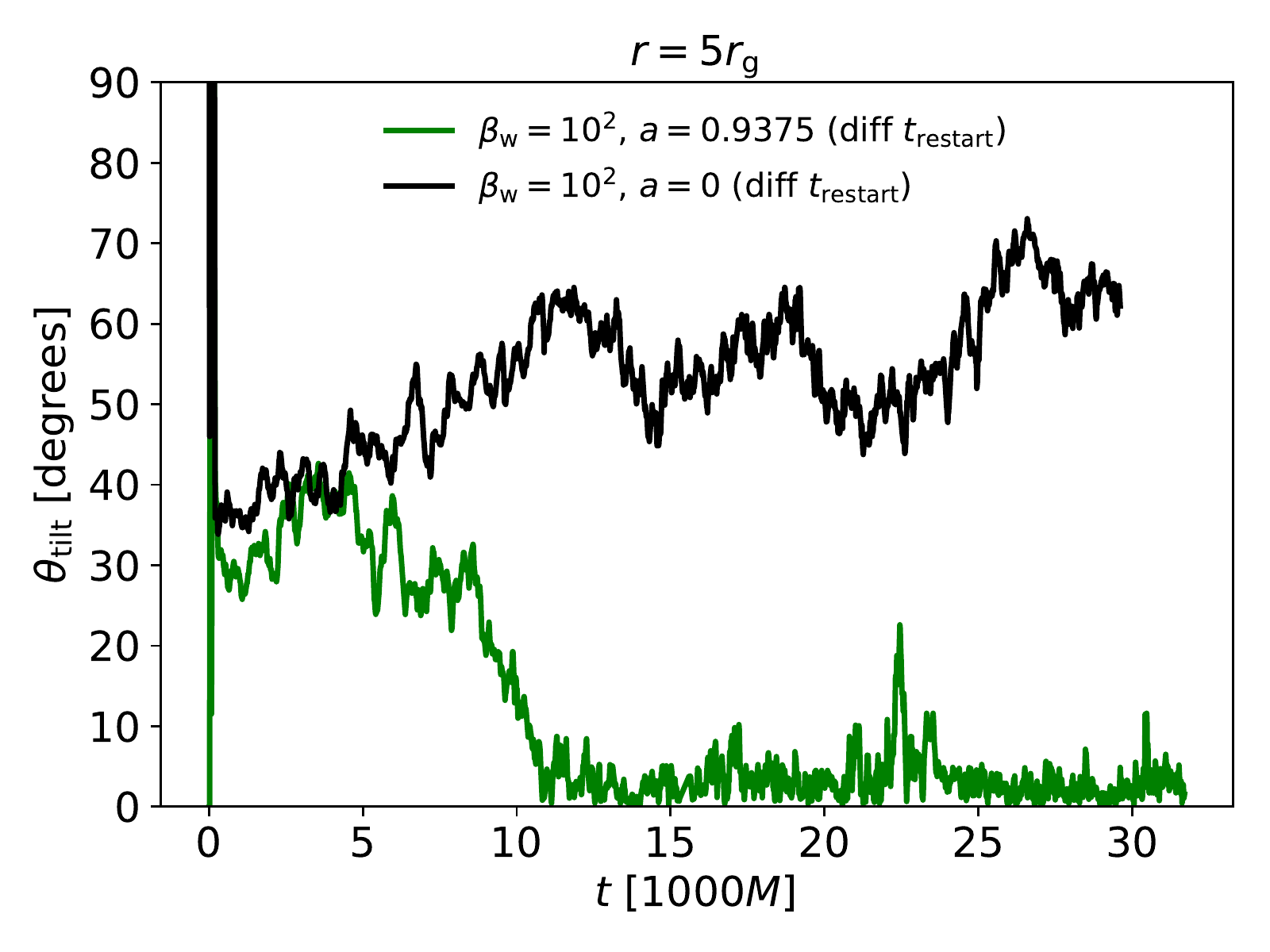}
\includegraphics[width=0.45\textwidth]{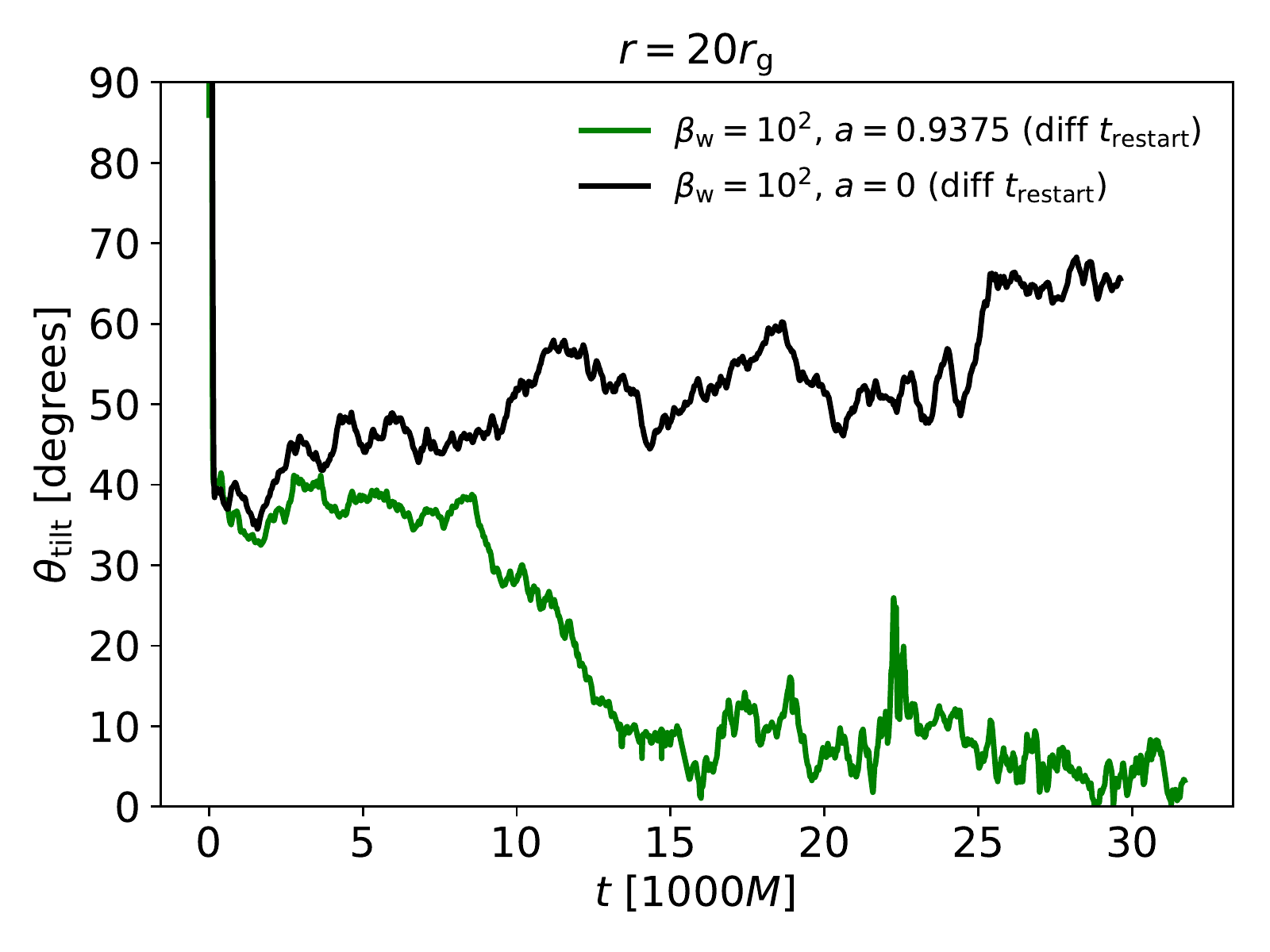}
\includegraphics[width=0.45\textwidth]{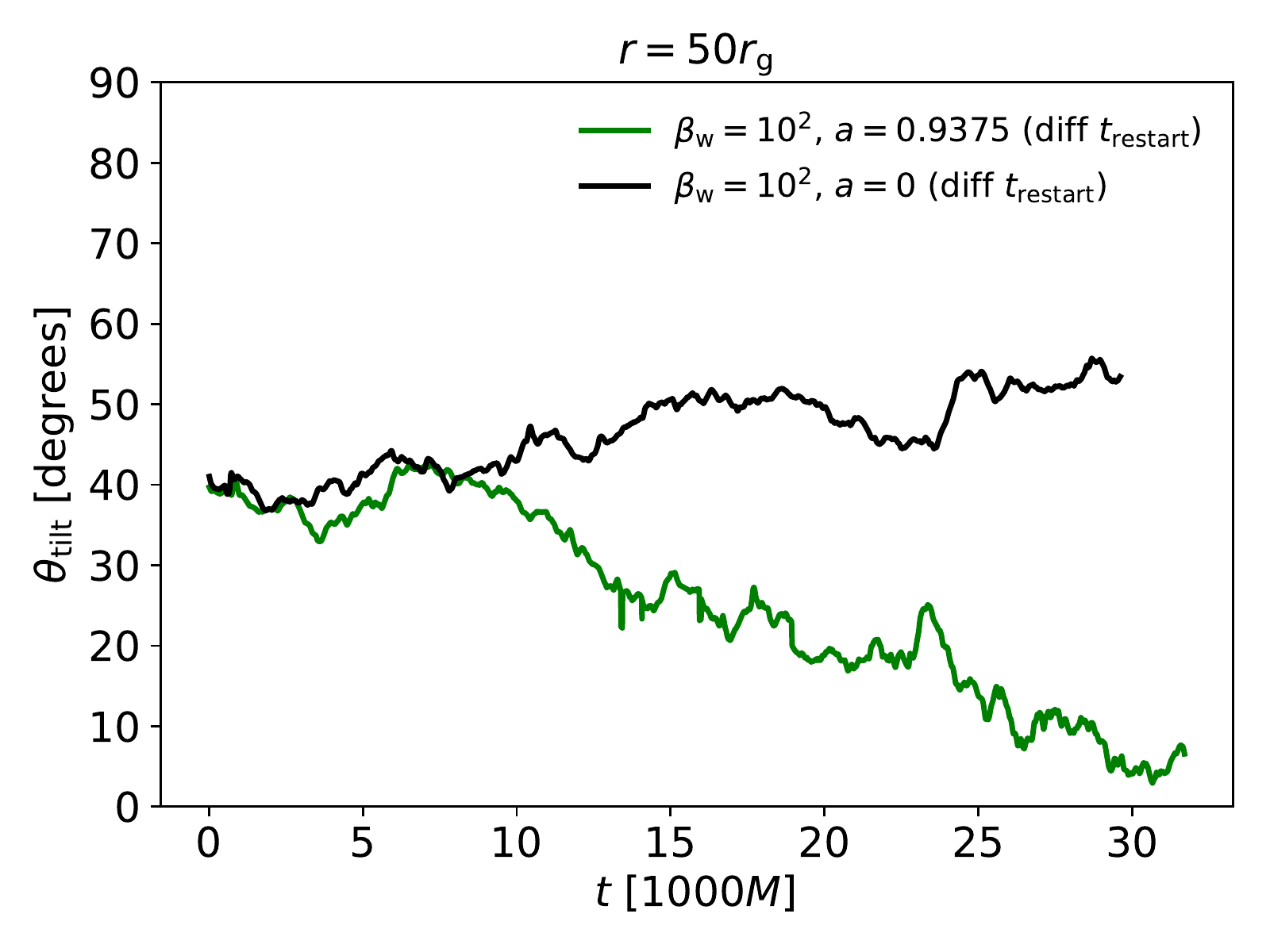}
\caption{Angle between the angle-averaged angular momentum of the gas and the black hole spin axis, $\theta_{\rm tilt}$, measured at $r=5 r_{\rm g}$ (top), $r=20 r_{\rm g}$ (middle), and $r=50 r_{\rm g}$ (bottom) for $\betaw=10^2$ simulations initialized using data from a different time compared to the fiducial $\betaw=10^2$ simulations.  For $a=0$, $\theta_{\rm tilt}$ represents the angle with the $z$-axis. 
The gas in the $a=0$ simulations has angular momentum tilted by $\sim$ 40--70$^\circ$ with respect to the $z$-axis.
The gas in the $a=0.9375$ simulation is completely reoriented once the MAD state is reached (compare with Figure \ref{fig:time_plots_113}), reducing $\theta_{\rm tilt}$ to $\approx$ 0$^\circ$ at $r= 5\ r_{\rm g}$,  $\approx$ 5$^\circ$ at $r= 20\ r_{\rm g}$, and $\approx$ 20$^\circ$  at $r = 50\ r_{\rm g}$.  }
\label{fig:th_tilt_113}
\end{figure}

In terms of orientation, the intermediate-scale simulation (and thus the $a=0$ GRMHD simulation) has an average angular momentum tilted by $\sim$ 30--40$^\circ$ with respect to the $z$-axis (the line of sight) in contrast to the fiducial intermediate-scale $\betaw=10^2$ simulation where the angular momentum is tilted by $\sim$ 80--90$^\circ$ with respect to the $z$-axis. 
This reflects the variable nature of the accretion flow's angular momentum direction at larger radii over $\sim$ decades (see, e.g., Figure 9 in \citealt{Ressler2020}) associated with the evolving positions and velocities of the innermost stellar winds.
For the GRMHD simulations, we show the $a=0$ and $a=0.9375$ values for $\theta_{\rm tilt}$ shown in Figure \ref{fig:th_tilt_113} at $r=5\ r_{\rm g}$, $r=20\ r_{\rm g}$, and $r=50\ r_{\rm g}$, where for $a=0$ $\theta_{\rm tilt}$ represents the angle between the gas angular momentum and the $z$-axis.
Similar to the intermediate-scale simulation, the $a=0$ GRMHD simulation starts at $\theta_{\rm tilt}$ $\approx$ 40$^\circ$ and evolves to 50--70$^\circ$ by the end of the simulation.  
Unlike the fiducial $\betaw=10^2$ simulations, the gas in the new $a=0.9375$ simulations completely aligns  ($\theta_{\rm tilt}\lesssim 5^\circ$) with the spin axis for $r=5\ r_{\rm g}$  ($\theta_{\rm tilt}\lesssim 5^\circ$) after $t \approx 11{, }000\ M$. $\theta_{\rm tilt}$ at $r = 20\ r_{\rm g}$ and $r=50\ r_{\rm g}$ also tends towards 0 starting at the same time, but does not fully align until $t \gtrsim 15{, }000\ M$ and $\gtrsim 30{, }000\ M$, respectively.
Since $t \approx 11{, }000\ M$ is approximately the same time as the MAD state is reached (see Figure \ref{fig:time_plots_113}), and the fiducial $\betaw=10^2$, $a=0.9375$ simulations were most aligned during periods where the dimensionless magnetic flux threading the horizon was close to the MAD state, we can reasonably conclude that in our models MAD accretion flows around rapidly spinning black holes align with the spin axis.
This is true even out to relatively large radii, though the greater the distance from the black hole, the longer this takes to achieve. 

\begin{figure}
\includegraphics[width=0.45\textwidth]{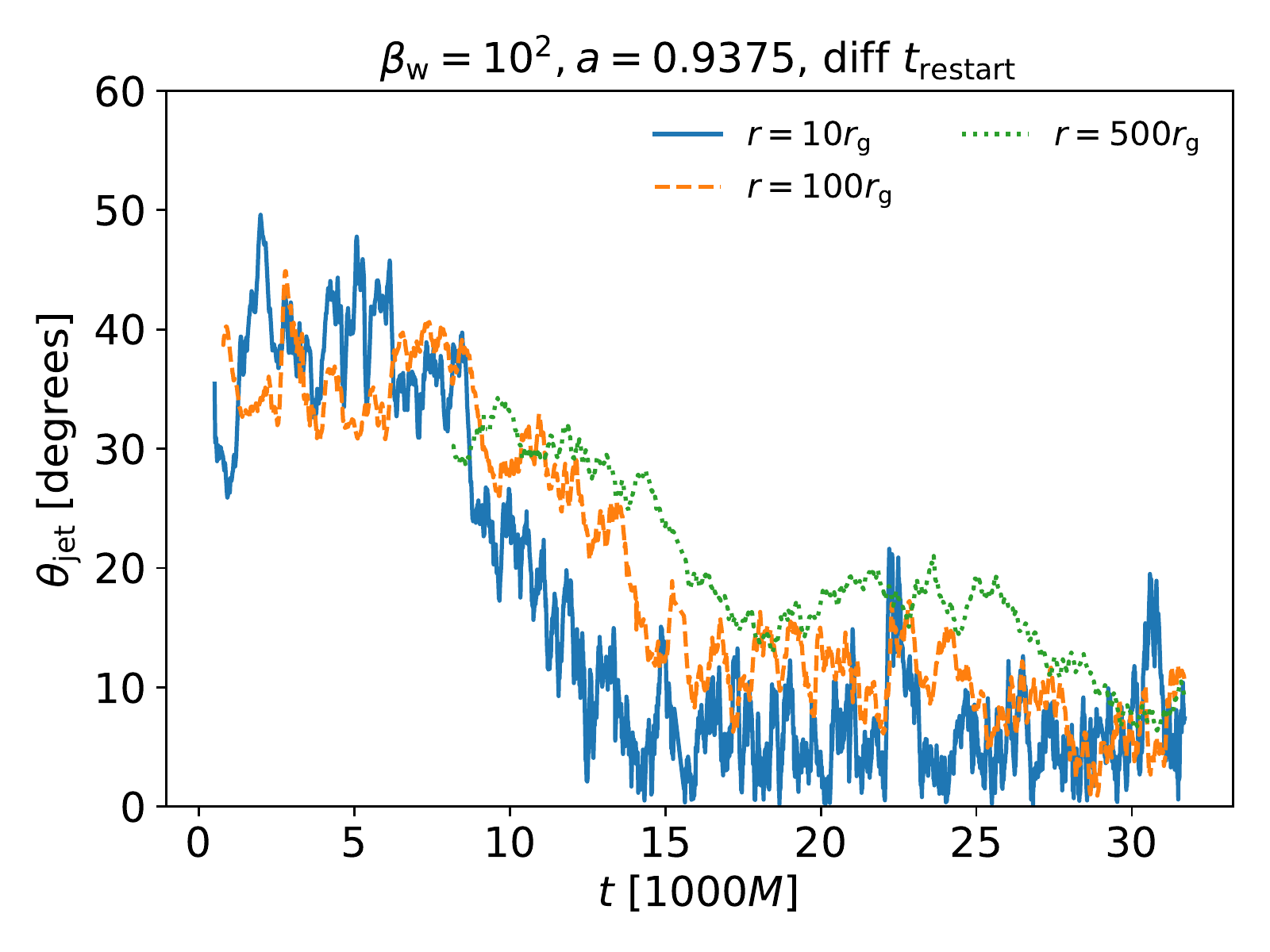}
\includegraphics[width=0.45\textwidth]{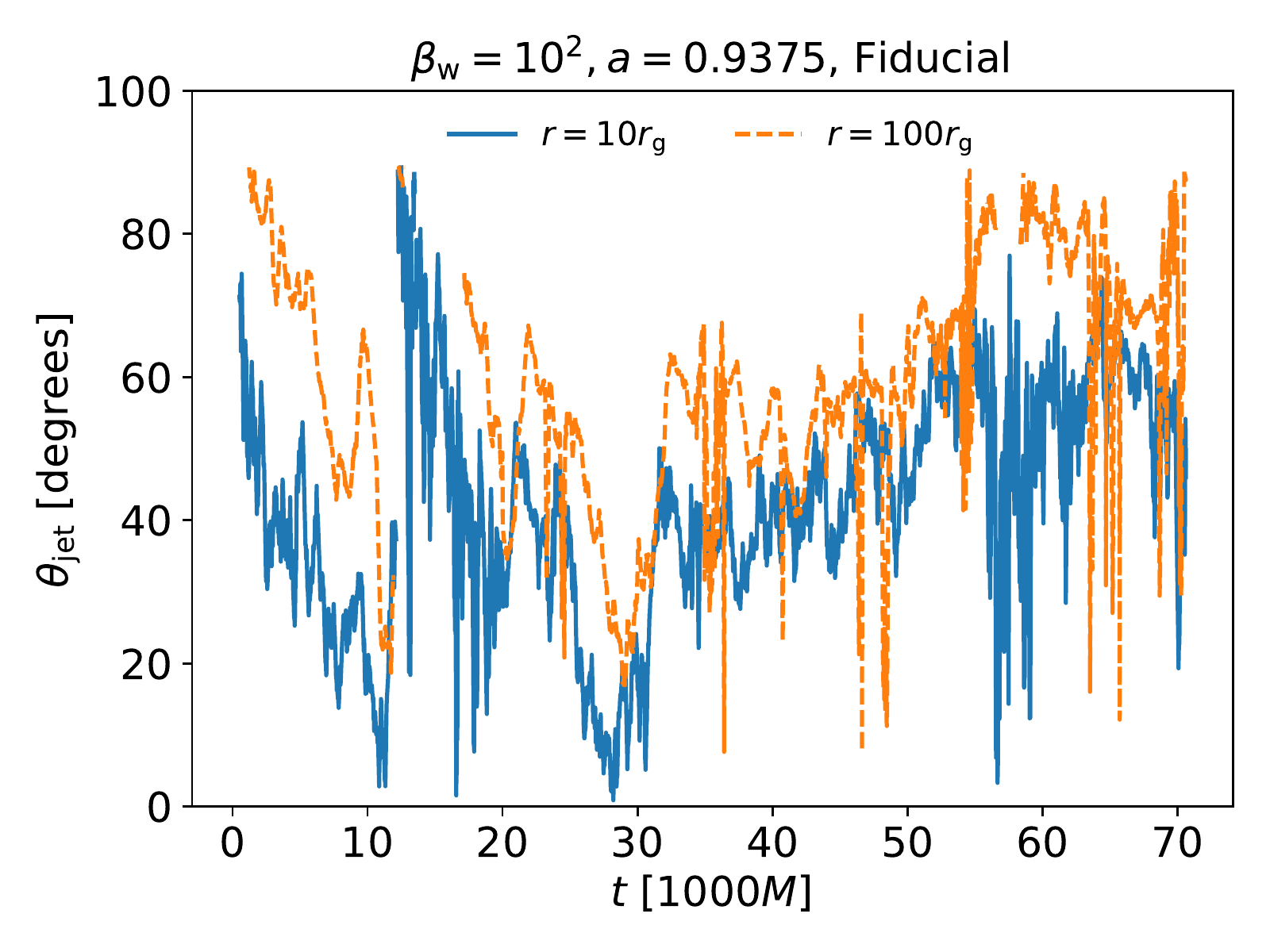}
\caption{
Angle between the jet and the black hole spin axis, $\theta_{\rm jet}$, as a function of time for $r=10\ r_{\rm g}$ (solid), $r=100\ r_{\rm g}$ (dashed), and $r=500\ r_{\rm g}$ (dotted), in our $a=0.9375, \betaw=10^2$ simulations.  
Top: simulation with a different $t_{\rm restart}$ described in \S \ref{sec:alt_real}. Bottom: fiducial simulation.
The jet in the fiducial simulation (which doesn't go MAD) tends to align with the spin axis of the black hole during periods of peak magnetic flux (compare with Figure \ref{fig:time_plots}) but otherwise tends to propagate closer to the angular momentum of the large-scale accretion flow (tilted by $\sim$ 60--90$^\circ$).  
The jet in the different $t_{\rm restart}$ simulation, on the other hand, aligns with the spin axis once the MAD state is reached at $t\sim 11{, }000\ M$.} 
\label{fig:th_jet}
\end{figure}

Similar alignment is seen in the relativistic jet.  In \S \ref{sec:jet} we described how the fiducial $\betaw=10^2$, $a=0.9375$ jets tended to align at smaller radii (10s of $r_{\rm g}$) only during periods of peak magnetic flux (close to the MAD value) and at larger radii tended to align with the angular momentum axis of the accretion flow.
Not so for the new $\betaw=10^2$, $a=0.9375$ simulations.
In Figure \ref{fig:th_jet} we plot the angle between the upper ($z>0$) jet and the black hole spin axis at three different radii ($r=10\ r_{\rm g}$, $r=100\ r_{\rm g}$, and $r=500\ r_{\rm g}$) in the new simulations. 
Also plotted for comparison are the same quantity in the fiducial $\betaw=10^2$, $a=0.9375$ simulation at two radii ($r=10\ r_{\rm g}$ and $r=100\ r_{\rm g}$).  
Note that the jet in the latter simulation is often not well defined at $r=500\ r_{\rm g}$.
Once the MAD state in the new simulation is reached the jet at all three radii aligns with the spin axis to within $\sim$ 10$^\circ$.  
For $r=10 \rg$, this happens in $\lesssim 5{, }000\ M$, while for $r=100 \rg$ and $r=500 \rg$ it takes $\gtrsim 10{, }000 \ M$ and $\gtrsim 25{, }000\ M$, respectively.  
That is, the longer the system remains in the MAD state the farther out the jet becomes aligned.
The jet in the new simulation also propagates to the edge of the computational domain by the end of the run.
In contrast, the jet in the fiducial $\betaw=10^2$, $a=0.9375$ simulation fluctuates in direction but tends to remain significantly tilted by $\gtrsim 40^\circ$ at both  $r=10\ r_{\rm g}$ and $r=100\ r_{\rm g}$ except for times when $\phi_{\rm BH}$ approaches the MAD value.
As seen in Figure \ref{fig:rjet} it also fails to propagate farther than $\sim$ 800 $\rg$ from the black hole.   
This suggests that in our models MAD jets around rapidly spinning black holes tend to align with the spin axis, and more so over time.
This is because the jets in the MAD state are powerful enough to forge their own path through the large-scale accretion flow instead of following the pre-existing cavity (as does the fiducial simulation that does not go MAD, see \S \ref{sec:jet} for a discussion).

These results for disk/jet alignment are relatively consistent with \citet{Mckinney2013}, where their rapidly spinning MAD simulations show alignment at $r=4 \rg$ but only partial alignment (corresponding to $\theta_{\rm jet}$ decreasing by $\sim$ half its value) at $r=30 \rg$. 
This difference may be caused by the fact that their simulations were run in a tilted state for only $8{,}000$--$12{, }000\ M$ so that the jet may not have had time to fully align at larger radii.

The RMs calculated from these simulations are strikingly different from those of the fiducial simulations, as shown in the RM versus time for the $a=0$ simulation (the $a=0.9375$ curve looks similar) in the bottom panel of Figure \ref{fig:RM_large_scale_beta_1e2_from_113}.  
Not only is the magnitude larger by a factor of $\sim$ 10, the sign is consistent for the entire duration of the simulation (except for a few brief instances in time).  
In fact, the magnitude of the predicted RM for these simulations is just above the mean value of observations, reaching $\approx 7 \times 10^5$ rad/m$^2$.  
The RM variability is at the $\lesssim$ 50\% level, significantly less than the fiducial simulations where the RM can vary by orders of magnitude. 
This result suggests that the behavior of the RM is sensitive to the amount of net magnetic flux available to accrete at large radii.


\section{Discussion}
\label{sec:disc}

\subsection{Rotation Measure}
\label{sec:disc_rm}
To gain insight into how the rotation measure predicted by our simulations might change over longer periods of time, we can compare the GRMHD simulation values to those predicted by the larger scale simulations.  For the latter, we can only approximate a rotation measure by assuming that the horizon scale emission is a point source and then calculating the integral (taken from \citealt{Mosci2017}; see also \citealt{Broderick2009}):
\begin{equation}
  \label{eq:pseudo_RM}
 \textrm{pseudo-RM} = \frac{10^4 \textrm{e}^3}{2 {\rm \pi} m_{\rm e}^2 c^4} \int\limits_{r_{\rm min}}^{r_{\rm max}} f(\Theta_{\rm e})  B_z n_{\rm e} dz,
\end{equation}
where $r_{\rm min}$ and $r_{\rm max}$ are the inner and outer edges of the simulations, respectively, and 
\begin{equation}
  f(x) = \begin{cases} 
      \log{(x)} \left(\frac{x-1}{2x^3}\right)+\frac{1}{x^2} & x> 1 \\
      1 & x\leq 1 
   \end{cases}
\end{equation}  
is an approximate relativistic correction term.  
For the GRMHD simulations, we calculate the RM as described in \S \ref{sec:emission}.

\begin{figure}
\includegraphics[width=0.45\textwidth]{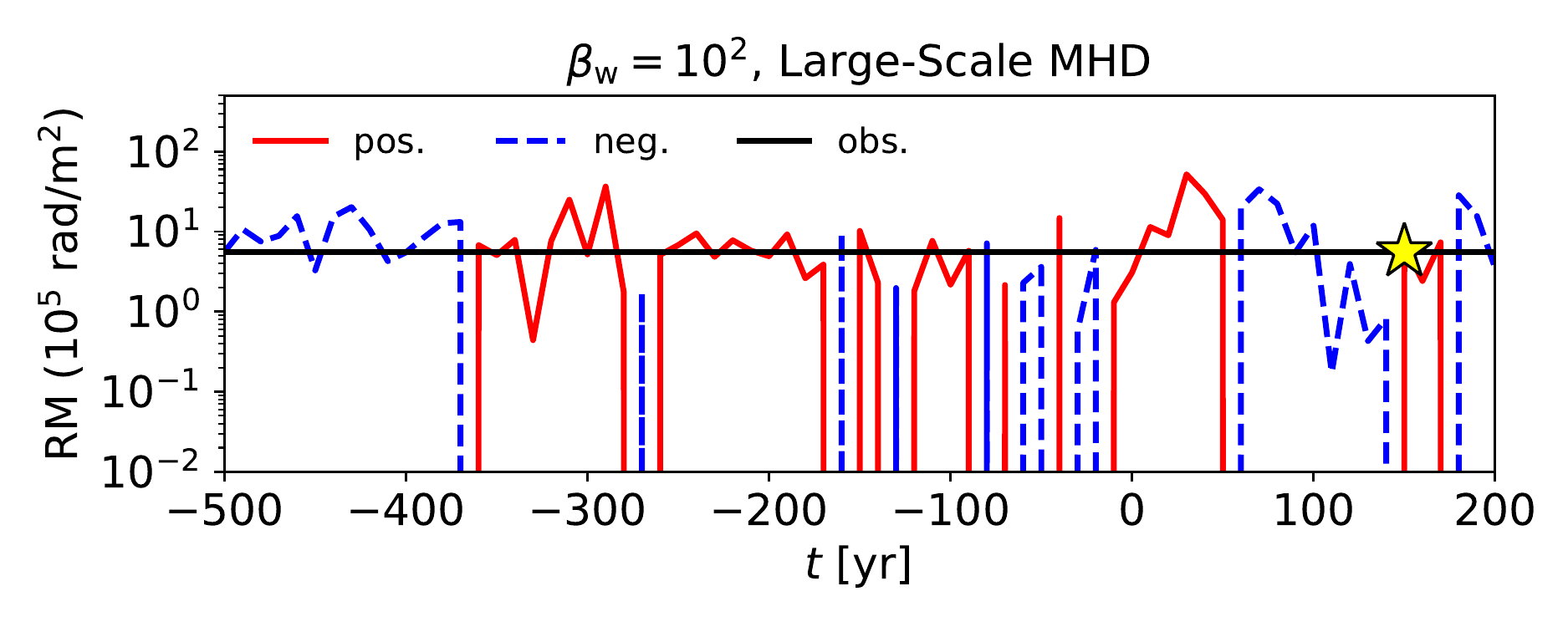}
\includegraphics[width=0.45\textwidth]{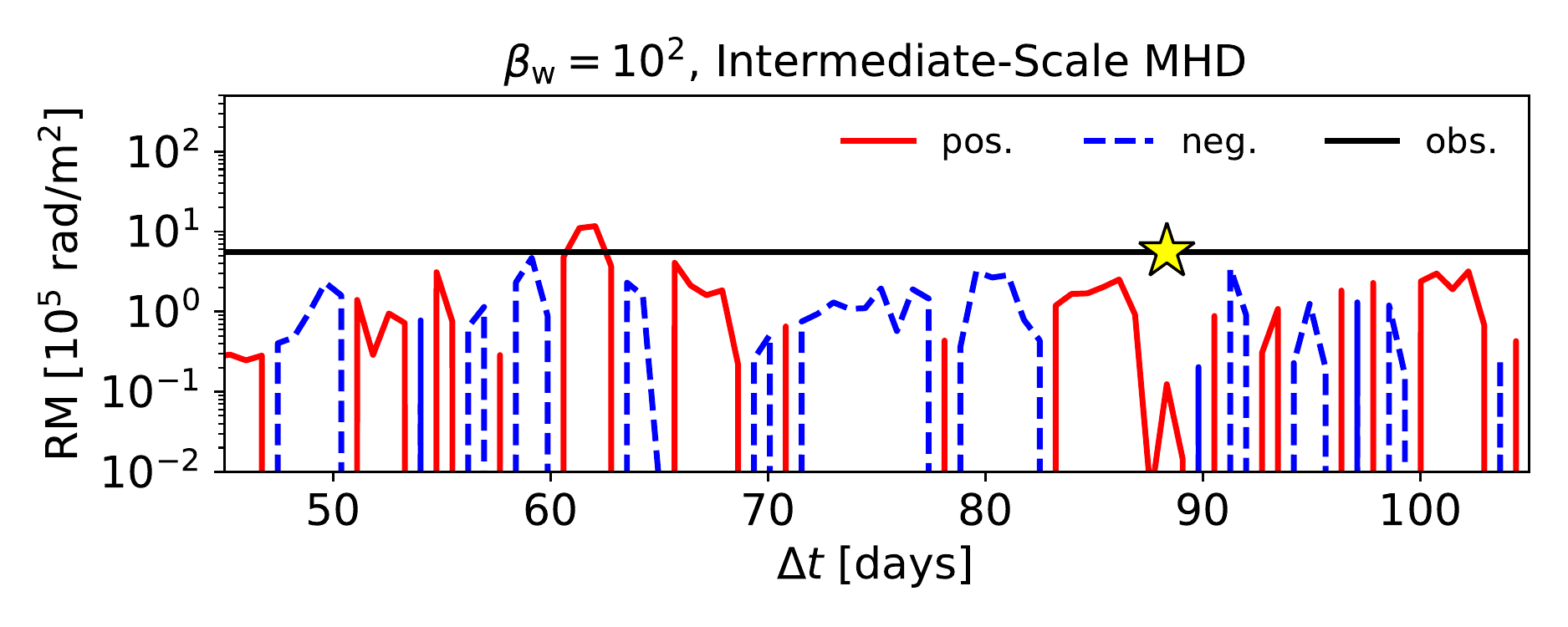}
\includegraphics[width=0.45\textwidth]{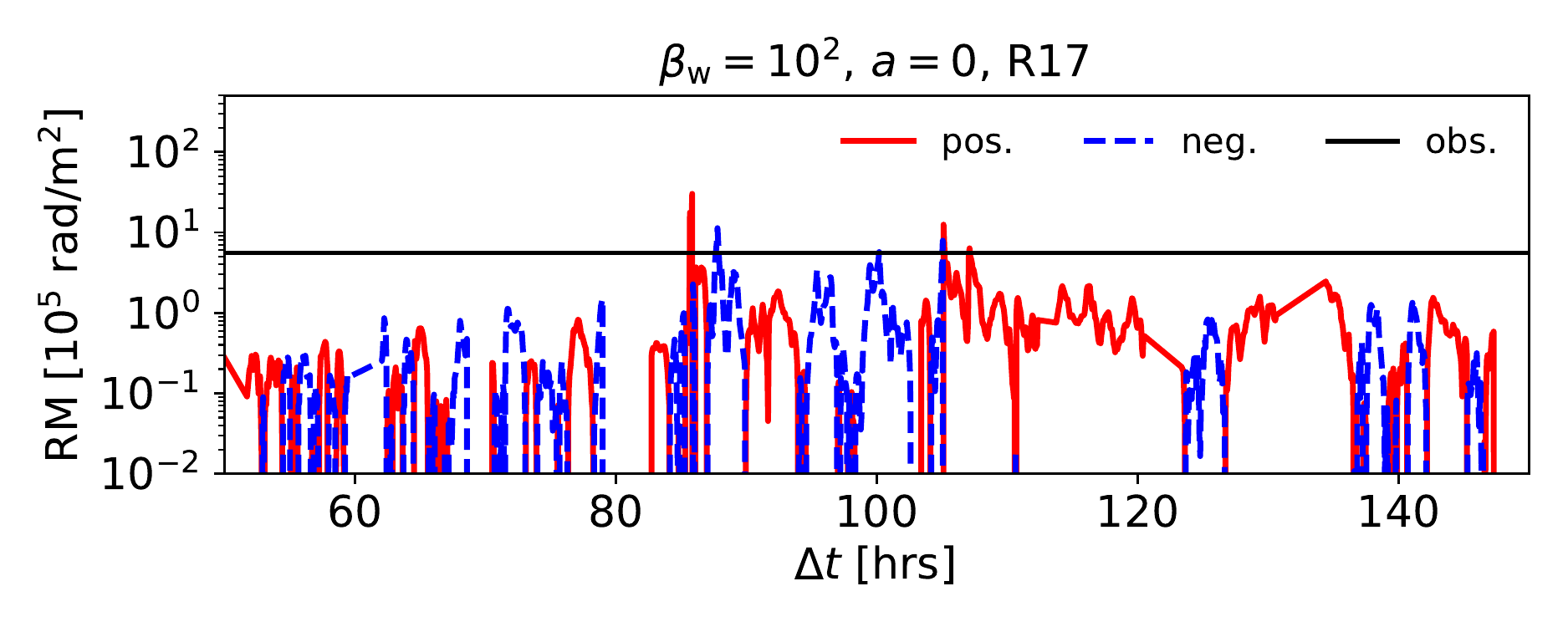}
\caption{Rotation measure as calculated from our $\betaw=10^2$ simulations.  Top: large-scale wind MHD simulation.  Middle: intermediate-scale MHD simulation.  Bottom: GRMHD simulation R17 electron heating.  Red solid lines represent positive rotation measure while blue dashed lines represent negative rotation measures.  The solid black line is the mean observed value for Sgr A*.  Yellow stars represent the time at which initial conditions are generated for the smaller simulation. 
Going from large to small scale simulations, the RM changes more rapidly and decreases in magnitude.  
This is a result of strong fields developing at smaller radii in the smaller-scale simulations providing cancellation in the line-of-sight integral.   }
\label{fig:RM_large_scale_beta_1e2}
\end{figure}

\begin{figure}
\includegraphics[width=0.45\textwidth]{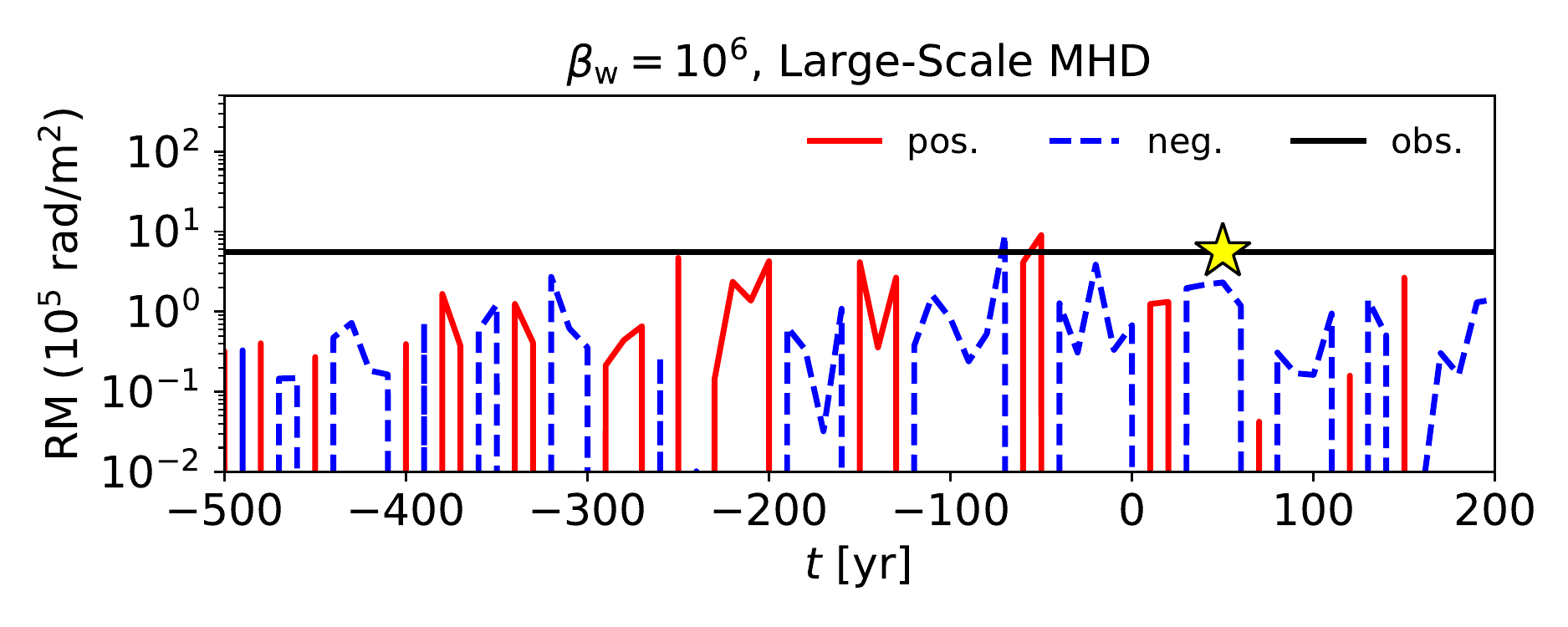}
\includegraphics[width=0.45\textwidth]{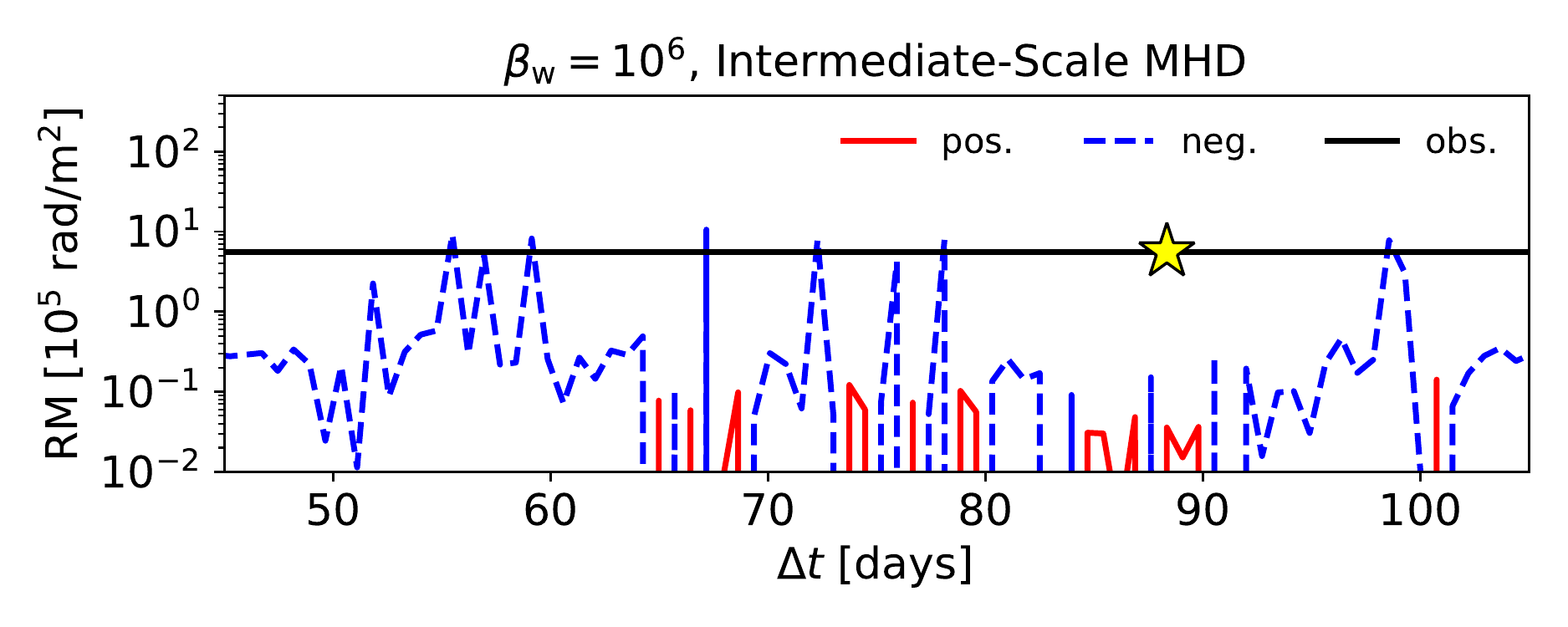}
\includegraphics[width=0.45\textwidth]{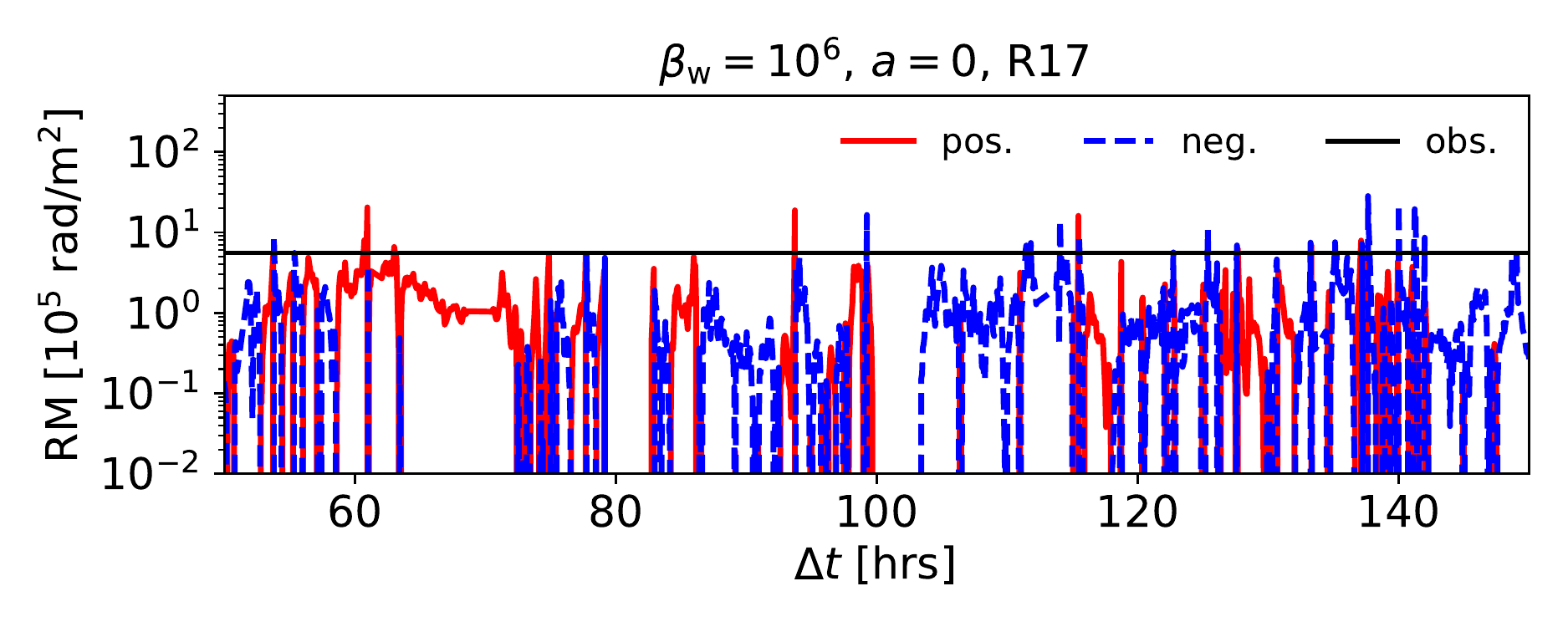}
\caption{Same as Figure \ref{fig:RM_large_scale_beta_1e2} but for $\betaw=10^6$.  In this simulation the field at larger scales is much weaker (see Figure \ref{fig:beta_comp_intermediate}) than the $\betaw=10^2$ simulation and so the predicted RM is much smaller for both the intermediate and large scale MHD simulations.  The GRMHD simulation, however, has a comparable or even slightly larger magnitude RM than its $\betaw=10^2$ counterpart as horizon scale magnetic flux is able to build up to the point of becoming MAD (see Figure \ref{fig:flux_vt}).}
\label{fig:RM_large_scale_beta_1e6}
\end{figure}

We show these calculations of the RM in Figures \ref{fig:RM_large_scale_beta_1e2} and \ref{fig:RM_large_scale_beta_1e6} for the fiducial $a=0$, $\betaw=10^2$ and $a=0$, $\betaw=10^6$ simulations, respectively. The time scales plotted are, in decreasing order, $\sim$  700 yr, $\sim$ 60 days, and $\sim$ 100 hrs.
For $\betaw=10^2$, the large-scale MHD wind simulation displays several times at which the magnitude of the RM is as large or larger than the mean observed value for Sgr A* and can have a consistent sign for $\gtrsim 100$ yrs.  
At the time chosen to use for initial conditions for the intermediate-scale simulation, the RM is close to the observed value but had just experienced a rapid sign change.  
As a result, the intermediate-scale MHD simulation has a RM that only occasionally has a magnitude comparable to observations and changes sign on time scales on the order of $\sim$ 10 dy.  
At the time chosen to use for initial conditions for the GRMHD simulation, the RM in this intermediate-scale simulation is particularly low, resulting in a GRMHD RM that is only very rarely comparable to observations in magnitude and that displays rapid sign changes on times scales that can be as short as $\lesssim $ hrs.

The large-scale MHD wind-fed $\betaw=10^6$ simulation has a weaker and less coherent magnetic field \citep{Ressler2020b} and so the RM is typically well below observations and has more frequent sign changes than its counterpart in the $\betaw=10^2$ simulation.
The intermediate-scale MHD simulation then has a comparable, if not slightly lower, magnitude RM with more frequent sign changes still.  
Finally, the GRMHD simulation typically has a larger magnitude RM than either of the larger scale simulations despite being initialized from a time in the intermediate-scale MHD simulation with a particularly low RM.

The fact that the RM for both sets of the fiducial GRMHD simulations changes on hour or less time scales implies that the horizon scale plasma is a significant contribution to this quantity given that $r_{\rm g}/c \sim 20$s for Sgr A*.  
The observed RM indeed has such rapid time scale variability in its magnitude \citep{Bower2018} but not in its sign, which has remained consistent for many years.  
This suggests that there is also a larger scale, coherent magnetic field that varies on much longer time scales and contributes a large fraction of the RM, enough so that there is a limit to how much horizon-scale variability can effect the RM.  
This is not seen in our fiducial set of simulations.  In the $\betaw=10^6$ simulations, the large-scale field is not strong enough (see Figure \ref{fig:beta_comp_intermediate}) to produce an RM comparable to observations and so it is dominated by smaller, horizon scales as the field builds up to the point of being MAD (see Figure \ref{fig:flux_vt}).
For $\betaw=10^2$, however, there is often plenty of large-scale field available to reproduce the observed RM as seen in the top panel of Figure \ref{fig:RM_large_scale_beta_1e2}.
Naively, we would expect that extending the simulations to smaller scales as we have done would then either not effect the RM very much (since the contribution to the RM is suppressed for relativistically hot electrons) or change it slightly.  
Instead we see a significantly decreased RM and much more rapid sign changes.
As we will now demonstrate, this is a result of the particular time chosen in the original large-scale MHD wind simulation where the RM had just experienced a rapid sign change, implying that there was less available supply of coherent magnetic field than typical (and as a result significant cancellation of the RM along the line of sight).  

\begin{figure}
\includegraphics[width=0.45\textwidth]{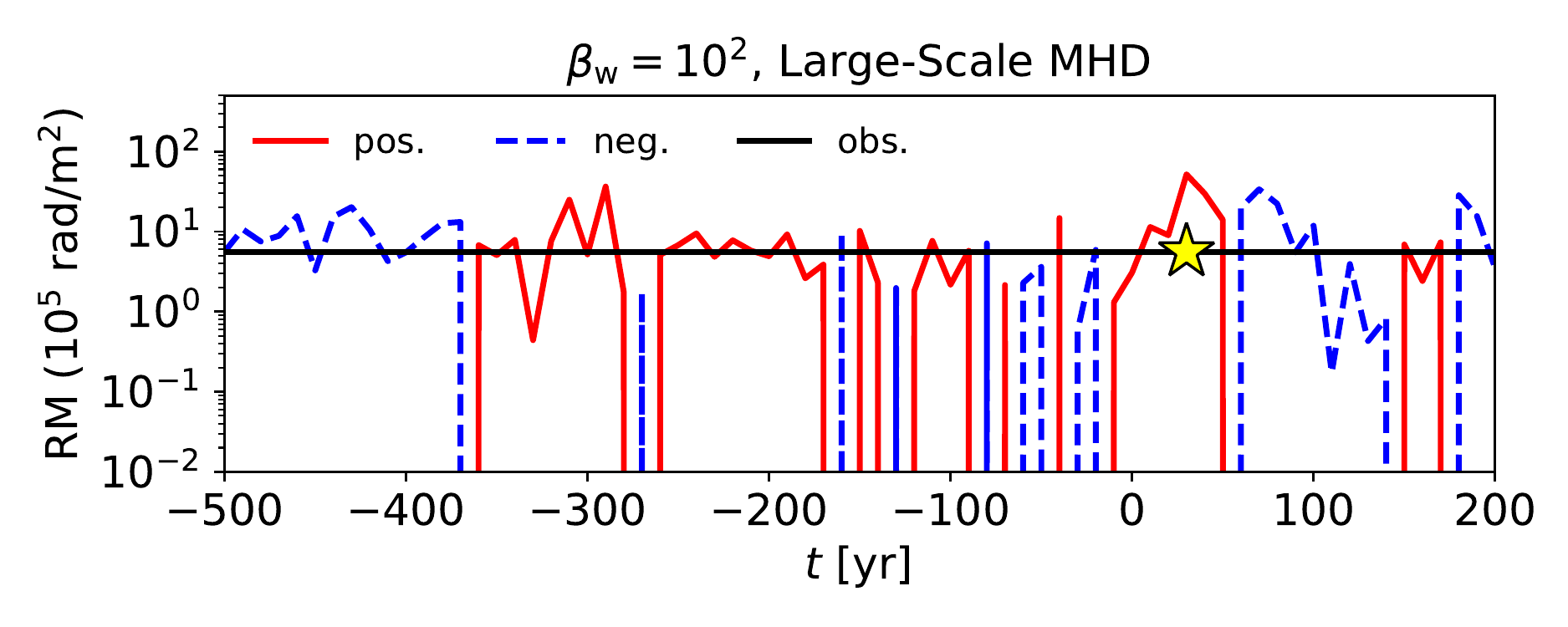}
\includegraphics[width=0.45\textwidth]{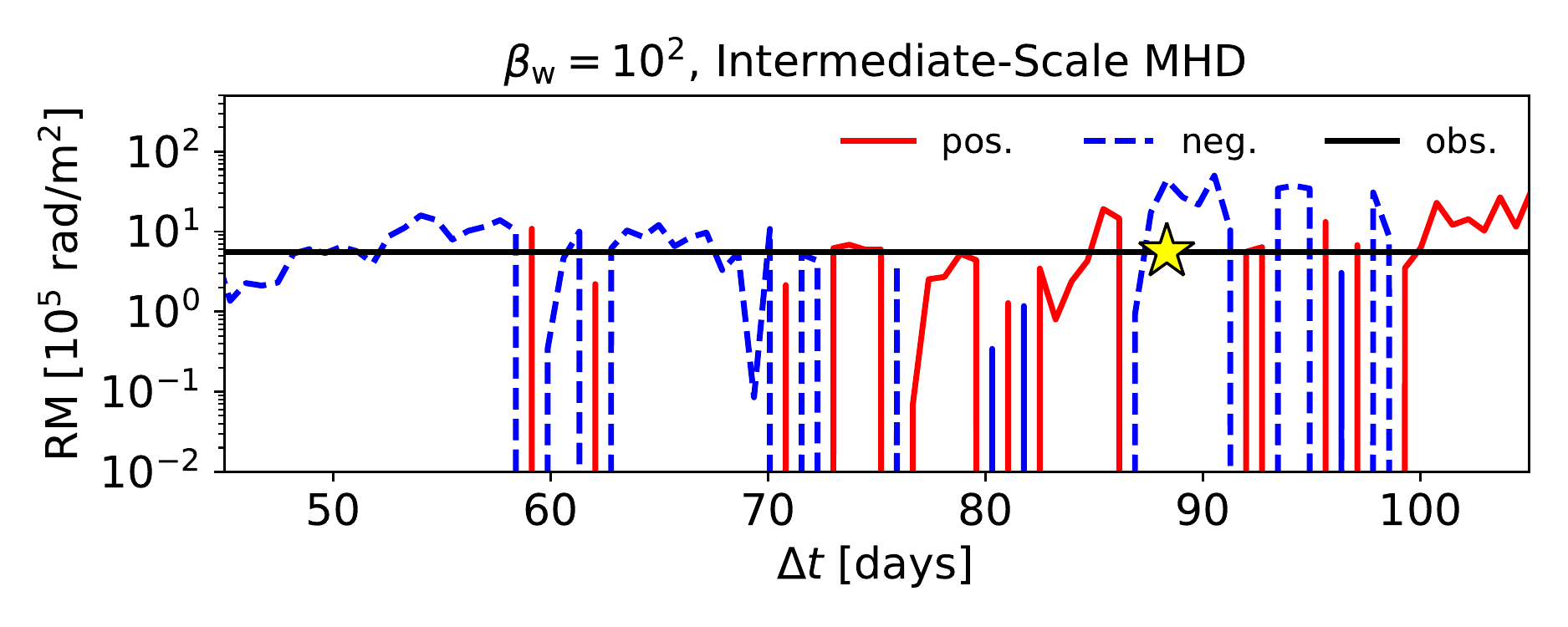}
\includegraphics[width=0.45\textwidth]{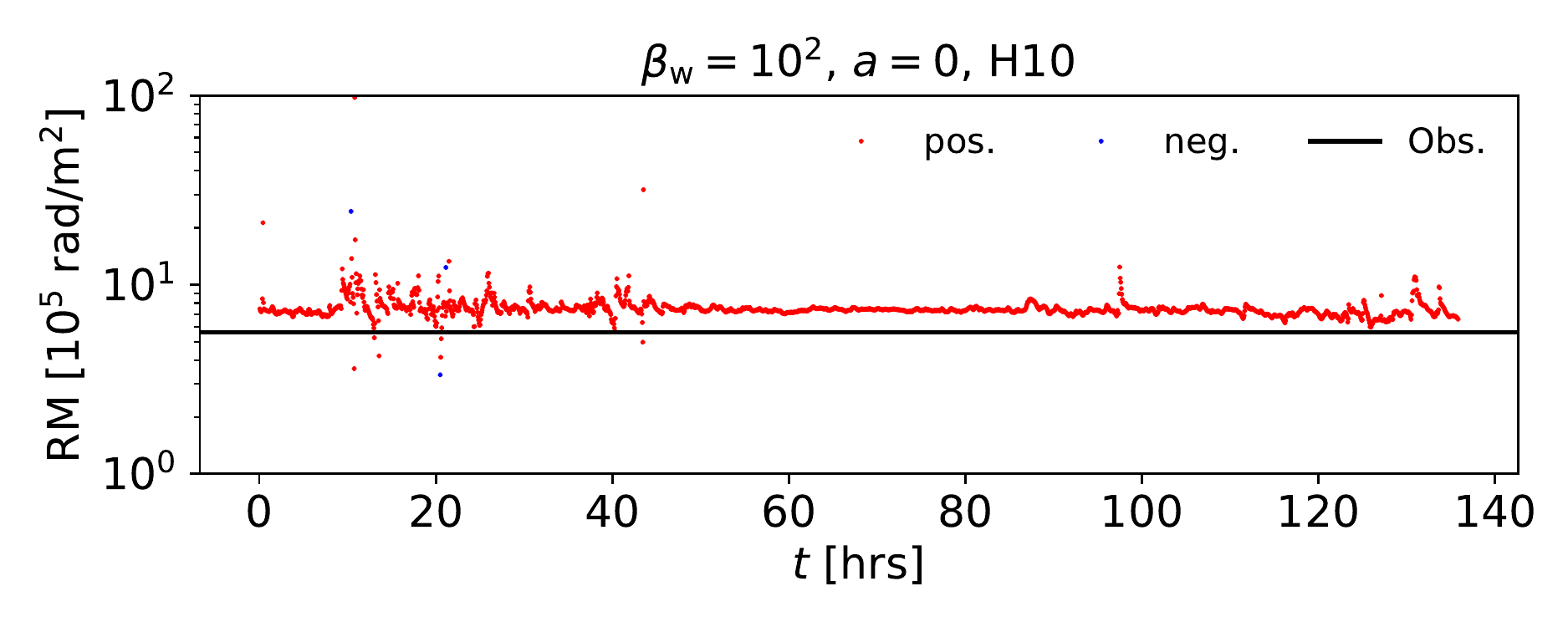}
\caption{Rotation measure as calculated from our $\betaw=10^2$ simulations using data from $t=30$ yr in the large scale simulation (a time particularly chosen for its large and consistent rotation measure).  
Top: large-scale wind MHD simulation.  
Middle: intermediate-scale MHD simulation.
Bottom: GRMHD simulation.   
Red solid lines represent positive rotation measure while blue dashed lines represent negative rotation measures.  The solid black line is the mean observed value for Sgr A*.  Yellow stars represent the time at which initial conditions are generated for the smaller simulations. Compared to the fiducial $\betaw=10^2$ simulation (Figure \ref{fig:RM_large_scale_beta_1e2}), the RM calculated from the intermediate scale simulation is consistently larger in magnitude, comparable to the mean observed value of Sgr A*.  The RM calculated from the GRMHD simulation is much larger in magnitude than that in the fiducial $\betaw=10^2$ simulation and even larger than observations.  The sign is consistently positive for the entire simulation except for a few instances in time and the RM is overall less variable (though still significant, at the $\sim$ 50\% level).}
\label{fig:RM_large_scale_beta_1e2_from_113}
\end{figure}

To show this, we consider the fifth set of simulations described in \S \ref{sec:alt_real} with $\betaw =10^2$ and $a=0$ in which the intermediate simulation is initialized from $t=30$ yr data in the original large-scale MHD simulation. 
This particular time was chosen for having a large ($>5\times 10^5$ rad/m$^2$) pseudo-RM with a consistent sign for over half a century.   
 This is seen in Figure \ref{fig:RM_large_scale_beta_1e2_from_113}, which again plots the RM from the large-scale MHD, intermediate-scale MHD, and GRMHD simulations.
The intermediate-scale ``pseudo-RM'' is consistently larger in magnitude than that from the fiducial $\betaw=10^2$ simulation (middle panel of Figure \ref{fig:RM_large_scale_beta_1e2}) and the sign changes are slightly more infrequent ($\sim$ 5--10 days).
The more dramatic difference is seen in the RM computed from the GRMHD simulation.  
Instead of underproducing the mean observed RM of Sgr A*, the new simulation's RM is actually larger in magnitude than observations and much less variable.  
In fact, the RM stays positive\footnote{The RM in the GRMHD simulation is, in fact, the opposite sign as the RM in the intermediate-scale MHD simulation.  This is because the RM in the latter is caused by both a strong positive magnetic field along the line of sight at $r\gtrsim 500 r_{\rm g}$ and a strong negative field along the line of sight at $r\lesssim 500 r_{\rm g}$. The RM contribution from the negative field at small radii is actually stronger than the positive field at large radii, causing an overall negative RM. The GRMHD simulation then provides a more accurate representation of the plasma at smaller radii (and a more accurate calculation of the RM using full polarized radiative transfer) in which the strong positive contribution to the RM from large radii remains but the strong negative contribution from smaller radii is no longer present.}  essentially the entire simulation except for momentary snapshots in time during the transient initial $\sim$ 20 hr ($\sim$ a few 1000 $M$).
The RM as calculated is dominated by scales $\gtrsim$ 1600 $r_{\rm g}$, i.e., scales in the intermediate-scale MHD simulation.  
Since the solution in this region is assumed constant during the entire GRMHD simulation by construction, all variability in the GRMHD RM is provided by plasma on scales closer to the horizon.  
If we take the RM computed from the large-scale MHD simulation as a proxy for strong, large-scale, coherent magnetic fields, these results provide compelling evidence that such fields are necessary to account for the observed properties of Sgr A*'s RM.

While this result is quite promising -- namely, that our wind-fed simulations can in fact produce a large enough rotation measure to explain observations and even have a consistent sign on horizon-scale time intervals -- there are still some unresolved issues.  
If we take the RM calculated from the intermediate-scale MHD alone at face value (middle panel of Figure \ref{fig:RM_large_scale_beta_1e2_from_113}), then our model still predicts sign changes on time-scales much shorter than decades ($\sim$ 5--10 days).  
This could be due to the inaccuracy of using the approximate Equation \eqref{eq:pseudo_RM} on event-horizon scales where the emission is no longer well approximated by a point source.
Unfortunately this remains speculative; without the ability to run GRMHD simulations for much longer times and out to much larger radii we cannot fully predict the long term evolution of the RM.
Nonetheless, it is reassuring that the more accurate calculation using GR ray tracing from the GRMHD simulation qualitatively agrees with observations on the $\sim$ several day timescales it is able to model.


It is important to note here that achieving such a high magnitude and consistent sign RM in the GRMHD simulation is likely only possible for lower $\betaw$.  For higher $\betaw$, e.g., $\betaw=10^6$, there is never (or at least very rarely) strong, coherent magnetic fields present at large radii (see Figures \ref{fig:beta_comp_intermediate} and \ref{fig:RM_large_scale_beta_1e6}).  
Although determining the maximum value for $\betaw$ at which this result is possible is beyond the scope of this work, we can reasonably conclude that the accreting WR stellar winds should have $\betaw \lesssim 10^6$ in order for our model to at least qualitatively reproduce the RM properties of Sgr A*.
That is of course, assuming they are the only source of magnetic fields in the accretion flow.
Another possibility is that there is an additional source of large-scale magnetic fields (e.g., material accreted from the CND \citealt{Blank2016,Hsieh2018}) that could easily produce a large and stable RM.

It is also worth noting that the consistency of the sign of the RM for at least two decades suggests that the physical mechanism responsible for sign changes happens on scales $\gtrsim$ $10^5 r_{\rm g}\ \approx\ 0.02$ pc.  Given that the WR stellar orbits are located at radii $\gtrsim 0.05$ pc, this is consistent with the hypothesis that large-scale magnetic fields provided by the stellar winds are the root cause of the RM.  
If so, sign changes in the RM would be caused by the movement of the nearby WR stars through their orbits causing different sections of the winds to contribute to accretion (see Appendix A of \citealt{Ressler2018}).
Assuming that the fields in the winds obey flux-freezing, they will be predominately toroidal in the frame of the star, as assumed here. 
This means that during a single orbit, the magnetic field direction of the lowest angular momentum portion of the wind (the portion on the side of the wind opposite to the direction of motion) will rotate a full 360$^\circ$.
Thus, every half an orbit the field in the accreting wind material will change sign. 
For 16C (AKA E20, \citealt{Paumard2006,Cuadra2008}) and 16SW (AKA E23), the two nearest WR stars, this timescale corresponds to $\sim$ 740 and $\sim$ 1270 yrs, respectively.

\subsection{Comparison to \citet{Ressler2020b}}
\label{sec:R20}
In principle, the dynamics of the $a=0$ simulations presented in \citeauthor{Ressler2020b} (\citeyear{Ressler2020b}, hereafter R20b) should be identical to the fiducial $a=0$ simulations in this work, since both sets started from an identical data dumps from the same large-scale simulations.  
The addition of electron thermodynamics via Equation \eqref{eq:electron_entropy} in the new simulations should not affect this result since it is solved without back-reacting on the flow.  
That said, we do see clear differences in the dynamical evolution of the two pairs of simulations, especially in magnetic flux.  
Both R20b simulations ($\betaw=10^2$ and $\betaw=10^6$) became magnetically arrested by $\approx$ $5{,}000\ M$, as seen in the saturation of $\phi_{\rm BH}$ around $\approx$ 50 shown in Figure 2 of that work.
Our $\betaw=10^6$, $a=0$ simulation does not go MAD until much later, $\approx 45{,}000\ M$, while our $\betaw=10^2$ simulation does not seem to go MAD at all within the $\approx\ 70{,}000\ M$ runtime.

So, what changed?  
Physically, nothing.  Numerically, however, a few things.  R20b also used piecewise-linear reconstruction (plm) in their GRMHD simulations, while these new simulations use parabolic piecewise reconstruction (ppm) for the GRMHD simulations (the MHD simulations in both R20b and here used plm).  
R20b used an older version of {\tt Athena++} without passive scalars.  
The process of implementing passive scalars required a reordering of certain parts of the equation-solving algorithm and some altered data structures, both of which can change the specific roundoff errors at each location and time.  
This is true of both the GRMHD simulations and the intermediate-scale MHD simulations (we used the old version of the code for the largest scale simulations that directly include the stellar winds). 
Still, none of these numerical differences in themselves, one would hope, should lead to such obvious differences in the resulting flow. 
Nonetheless, there are two important things to consider.  
First, the flow is chaotic, as we have verified by running the exact same simulation using two different compilers (which changes only roundoff errors), finding that quantities like the magnetic flux near the inner boundary can be non-neglibibly different in the two cases.
It is known that chaotic flows can be sensitive in this way even to roundoff errors, but one would still expect/hope the statistical behavior over time of quantities like $\phi_{\rm BH}$ to be similar.
But that is where the second important thing to consider comes in: the nature of multi-scale simulations. 
Small differences in quantities at large radii can propagate to large differences in quantities at small radii since the cell size itself depends on radius. 
Combined with the chaotic nature of the system, it is thus not improbable that the difference in roundoff/truncation error induced by algorithmic and reconstructive changes could lead to noticeably different magnetic fluxes threading the event horizon. 

We provide strong evidence of this hypothesis in Appendix \ref{app:num_tests} by running additional simulations that use different reconstruction methods and/or have electron thermodynamics turned off. 
In summary, we find that when these small numerical changes are applied to the GRMHD simulations alone they lead to only slightly different realizations of the flow (as measured by $\phi_{\rm BH}$).
On the other hand, when they are applied to the intermediate-scale MHD simulations from which the GRMHD simulations are restarted, then more significant departures from the fiducial $\phi_{\rm BH}$ are seen, on the order of at least $\sim$ 100\%.  

It should also be noted that the quantitative difference in $\phi_{\rm BH}$ between, say, the $\betaw=10^2$ simulation in R20b and the fiducial $\betaw=10^2$, $a=0$ simulation presented here is only a factor of 2--3.  
That factor of 2--3 just so happens to crucial for determining whether the flow becomes MAD or not.  
Additionally, it is possible that our $\betaw=10^2$ simulations will eventually become MAD if they were run for longer times.  

Finally, we note as well that by choosing a different time in the large-scale MHD wind-fed simulation at which to generate initial conditions for the smaller scale simulations, the new $\betaw=10^2$, $a=0$ simulations do in fact go MAD as shown in \S \ref{sec:alt_real}.  

\subsection{Limitations of the Simulations}

\subsubsection{Electron Temperature Concerns}

We have used a particular method for evolving electron temperatures that calculates total heating from the GRMHD simulations and then gives a fraction of that heat to the electrons via several functions derived from kinetic calculations/simulations.  
There are two open questions regarding this method.  
1) How well does this method capture the irreverisble heating implied by the fluid dynamics? 
2) How well do these local formulations of $f_{\rm e}$ capture the small scale plasma heating physics?
As discussed in \citet{Ressler2017}, 1) is particularly a concern near the disc-jet boundary where there is a gradient of several orders of magnitude in total gas entropy.  
This can lead to an artificial calculation of negative heating in these regions which likely influences the resulting electron temperatures.  One solution would be to use an explicit resistivity so that the dissipative scale could be resolved \citep{Ripperda2019}. Simply going to higher resolution would not fully solve the problem because dissipation would still primarily happen at the (always unresolved) grid scale.
2) is the subject of active research, particularly in the PIC community.  
Ideally there would be a prescription that could determine whether the heating in a particular cell is caused by magnetic reconnection or turbulence and then apply the most appropriate version of $f_{\rm e}$.
Nevertheless, the current implementation of electron thermodynamics represents the state-of-the-art for non-radiative systems and an improvement in the predictive power for GRMHD simulations compared to models that simply assign electron temperatures to the flow.

\subsubsection{No Feedback to Large Scales}
\label{sec:no_feed}
Communication from one simulation to another in our models happens in only one direction: large-to-small scales.  
Although this is likely appropriate for cases in which there is no strong outflow or jet, e.g., our $a=0$ simulations, or where the jet does not reach large radii, e.g., our $ a= 0.9375$, $\beta_{\rm w}=10^2$ simulation, if there \emph{is} a powerful jet reaching to large radii as in our $ a= 0.9375$, $\beta_{\rm w}=10^6$ simulation than the larger scale flow should be affected by its presence.
The jet could alter the flow structure at large radii, changing how the accretion flow is fed, and then change the properties of the jet in a non-linear feedback cycle.
The method used here of connecting multiple simulations across several orders of magnitude in radius can in principle be used to study this as it was for MHD simulations in \citet{Yuan2012} by allowing the smaller scale simulations to influence the larger scale simulations.  
This, however, is beyond the scope of the present work.
In general ``feedback'' from small-to-large scales is itself an active area of research in the broader black hole accretion community \citep{Fabian2012,Heckman2014,Somerville2015}.  Its influence is likely significant for many systems and yet is still not well understood.

\subsection{The Robustness of Our Results}
\label{sec:robust}
Given that there are clear differences in simulations with different $\betaw$ (an unknown physical parameter) and even simulations with the same $\betaw$ and different numerical methods as discussed in \S \ref{sec:R20}, an important topic to address is the robustness of our conclusions.     In other words, which of our results are generic predictions for observationally-informed, wind-fed simulations of Sgr A* and which are specific to the realizations of such a flow presented here?
For example, the question of whether or not the flow goes MAD within a certain amount of time does not seem to be something that our simulations can robustly predict.  
On the other hand, things like plasma $\beta$ on horizon scales and the radial density profiles of fluid quantities do consistently agree across different realizations of our simulations.

Below we list quantities that we can reliably conclude are robustly determined by the model and those that are more strongly effected by the specific parameters and realizations.  
\\
\\
Robust:
\begin{itemize}
  \item Radial profiles of fluid quantities such as $\rho\ \tilde \propto\ r^{-1}$, $T_{\rm g}\ \tilde \propto\ r^{-1}$, and $v_\varphi \approx 0.5 v_{\rm kep}$ across the dynamic range of accretion (see \citealt{Ressler2020} and R20b) 
  \item $\beta$ $\sim$ 2 near the horizon
 \item The presence of significant low-angular momentum gas at all angles 
 \item Total X-ray luminosities at large radii \citep{Ressler2018}
 \item Accretion rates and mm fluxes similar to observations
 \item Accretion flows that tend to align with black hole spin on horizon scales for rapidly spinning black holes during periods of higher magnetic flux (i.e., close to the MAD limit), but that can also be significantly tilted at times of lower magnetic flux
 \item Jets that tend to align with the angular momentum of SANE accretion flows at large radii when the black hole is rapidly spinning
  \item Jets that tend to align with the angular momentum of MAD accretion flows at large radii when the black hole is rapidly spinning 

\end{itemize}

Not Robust:
\begin{itemize}
  \item When (or if) the accretion flow becomes MAD
  \item The orientation of the accretion flow for non-spinning black holes on horizon scales\footnote{Simulations with higher wind magnetization (e.g., $\betaw=10^2$) tend to more often have orientations aligned with the Galactic Centre's clockwise stellar disc, but simulations with lower magnetization tend to have a more random orientation \citep{Ressler2020}.}
  \item The rotation measure and the amount of strong magnetic field present at larger radii
  
\end{itemize}

\section{Conclustions}
\label{sec:conc}
We have presented the results of 3D, wind-fed GRMHD simulations of Sgr A* evolved with electrons as a separate fluid.  
These simulations are initialized from larger-scale MHD simulations of the winds of the $\sim$ 30 WR stars in the Galactic Centre that include known observational constraints.
The result is a set of observationally-motivated simulations that include both MAD and SANE flows, as well as tilted and aligned flows with respect to black hole spin, coupled with both reconnection heating-based and turbulent heating-based electron heating models.
We have analyzed the resulting accretion flows, electron temperatures, relativistic jets, and predicted unresolved emission properties at 230 GHz.  A follow-up work will analyze the resolved images and multi-frequency emission in greater detail.

In principle, there is a single correct time to use in the larger-scale MHD simulations for the purpose of initializing the smaller scale simulations: that which corresponds to the present day (minus the length of time for which the smaller scale simulations are run). 
However, as we demonstrate in Appendix \ref{app:num_tests}, the results of the GRMHD simulations are so sensitive to small differences in the larger-scale simulations that there is a high degree of stochasticity.  
Thus we sample different times in the larger-scale simulations as a proxy for obtaining a better sample of all possible realizations of the flow. 
As a result, our framework cannot result in a specific simulation that we consider fully predictive in all the particular details.
Instead, it produces a suite of simulations that describe different possible states of the Sgr A* accretion flow, some of which may be achieved in reality at different times.
From this we obtain governing principles of the system, determine those features that are robust (\S \ref{sec:robust}), and connect these to observations. 


The main dynamical parameters in our model are black hole spin (magnitude and direction) and $\betaw$, which quantifies the ratio between the magnetic and ram pressure in the stellar winds.  We find that the horizon scale accretion flow reaches $\beta\approx 2$ even for relatively weakly magnetized winds ($\betaw=10^6$) due to compression/flux freezing (Figure \ref{fig:beta_comp_intermediate}). 
Despite this relatively large amount of magnetic field in the resulting flow, however, only some of our simulations go MAD by the end of their runs ($\approx 70{,}000\ M$). 
Depending on which time is chosen from the large-scale MHD wind-fed simulation as initial conditions, our $\betaw=10^2$ simulations can be either SANE or MAD, while the fiducial time chosen for the $\betaw=10^6$ simulations results in a MAD state being reached after $\sim\ 50{, }000\ M$. 
Even the SANE $\betaw=10^2$ simulations, however, have relatively large dimensionless flux values of $\phi_{\rm BH} \gtrsim 20$ (in units where the MAD state corresponds to 50--70).

We find that rapid black hole spin combined with a large amount of magnetic flux can force alignment of the horizon-scale accretion flow with the black hole spin axis (Figures \ref{fig:rho_contour}, \ref{fig:tilt_angles}, and \ref{fig:th_tilt_113}), even when large scale angular momentum of the gas is nearly perpendicular to the black hole spin axis. 
In general, the amount of alignment is correlated with horizon-penetrating magnetic flux (Figure \ref{fig:time_plots}).
When $\phi_{\rm BH}$ is close to the MAD limit, the tilt angles for $r\lesssim 5\ \rg$ are typically $\lesssim\ 30^\circ$ regardless of large scale orientation (Figures \ref{fig:tilt_angles} and \ref{fig:th_tilt_113}). 
When the flux is lower, however, the tilt angles at this radius can be as high as 80$^\circ$.
This is even more true for larger radii, where the orientation of the accretion flow is only strongly affected when magnetic flux is highest, though during times of peak flux the flow can align out to $r \gtrsim 20\ \rg$.  
At other times the black hole spin has a non-negligible effect on the orientation of the gas (unless it is already aligned) but is limited to changes of $\lesssim 30^\circ$.  
In the most extreme case of a MAD flow, the angular momentum of the gas almost completely aligns with black hole spin out to $r \gtrsim 20\ \rg$ and becomes more aligned with time at large radii.

We find that the angular momentum direction of the large-scale accretion flow is the primary factor in determining the jet orientation at large radii (Figure \ref{fig:jet_3D_comp}) for SANE flows.
This direction can even be perpendicular to the black hole spin axis when the jet power is relatively weaker (Figure \ref{fig:rjet}).
For MAD flows, where the jets are more powerful, the black hole spin axis is the primary factor in determining the jet orientation at large radii (Figure \ref{fig:th_tilt_113}).
The further the distance from the black hole, the longer it takes for this alignment to happen.
Despite the significant presence of low angular momentum gas (Figure 
\ref{fig:intermediate}), we find that powerful jets can still efficiently propagate to large radii if the large-scale accretion flow and black hole spin axis are closely aligned and/or the jet power is high (i.e., when the flow is in the MAD state).  
Misaligned, SANE jets do not reach past $\sim$ $800\ \rg$ and frequently regress in length do to their weak power.
It is possible that this result may depend on resolution; higher resolution will reduce numerical dissipation at the jet wall boundary and perhaps allow the jet to propagate further.

Our results suggest that even if Sgr A* is MAD most of the time or in a time-averaged sense, there could still be periods of ``quiescence'' where the horizon-penetrating flux is lower. 
We demonstrated this by considering smaller-scale simulations initialized from different times in the same large-scale MHD wind-fed simulation (Figure \ref{fig:time_plots_113}).
Both of these simulations reached $\beta\sim 2$ near the horizon but only one went MAD by $70{, }000$ $M$.
Our $\betaw=10^6$ simulations also went MAD even though they start with significantly less available supply of magnetic flux (Figure \ref{fig:beta_comp_intermediate}).
Studying the long term evolution of the accretion flow, particularly in regards to whether or not MAD states could be intermittent when fed by random or semi-random magnetic fields, is of great interest (e.g., for application to NIR/X-ray flares) but difficult without simulations that have much longer runtimes.

Electrons in our simulations, on average, become relativistically hot around $r \approx $ 200--400 $\rg$ depending on whether the electron heating model is reconnection-inspired or turbulent heating-inspired (Figure \ref{fig:Te_Tg}).  
The resulting angle-averaged electron-to-total temperature ratio near the horizon is 0.1--0.3.
Our simulations suggest that an appropriate initialization of $T_{\rm e}/T_{\rm g}$ for simulations that start with a torus at $20--100$ $\rg$ is closer to $T_{\rm e}/T_{\rm g}\approx$ 0.2--0.3 than $T_{\rm e}/T_{\rm g}=1$.

In spite of the fact that our simulations have limited free parameters (namely, the magnetization of the winds, black hole spin, and the heating model), a few of our model combinations predict 230 GHz fluxes that fluctuate about the observed range of 2--4.5 Jy for Sgr A* with a factor of $\sim$ 2 variability (Figure \ref{fig:flux_vt}).  These tend to be models with reconnection-based heating where the maximum electron temperatures are lower.   
The other simulations/models tend to have significant portions of their 230 GHz light curves that either overpredict the emission by factors of as much as 10 or underpredict the emission by factors of $\gtrsim$2.  
All simulation/model combinations tend to have higher than observed temporal variability, with RMS variability fractions between 50--60\% (compared with the observed 20--40\%) unless the analysis is restricted to the latter part of the $\betaw=10^6$ simulations when the MAD state has been reached, in which case the variability can be as low as 30--40\%. 
How much the initial transient evolution of the simulations affect the analysis of the variability is impossible to know without running them for much longer times.  
In terms of un-resolved polarization, $\betaw=10^6$ simulations consistently predict a linear polarization fractions that are within the observed limits for all heating models (Figure \ref{fig:LP_vt}).  $\betaw=10^2$ simulations tend to have higher-than-observed LP fractions.

Our model can also produce an RM large enough to explain observations while having a consistent sign for the duration of the GRMHD simulation (Figure \ref{fig:RM_large_scale_beta_1e2_from_113}). 
This requires that the smaller scale simulations be initialized from a time in the large-scale MHD wind simulation where a large-scale, strong, and coherent magnetic field varying only on relatively long timescales is present.  
Such a field is likely required to explain the observed properties of Sgr A*'s RM, which suggests that the WR stellar winds in the Galactic Centre have a relatively high magnetization or that the magnetic field is provided by another source.  
Without such a field, our models predict RMs that are too small in magnitude and change sign on the order of $\sim$ 20 hours (Figures \ref{fig:rm}, \ref{fig:RM_large_scale_beta_1e2}, and \ref{fig:RM_large_scale_beta_1e6}) while the RM of Sgr A* has never been observed to change sign for almost 20 years.


Several of our findings are of interest not just for applications to Sgr A* but for low luminosity black hole accretion in general.
For instance, it is interesting that even though our simulations consistently show $\beta \sim 2$ on horizon scales, only some of our simulations become MAD.  
This shows that SANE flows do not require the gas to be weakly magnetized but can have dynamically important magnetic fields.  Understanding the statistics of how the horizon-penetrating magnetic flux behaves over time and how/if flows alternate between SANE and MAD states requires simulations run for much longer periods of time. 
Our results on tilt/alignment for the jet/accretion disk in SANE and MAD flows are generally consistent with what is found in torus simulations (e.g., McKinney, Tchekhovskoy \& Blandford 2013).
This has important implications for, e.g., inferring black hole spin directions from jet directions. 
Our results can also help motivate better initial conditions for $T_{\rm e}/T_{\rm g}$ in torus simulations of Sgr A*.
Since we find that $T_{\rm e}/T_{\rm g} \sim$ 0.2--0.4 for $r \lesssim 10^3 \rg$, initializing simulations with $T_{\rm e}/T_{\rm g}=1$ at this scale could overpredict the resulting electron temperature.
A more detailed comparison of multi-scale simulations to torus simulations would be valuable for clarifying the similarities and differences between the two.

We are actively working on a follow up work that will involve a more detailed anlysis of the observational predicitons of our simulation including resolved images, spectra, and resolved polarization properties.

\section*{Acknowledgments}
We thank S. Gillessen for kindly providing the data for the observational estimates of the number density used in Figure \ref{fig:gillessen_comp}.
We thank J. Stone for useful comments on the manuscript and the original inspiration for this work.
SMR was supported by the Gordon and Betty Moore Foundation through Grant GBMF7392.  
EQ was supported in part by a Simons Investigator award from the Simons Foundation
This research was supported in part by the National Science Foundation (NSF) under Grant No. NSF PHY-1748958, and by the NSF
through XSEDE computational time allocation TG--AST200005 on Stampede2 and Bridges-2.  This work was made possible by computing time granted by UCB on the Savio cluster.

\section*{Data Availability}
The data underlying this paper will be shared on reasonable request
to the corresponding author.

\bibliographystyle{mn2efix}
\bibliography{restart_grmhd}

\appendix

\section{Sensitivity of GRMHD Simulation to Intermediate MHD Simulations}
\label{app:num_tests}
In this Appendix we explore how sensitive our results are to small algorithmic changes in our simulations.  To do this, we run three additional sets of GRMHD/intermediate scale MHD simulations, varying the reconstruction method and turning electron thermodynamics off and on.  Higher order reconstruction can increase the effective resolution of the simulations and reduce numerical diffusion.  Therefore it is expected that changing the reconstruction method can quantitatively change simulation results to a modest degree.  However, turning electron thermodynamics on or off should ideally have no effect on the simulation results because the electron variables do not back-react on the flow and their equations are solved independently of the rest of the GRMHD/MHD equations.  Inclusion of the extra electron equations thus can only affect the total fluid variable evolution at the roundoff error level by altering the exact order of computations in the algorithm.
In chaotic flows, even such small differences can ultimately lead to divergence of results.

\begin{figure}
\includegraphics[width=0.45\textwidth]{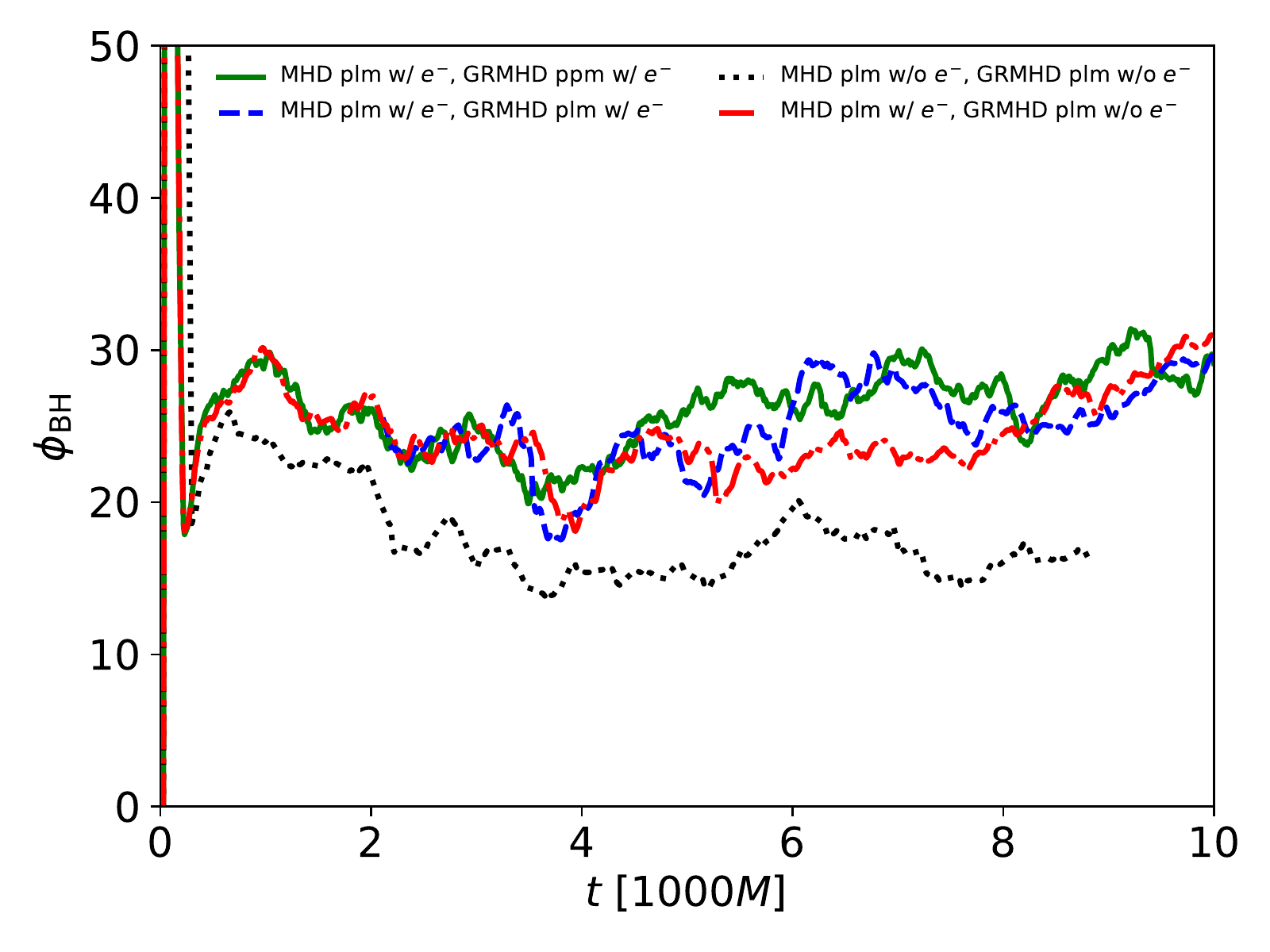}
\caption{Effect of numerical choices on the net flux threading the black hole in $\beta_{\rm w}=10^2$, $a=0$ GRMHD simulations. The four curves represent different combinations of the GRMHD/MHD simulations having plm vs. ppm reconstruction and electron thermodynamics turned on/off.  Small changes in the MHD simulation caused by simply turning off electron thermodynamics lead to significant changes in the flux threading the black hole in the GRMHD simulation.  This is true even though there is no back reaction from the electrons to the total fluid.   }
\label{fig:electron_on_off}
\end{figure}

Figure \ref{fig:electron_on_off} shows $\phi_{\rm BH}$ for four different GRMHD simulations.  One is the fiducial $\beta_{\rm w}=10^2$, $a=0$ simulation described in the main text, with the ppm GRMHD simulation with electrons restarted from a plm MHD simulation with electrons.  Another is a plm GRMHD simulation with electrons restarted from the same plm MHD simulation with electrons.  Another is a plm GRMHD simulation without electrons restarted from the same plm MHD simulation with electrons.  Finally, another is a plm GRMHD simulation with electrons restarted from the plm MHD simulation without electrons. 
The three simulations that start from the same MHD simulation show very similar behavior in $\phi_{\rm BH}$. At early times ($\lesssim 3500 M$) the curves are essentially identical, but $\sim$ 10--20\% relative differences are seen as time progresses in the simulation.  
The simulation that starts from an MHD simulation without electron thermodynamics, however, has flux values lower by up to a factor of $\sim$ 2 from the other three simulations.
This strongly suggests that the resulting $\phi_{\rm BH}$ is much more sensitive to changes in the MHD simulation's numerical parameters at larger scales than to changes in the GRMHD simulation's numerical parameters.
We suspect that this is a generic feature of modeling a large range of scales in chaotic flows via simulations:  small differences at large scales can correspond to non-negligible differences at small scales.

\end{document}